\documentclass[sigconf]{acmart}

\AtBeginDocument{%
  }

\copyrightyear{2025}
\acmYear{2025}
\setcopyright{acmlicensed}
\acmConference[WWW '25] {Proceedings of the ACM Web Conference 2025}{April 28--May 2, 2025}{Sydney, NSW, Australia.}
\acmBooktitle{Proceedings of the ACM Web Conference 2025 (WWW '25), April 28--May 2, 2025, Sydney, NSW, Australia}
\acmISBN{979-8-4007-1274-6/25/04}
\acmDOI{10.1145/3696410.3714915}

\usepackage{soul}
\usepackage{url}
\usepackage{amsthm}
\usepackage{booktabs}
\usepackage{amsfonts}
\usepackage{algorithm}
\usepackage{algorithmic}
\usepackage{amsmath} %,amsfonts,amsthm,bm
\usepackage{color}
\usepackage{colortbl}
\usepackage{xcolor}
\usepackage[english]{babel}
\usepackage{soul}
\usepackage{subfigure}
\usepackage{graphicx}
\newtheorem{theorem}{Theorem}
\newtheorem{lemma}{Lemma}
\newtheorem{Definition}{Definition}

\newtheorem{problem}{Problem}
\newtheorem{proposition}{Proposition}

\usepackage{pifont}
\usepackage{tikz}

\usepackage{enumitem}

\begin{document}

\title{Robust Deep Signed Graph Clustering via Weak Balance Theory}
%\title{Robust Deep Signed Graph Clustering via Weak Balance}

% \affiliation{%

\author{Peiyao Zhao}
\affiliation{%
  \institution{Beijing Institute of Technology}
  \city{Beijing}
  \country{China}}
\email{peiyaozhao@bit.edu.cn}

\author{Xin Li}
\authornote{corresponding author.}
\affiliation{%
  \institution{Beijing Institute of Technology}
  \city{Beijing}
  \country{China}}
\email{xinli@bit.edu.cn}

\author{Zeyu Zhang}
\affiliation{%
  \institution{Huazhong Agricultural University}
  \city{Wuhan}
  \country{China}}
  \email{zzha669@aucklanduni.ac.nz}

\author{Mingzhong Wang}
\affiliation{%
  \institution{University of the Sunshine Coast}
  \city{Queensland}
  \country{Australia}}
\email{mwang@usc.edu.au}

\author{Xueying Zhu}
\affiliation{%
  \institution{Beijing Institute of Technology}
  \city{Beijing}
  \country{China}}
\email{zhuxueying@bit.edu.cn}

\author{Lejian Liao}
\affiliation{%
  \institution{Beijing Institute of Technology}
  \city{Beijing}
  \country{China}}
\email{liaolj@bit.edu.cn}

\renewcommand{\shortauthors}{Zhao et al.}

%% to facilitate the denoising by refining the graph topology,
\begin{abstract}

Signed graph clustering is a critical technique for discovering community structures in graphs that exhibit both positive and negative relationships. We have identified two significant challenges in this domain: i) existing signed spectral methods are highly vulnerable to noise, which is prevalent in real-world scenarios; ii) the guiding principle ``an enemy of my enemy is my friend'', rooted in \textit{Social Balance Theory}, often narrows or disrupts cluster boundaries in mainstream signed graph neural networks. Addressing these challenges, we propose the \underline{D}eep \underline{S}igned \underline{G}raph \underline{C}lustering framework (DSGC), which leverages \textit{Weak Balance Theory} to enhance preprocessing and encoding for robust representation learning. First, DSGC introduces Violation Sign-Refine to denoise the signed network by correcting noisy edges with high-order neighbor information. Subsequently, Density-based Augmentation enhances semantic structures by adding positive edges within clusters and negative edges across clusters, following \textit{Weak Balance} principles. The framework then utilizes \textit{Weak Balance} principles to develop clustering-oriented signed neural networks to broaden cluster boundaries by emphasizing distinctions between negatively linked nodes. Finally, DSGC optimizes clustering assignments by minimizing a regularized clustering loss. Comprehensive experiments on synthetic and real-world datasets demonstrate DSGC consistently outperforms all baselines, establishing a new benchmark in signed graph clustering. 
\end{abstract}

\ccsdesc[500]{Information systems~Social networks}
\ccsdesc[500]{Information systems~Clustering}

\keywords{Representation learning; Balance theory; Signed graph clustering}

%   \includegraphics[width=\textwidth]{Figures/framework v1.pdf}


\maketitle

% \begin{figure}[ht!]
% 	\centering
% 		\subfigure[]{
% 			\includegraphics[width=0.22\textwidth]{figures/conv_pos_intro.pdf}
% 	}
% 		\subfigure[]{
% 			\includegraphics[width=0.22\textwidth]{figures/conv_neg_intro.pdf}
% 	}
% 	\caption{}
% \end{figure}\label{fig:traditional_conv}
% v_1 resp. 空白太, 一个图，间距变小，自己标a, b, 字体：
% Illustration of unclear cluster boundary driven by ``\textit{EEF}''. Red and green solid lines indicate negative and positive edges with arrows indicating message-passing, resp.. (a) Node $i$ aggregates its positive neighbor $1$ (recognized by ``FFF'') and $2$ (recognized by ``\textit{EEF}'') to obtain positive representation vector $\mathbf{Z}^{+}_{i}$ in positive-aggregation mechanism. (b) Node $i$ aggregates its negative neighbors $3$ (recogized by ``\textit{EEF}'' and ``\textit{EFE}'') and $4$ to obtain negative representation vector $\mathbf{Z}^{-}_{i}$ in negative-aggregation mechanism. Aggregating neighbor $2$ causes  thus hard to be clustered correctly. $\mathbf{Z}^{-}_{i}$ exhibits a similar trend.,  on online social platforms, such as \textit{Epinions} and \textit{Slashdot}

\section{Introduction}

\begin{figure}[t]
	\centering
  		\subfigure[Flipping signs]{
			\includegraphics[width=0.22\textwidth]{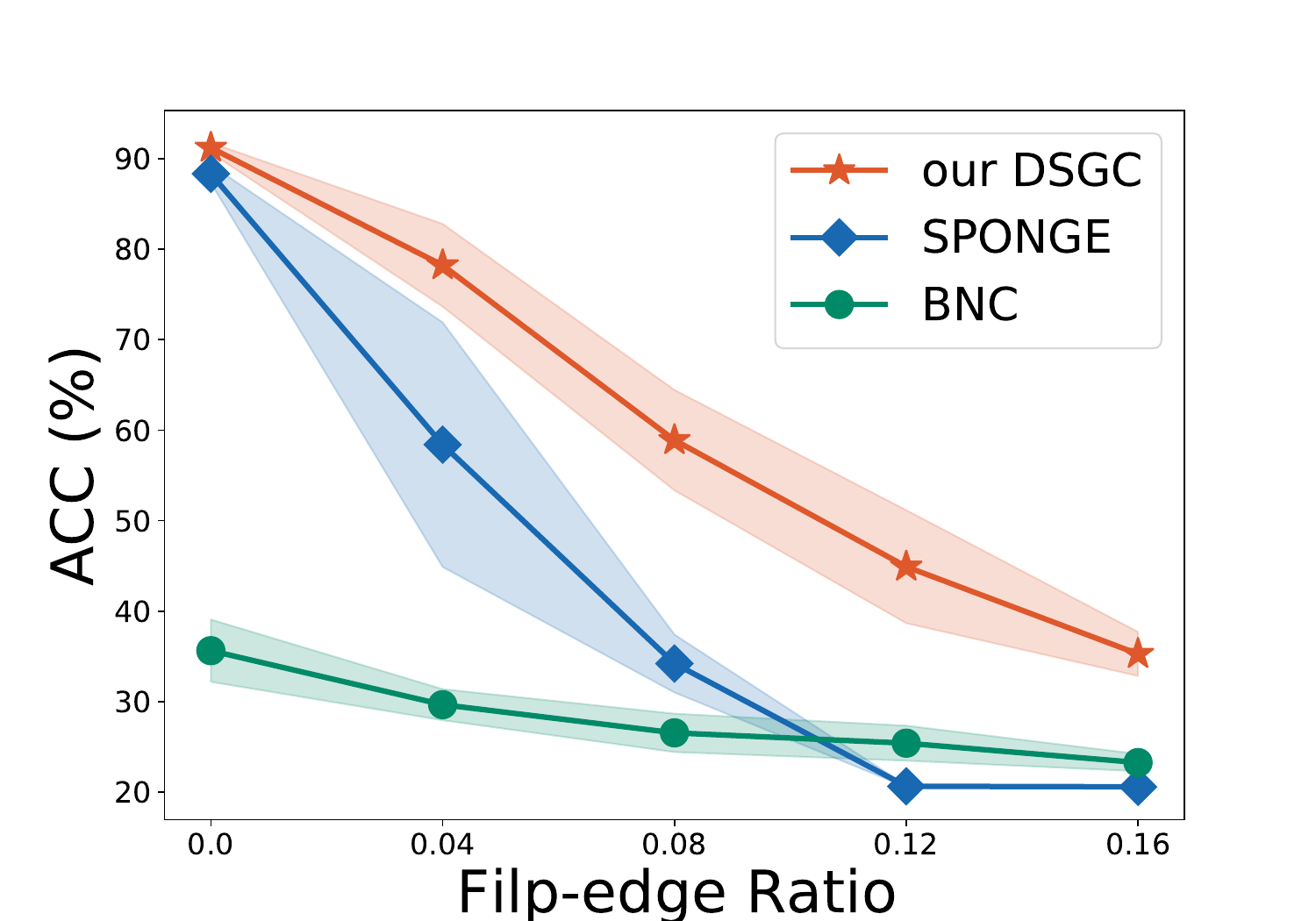}
	}
		\subfigure[Randomly adding edges]{
	\includegraphics[width=0.22\textwidth]{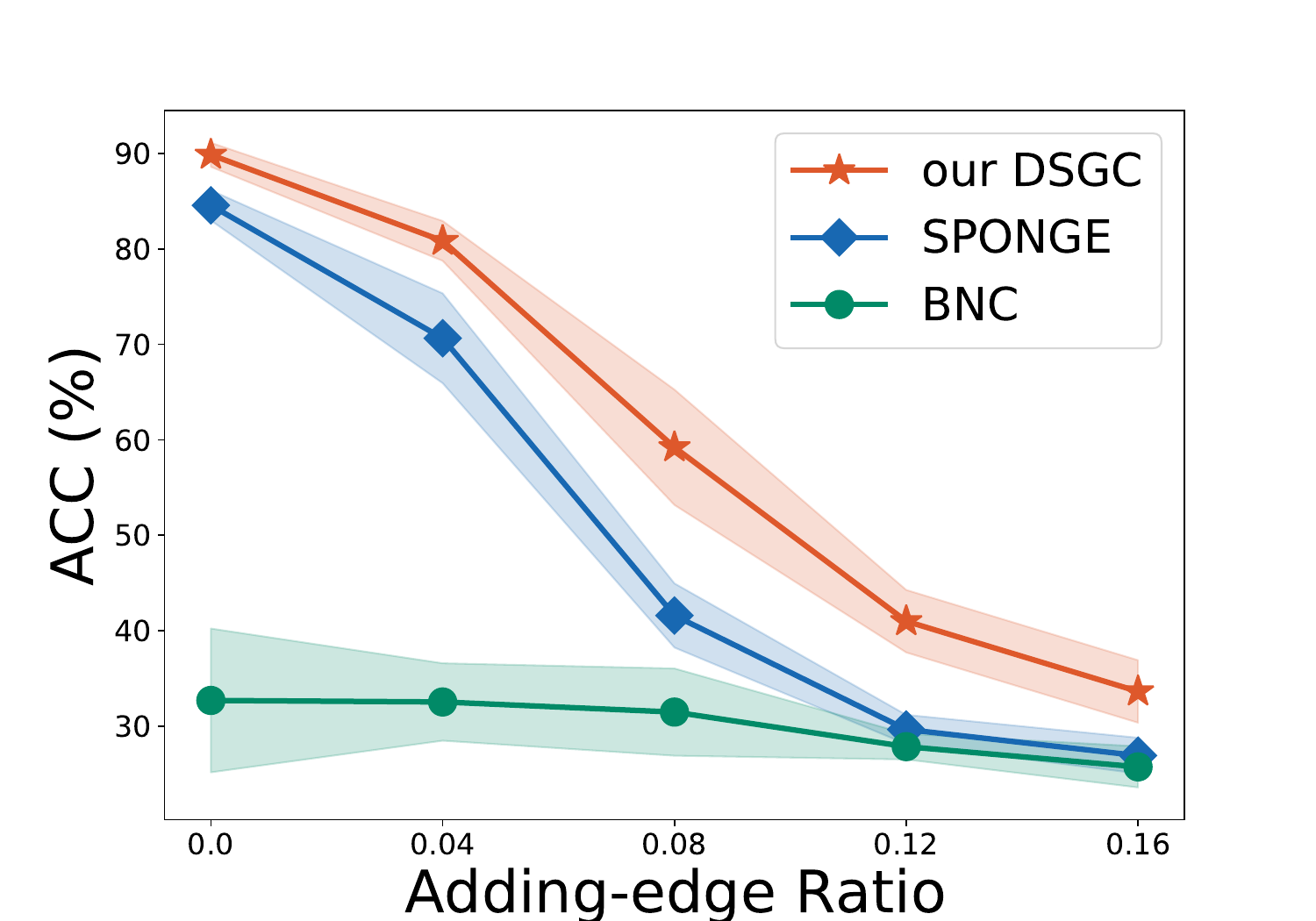}
	}
	\caption{Effects of different perturbations, including flipping signs and randomly adding negative edges, on the clustering performance of popular spectral methods in signed graphs.}
        \label{fig:eta_acc_intro}
\end{figure}

Deep graph clustering has emerged as a pivotal technique for uncovering underlying communities within complex networks. However, existing methods~\cite{PanHLJYZ18, wang2019attributed, LiuTZLSYZ22, bo2020structural, tu2021deep} predominantly target unsigned graphs, which represent relationships solely with ``non-negative'' edges and inherently fail to capture conflicting node interactions, such as friendship versus enmity, trust versus distrust, and approval versus denouncement. Such dynamics are commonplace in social networks and can be effectively modeled by signed graphs that incorporate both positive and negative edges. Although significant work has studied link prediction tasks in deep signed graphs~\cite{li2020learning, huang2019signed, WangATL17, DerrAT18, ZhangLZWHW2023RSGNN, HuangSHC21, zhang2023contrastive}, deep signed graph clustering remains substantially unexplored. In this paper, we aim to develop a deep signed graph clustering method that enhances the robustness of graph representations, facilitating more distinctive clusters and better reflecting the intricate relationships within signed graphs.

Signed graph clustering is broadly applied in the analysis of social psychology~\cite{Mercado2016clustering, davis1967clustering, kunegis2010spectral}, biologic gene expressions~\cite{FujitaSKSPM12,pan2024csgdn}, etc. Recent studies have predominantly focused on spectral methods~\cite{davis1967clustering, Mercado2016clustering, chiang2012scalable, cucuringu2019sponge}, which design various Laplacian matrices specific to a given network to derive node embeddings, aiming to find a partition of nodes that maximizes positive edges within clusters and negative edges between clusters. However, these methods are vulnerable to random noise, a common challenge in real-world scenarios. For instance, on shopping websites, the signed graph encoding user-product preferences often includes noisy edges, typically when customers unwillingly give positive ratings to items in exchange for meager rewards or coupons. 
Fig.~\ref{fig:eta_acc_intro} illustrates the significant impact of noise on signed spectral methods like BNC and SPONGE ~\cite{chiang2012scalable, cucuringu2019sponge}. As perturbation ratios increase, which indicates a higher percentage of randomly flipped edge signs or inserted negative edges in a synthetic signed graph with five clusters, these methods suffer a sharp decline in clustering accuracy. Therefore, denoising the graph structure is essential to enhance robust representation learning in deep signed clustering.

%As illustrated in Fig.~\ref{fig:eta_acc_intro}, signed spectral methods BNC and SPONGE ~\cite{chiang2012scalable, cucuringu2019sponge} are vulnerable to increasing perturbation ratios, exhibiting a sharp decline in signed clustering accuracy. In Fig.~\ref{fig:eta_acc_intro} (a), BNC~\cite{chiang2012scalable} exhibits a small variance due to mode collapse, where nearly all nodes are assigned to few or even a single cluster, leading to rather low accuracy around $20\%$. Therefore, denoising graph structure is essential to enhance robust representation learning for deep signed clustering.

\begin{figure}
    \centering
    \includegraphics[width=0.8\linewidth]{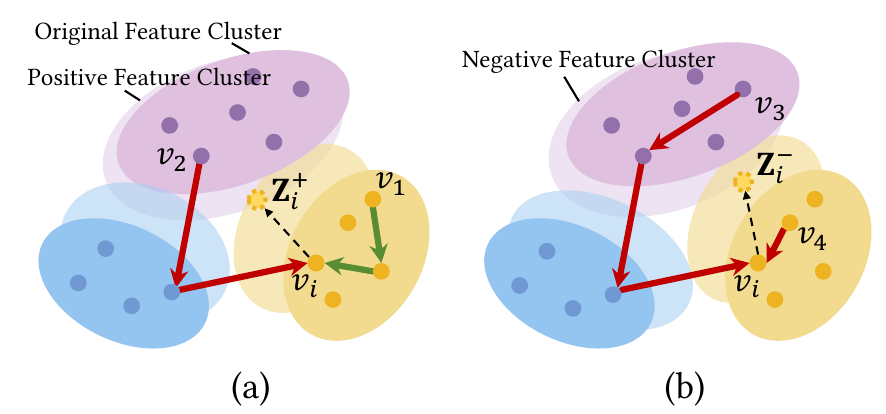}
    \caption{Illustration of ``\textit{an Enemy of my Enemy is my Friend (EEF)}'' narrowing cluster boundaries. Aggregating positive (/ negative) neighbors $1\& 2$ (/ $3 \& 4$) causes $\mathbf{Z}^{+}_{i}$ (/ $\mathbf{Z}^{-}_{i}$) mapped far from its clusters or even cross the boundary, where positive (/ negative) neighbors $2$ (/ $3$) are defined by ``\textit{EEF}''.}
    \label{fig:traditional_conv}
\end{figure}

Furthermore, the investigation on deep signed graph neural networks (SGNNs) reveals that existing SGNNs — which are mostly developed for link prediction~\cite{li2020learning, huang2019signed, WangATL17, DerrAT18, ZhangLZWHW2023RSGNN, HuangSHC21} — do not adapt well to signed clustering. Specifically, mainstream SGNNs models typically leverage principles from the well-established \textit{Social Balance Theory}~\cite{harary1953notion} (or Balance Theory) to design their messaging-passing aggregation mechanisms, including the classical principle ``\textit{an Enemy of my Enemy is my Friend (EEF)}'', ``\textit{a Friend of my Friend is my Friend (FFF)}'', ``\textit{an Enemy of my Friend is my Enemy (EFE)}''. However, ``\textit{EEF}" implies an assumption that a given signed network has only $2$ clusters, which is not directly applied to signed graphs with $K$ ($K>2$) clusters. Specifically, as illustrated in Fig.~\ref{fig:traditional_conv}, ``\textit{EEF}'' can narrow cluster boundaries, leading to more nodes being located at the margins of clusters, which makes it difficult to assign them to the correct clusters and thus results in poor performance. For example, node $v_i$ aggregates its positive neighbor $v_2$ (recognized by ``\textit{EEF}'' but inconsistent to the real semantic relationship in clusters), which causes its positive representation $\mathbf{Z}^{+}_{i}$ mapped closer to the cluster of node $v_2$, thus leading to narrowed or even overlapped cluster boundaries. In contrast, \textit{Weak Balance Theory}~\cite{davis1967clustering} (or Weak Balance), introducing a new principle, ``\textit{an enemy of my enemy might be my friend or enemy}'', can generalize Balance Theory to $K$-way ($K>2$) clustering situation but remains underexplored.

% Therefore, we face two major issues that need to be addressed: \textit{(i)} existing spectral methods are vulnerable to noisy edges; \textit{(ii)} it is essential to invoke suitable principles to design task-oriented SGNNs for the signed clustering task. To address the first issue, we design two rewiring strategies to denoise and augment signed graph structure, including \textit{Violation Sign-Refine}, which can identify and correct noisy edges with long-range neighbor relationships, and \textit{Density-based Augmentation}, which inserts new positive edges to increase positive density within clusters and negative edges to increase negative density across clusters. Such refined graph topology can promote signed encoders to enhance the robustness of node representations. To address the second issue, we invoke a new principle, ``\textit{an enemy of my enemy might be my friend or enemy}'' from a significant yet underexplored \textit{Weak Balance Theory} (or Weak Balance) that generalized Balance Theory to $K$-way ($K>2$) clustering situation. We leverage the principles in Weak Balance to design clustering-specific neighbor aggregation mechanism for enhancing the discrimination among node representations, specifically for nodes with negative edges to widen cluster boundaries. 

To address these challenges, we propose  Deep Signed Graph Clustering (DSGC) for $K$-way clustering, designed to enhance representations' robustness against noisy edges and reduce the impact of the ill-suited principle on cluster boundaries. 
DSGC first introduces the \underline{S}igned \underline{G}raph \underline{R}ewiring module (SGR) in the preprocessing stage for denoising and graph structure augmentation. SGR provides two rewiring strategies, including \textit{Violation Sign-Refine}, which can identify and correct noisy edges with long-range neighbor relationships, and \textit{Density-based Augmentation}, which follows Weak Balance principles to insert new positive edges to increase positive density within clusters and negative edges to increase negative density across clusters. Such refined graph topology can promote signed encoders to enhance the robustness of node representations. DSGC then constructs a clustering-oriented signed neural network that utilizes Weak Balance. This helps design clustering-specific neighbor aggregation mechanism for enhancing the discrimination among node representations, specifically for nodes with negative edges to widen cluster boundaries. 
Finally, DSGC designs a $K$-way clustering predictor that optimizes a non-linear transformation function to learn clustering assignments. This framework is designed to refine the clustering process by correcting noisy edges and enhancing the discriminative capability of node representations, ultimately leading to more accurate clustering outcomes.

%\vspace{4pt}
Overall, our major contributions are as follows:
\begin{itemize}
    \item We develop DSGC, the first Deep Signed Graph Clustering framework,  by leveraging Weak Balance Theory.
    \item We design two graph rewiring strategies to denoise and augment the overall network topology.
    \item We propose a task-oriented signed graph encoder to learn more discriminative representations, particularly for nodes connected by negative edges.
    \item Extensive experiments on synthetic and real-world datasets demonstrate the superiority and robustness of DSGC.
\end{itemize}

\section{Related Work} 
In this section, we succinctly review existing studies for signed graph neural networks and signed graph clustering.

\textbf{Signed Graph Neural Networks (SGNNs)}
, which maps nodes within a signed graph to a low-dimensional latent space, has increasingly facilitated a variety of signed graph analytical tasks, including node classification~\cite{bosch2018node}, signed link prediction~\cite{islam2018signet,xu2019link,zhang2024dropedge,jung2020signed}, node ranking~\cite{chung2013dirichlet,shahriari2014ranking}, and signed clustering~\cite{tzeng2020discovering, kunegis2010spectral, chiang2012scalable,cucuringu2019sponge,he2022sssnet}. 
 % This theory, rooted in psychology, reasons that \textit{"the enemy of my enemy is my friend, the friend of my friend is my friend, the friend of my enemy is my enemy"}, reflecting the intricate relationships within signed networks. signed clustering
Most works of signed graph center around integrating \textit{Social Balance Theory} to signed convolutions into Graph Neural Networks (GNNs). As the pioneering work, SGCN~\cite{derr2018signed} adapts unsigned GNNs for signed graphs by aggregating and propagating neighbor information with Balance Theory. 
% Thereafter, other work has integrated additional social-psychological theories. 
\cite{chen2018bridge} appends the status theory, which is applicable to directed signed networks, interpreting positive or negative signs as indicators of relative status between nodes.
%applies Balance Theory and the status theory to model triangles and "bridge" edges (edges not included in any triangles). The status theory is applicable to directed signed networks, interpreting positive or negative signs as indicators of relative status between nodes. 
SiGATs~\cite{huang2019signed}, which extends Graph Attention Networks (GATs) to signed networks, also utilizes these two signed graph theories to derive graph motifs for more effective message passing. %SNEAntroduces masked self-attentional layers to estimate the importance coefficient of node pairs and then aggregates different importance of neighbors based on Balance Theory.
SiNEs~\cite{wang2017signed} proposes a signed network embedding framework guided by the extended structural balance theory. 
% SGDNET leverages a random walk technique specifically tailored for signed graphs, effectively diffusing hidden node features in line with Social Balance Theory. 
GS-GNN~\cite{liu2021signed} applies a dual GNN architecture that combines a prototype-based GNN to process positive and negative edges to learn node representations. SLGNN~\cite{li2023signed}
%the spectral graph theory and graph signal processing to design a spectral-based signed graph neural network, utilizing 
especially design low-pass and high-pass graph convolution filters to capture both low-frequency and high-frequency information from positive and negative links. 

%\vspace{-8pt}
\textbf{Signed Graph Clustering.}
% The study of signed graph clustering traces its roots back to the 1950s, beginning with Cartwright and Harary's introduction of the balance concept which has been applied to diverse configurations, such as communication networks, sociometric structures, and neural networks. Subsequently, Harary articulated this concept more formally as \textit{Social Balance Theory}~\cite{harary1953notion}.
Its study has its roots in Social Balance Theory~\cite{cartwright1956structural}, which is equivalent to the $2$-way partition problem in signed graphs~\cite{Mercado2016clustering}. 
Building upon this foundational concept, \cite{kunegis2010spectral} propose a signed spectral clustering method that utilizes the signed graph Laplacian and graph kernels to address the $2$-way partition problem.  However, \cite{yang2007community} argues that community detection in signed graphs is equivalent to identifying $K$-way clusters using an agent-based heuristic. The \textbf{\textit{Weak Balance Theory}}~\cite{davis1967clustering} relaxes Balance theory to enable $K$-way clustering. Following ~\cite{kunegis2010spectral}, ~\cite{chiang2012scalable} proposed the ``Balanced Normalized Cut (BNC)'' for $K$-way clustering, aiming to find an optimal clustering assignment that minimizes positive edges between different clusters and negative edges within clusters with equal priority. SPONGE~\cite{cucuringu2019sponge} transforms this discrete NP-hard problem into a continuous generalized eigenproblem and employs LOBPCG~\cite{knyazev2001toward}, a preconditioned eigensolver, to solve large positive definite generalized eigenproblems. In contrast to the above $K$-way complete partitioning, \cite{tzeng2020discovering} targets detecting $K$ conflicting groups, allowing other nodes to be neutral regarding the conflict structure in search. This conflicting-group detection problem can be characterized as the maximum discrete Rayleigh's quotient problem.

While GNNs have been extensively applied to unsigned graph clustering~\cite{PanHLJYZ18,wang2019attributed,LiuTZLSYZ22,bo2020structural,tu2021deep}, their adoption in signed graph clustering remains overlooked. A notable exception is the Semi-Supervised Signed NETwork Clustering (SSSNET)~\cite{he2022sssnet}, which simultaneously learns node embeddings and cluster assignments by minimizing the clustering loss and a Cross-Entropy classification loss. In contrast, our work develops an unsupervised method for signed graph clustering, eliminating the reliance on ground truth labels.

\section{Preliminaries}
% This section introduces the mathematical notations and definitions essential for discussing Social Balance Theory, Weak Balance Theory, and the problem of deep signed graph clustering. 

\subsection{Notations}~\label{notations}
We denote an undirected signed graph as $\mathcal{G}=\{\mathcal{V},\mathcal{E}, \mathbf{X}\}$, where $\mathcal{V}=\{v_1,v_2,\dots,v_n\}$ is the set of nodes, $\mathcal{E}$ is the set of edges, and $\mathbf{X}\in\mathbb{R}^{\left | \mathcal{V} \right | \times d_0}$ is the $d_0$-dimensional node attributes. Each edge $e_{ij}\in \mathcal{E}$ between $v_i$ and $v_j$ can be either positive or negative, but not both. $\mathbf{A}$ is the adjacency matrix of $\mathcal{G}$, where $\mathbf{A}_{ij}=1$ if
$v_i$ has a positive link to $v_j$; $\mathbf{A}_{ij}=-1$ if $v_i$ has
a negative link to $v_j$; $\mathbf{A}_{ij}=0$ otherwise. The signed graph is conceptually divided into two subgraphs sharing the common vertex set $\mathcal{V}$: $\mathcal{G}=\{\mathcal{G}^{+},\mathcal{G}^{-}\}$, where $\mathcal{G}^{+}=\{\mathcal{V}, \mathcal{E}^+\}$ and $\mathcal{G}^{-}=\{\mathcal{V}, \mathcal{E}^-\}$ contain all positive and negative edges, respectively.
Let $\mathbf{A}^{+}$ and $\mathbf{A}^{-}$ be the adjacency matrices of $\mathcal{G}^{+}$ and $\mathcal{G}^{-}$ with $\mathbf{A}=\mathbf{A}^{+}-\mathbf{A}^{-}$, where $\mathbf{A}^{+}_{ij}=max(\mathbf{A}_{ij},0)$ and $\mathbf{A}^{-}_{ij}=-min(\mathbf{A}_{ij},0)$. 

\subsection{Relaxation of Social Balance}~\label{sec: Relaxation of Social Balance}
Balance and Weak Balance Theories, essential for signed graph clustering, are briefly explained here; more details are in App. ~\ref{app:sociological_theories}.

\textbf{Balance Theory}~\cite{harary1953notion} 
%defines a signed network is balanced if and only if its node set can be partitioned into two mutually exclusive subsets, such that all edges within the same subset are positive and all edges between two subsets are negative. Balance Theory 
consists of four fundamental principles: ``\textit{the friend of my friend is my friend}'', ``\textit{the enemy of my friend is my enemy}'', ``\textit{the friend of my enemy is my enemy}'', and ``\textit{the enemy of my enemy is my friend (\textit{EEF})}''. A signed network is balanced if it does not violate these principles. Theoretically, the Balance Theory is equivalent to $2$-way clustering on graphs~\cite{Mercado2016clustering}.

%For example, triads with an even number of negative edges are balanced, as shown by the first two triads in Fig.~\ref{fig:triads_theories}(a), which have 0 and 2 negative edges, respectively. These principles are traditionally applied to $2$-way clustering.
%TODO: 小心使用 vspace，有的会议会认定不符合格式要求。待查
%\vspace{4pt}
\textbf{Weak Balance Theory}~\cite{davis1967clustering} relaxes Balance Theory to accommodate $K$-way clustering, % It allows $K\in \mathbb{N}^{+}$ disjoint sets, such that positive edges exist only within clusters, and negative edges exist only between clusters. 
by replacing the ``\textit{EEF}'' principle with ``\textit{the enemy of my enemy might be my enemy (EEE)}''. 
%Accordingly, the first three triads in Fig.~\ref{fig:triads_theories}(b) are considered weakly balanced. 
This principle allows nodes in a triangle to belong to three different clusters, e.g., the blue triangle in Fig.~\ref{app:balance_theory} (b), thus relaxing Social Balance Theory.
% (a) $2$-way clustering. (b) $K$-way ($K=3$) clustering.
% The blue triangle proves the principle "\textit{the enemy of my enemy might be my enemy}"
% The orange triangle proves the principle "\textit{the enemy of my enemy is my friend}"
% Red and green lines indicate negative and positive edges, respectively. Black dashed lines denote the boundaries between clusters.
% \vspace{4pt}
% \noindent\textbf{Formulating Balance Theory and Weak Balance Theory}\\
The partition $\{\mathcal{C}_1,\dots,\mathcal{C}_K\}$ of a signed graph $\mathcal{G}$ satisfying either theory can be uniformly defined as the following conditions: 
\begin{equation}
    \begin{cases}
       \textbf{A}_{ij}>0 & (e_{ij}\in\mathcal{E}) \cap(v_i \in \mathcal{C}_k) \cap (v_j \in \mathcal{C}_k)\\
       \textbf{A}_{ij}<0 & (e_{ij}\in\mathcal{E}) \cap(v_i \in \mathcal{C}_k) \cap (v_j \in \mathcal{C}_l) (k\ne l)\\
    \end{cases},
\end{equation}
where $\mathbf{A}_{ij}$ is the weight of edge $e_{ij}$ and $0<k,l<K$.
% Weak Balance Theory in Theorem \ref{Weak Balance Theory} generalized \textit{Social Balance} to $K$-way ($K>2$) clustering, relaxing the node relationship by \textit{allowing} "the enemy of my enemy is still my enemy". 
% However, the "\textit{EEE}" principle is specific to $K$-clusterable ($K>2$) networks (e.g., Fig.~\ref{balance theory}(b)) and does not appear in $2$-clusterable systems (Fig.~\ref{balance theory}(a)). 

% Recent literature~\cite{li2020learning,huang2019signed,li2020learning, WangATL17, DerrAT18, ZhangLZWHW2023RSGNN, HuangSHC21} has primarily leveraged Social Balance Theory principles to improve node representations for signed graphs, potentially overlooking the broader applicability of Weak Balance Theory in datasets with more than $2$ antagonistic groups, especially when explicit labels are lacking. Our work aims to fully explore Weak Balance Theory and its principles in the design of a signed graph encoder for $K$-way clustering.
% For example, the positive representation is partially obtained by aggregating the enemies ($v_k$) of the central node ($v_i$)'s enemy ($v_j$), which will decrease their discrimination in the latent space.

\subsection{Problem Definition}
This paper aims to leverage the capabilities of deep representation learning to enhance robust graph signed clustering. Unsupervised 
\textbf{Deep Signed Graph Clustering} is formally defined below.
\begin{problem}
Given a signed graph $\mathcal{G}=\{\mathcal{V}, \mathcal{E}, \mathbf{X}\}$, deep signed graph clustering is to train a function $f(\mathbf{A}, \mathbf{X})\longrightarrow \mathbf{Z}$ that transforms each node $v\in \mathcal{V}$ into a low-dimensional vectors $\mathbf{Z}_v\in \mathbb{R}^d$. It aims to optimize a partition to divide all nodes $\{\mathbf{Z}_i\}_{i=1}^{{|\mathcal{V}|}}$ into $K$ disjoint clusters $\mathcal{V}=\mathcal{C}_1 \cup \dots \cup \mathcal{C}_K$, by minimizing a signed clustering loss objection that makes as many as positive edges exist within clusters and as many as negative edges exist across clusters.
% the adjacency matrix $\mathbf{A}$ and node attributes $\mathbf{X}$ 
% derive a function $f(\mathbf{A}, \mathbf{X})\longrightarrow \mathbf{Z}$ that transforms the adjacency matrix $\mathbf{A}$ and node attributes $\mathbf{X}$ into a set of low-dimensional vectors $\mathbf{Z}_v\in \mathbb{R}^d$ for each node $v\in \mathcal{V}$, where $\mathbf{Z}\in\mathbb{R}^{|\mathcal{V}|\times d}$ represents the node embeddings. The function $f$ could be neural networks, such as SGNNs. The ultimate goal is to This clustering should be achieved by minimizing a designed signed clustering loss without relying on labels. 
\end{problem}

\section{Methodology}

\begin{figure*}[t]
    \centering
    \includegraphics[width=1\linewidth]{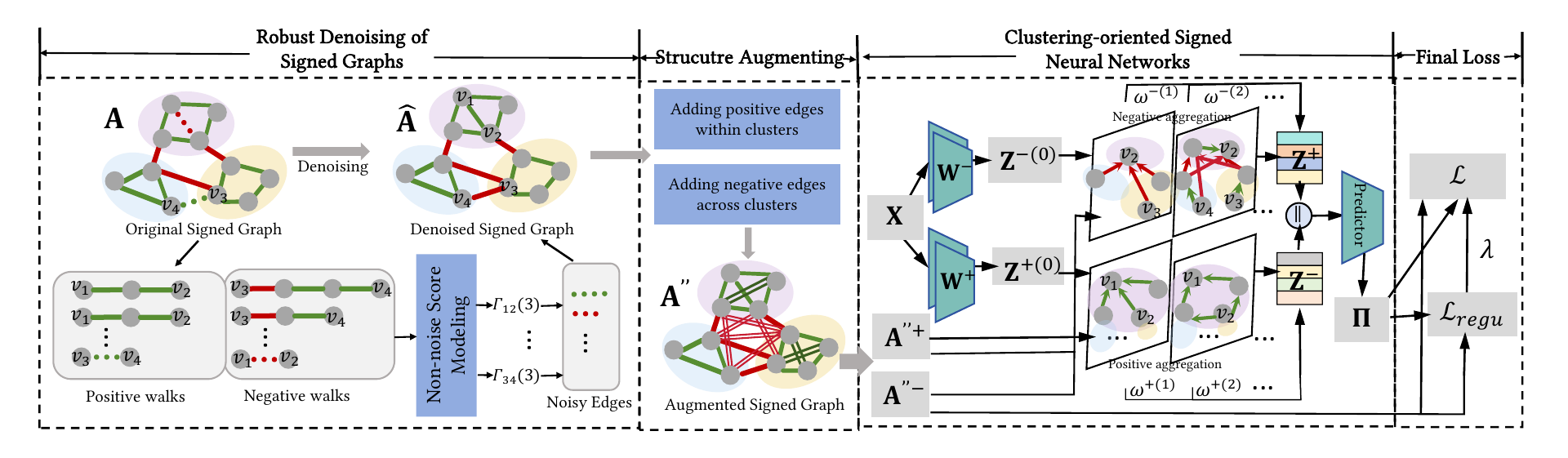}
    \caption{The overall framework of DSGC. The Violation Sign-Refine first computes non-noise scores to correct the signs of noisy edges. Then, the Density-based Augmentation adds positive edges within clusters and negative edges across clusters. These two rewiring methods generate a new adjacency matrix with reduced noise and enhanced semantic structures. Thereafter, clustering-specific signed convolutional networks can be trained by minimizing the differential clustering loss for learning and strengthening the discrimination among node representations linked negatively.}
    \label{fig:framework}
\end{figure*}

%This section outlines the Deep Signed Graph Clustering (DSGC) model (Fig.~\ref{fig:framework}). The model begins with graph rewiring, utilizing two methods: Violation Sign-Refine to correct noisy edges and Density-based Augmentation to add new edges based on connectivity patterns. Subsequently, a graph encoder, guided by Weak Balance principles, processes the rewired graph to generate discriminative node representations. These representations are then transformed via a non-linear function into clustering assignment vectors, optimized through a regularized signed clustering loss. Finally, clusters are inferred by identifying the maximum assignment probability for each node. Detailed explanations of each module are presented in the following subsections.
As illustrated in Fig.~\ref{fig:framework}, DSGC consists of 4 major components, including Violation Sign-Refine and Density-based Augmentation for graph rewiring, signed clustering encoder, and cluster assignment.

%methods to re-edit edges to generate a new adjacency matrix. The Violation Sign-Refine approach can recognize and correct the sign of noisy edges followed by a Density-based Augmentation approach that adds the new positive (resp. negative) edges between two reachable nodes through a positive (resp. negative) walk. With the denoised adjacency matrix and node features as the input, our graph encoder integrates Weak Balance principles to learn the positive and negative node representations, which are concatenated as final representations. Its important design is abandoning the "\textit{EEF}" principle in Balance Theory and the minus sign for the negative adjacency matrix to further separate far nodes connected negatively for learning clearer clustering boundary. The non-linear transformation maps node representation to the probability space, producing the clustering assignment vectors optimized by minimizing a regularized signed clustering loss. Finally, the clustering results are inferred by locating the maximal assignment probability. More details of the design are presented in the following sections.
 % 这个图要画满，, trim=0 30 0 50, clip

\subsection{Signed Graph Rewiring}
%In ideal scenarios, negative edges in balanced signed graphs (Figure \ref{balance theory}) act as the bridges connecting distinct clusters. 
In real-world signed graphs, noisy edges(violations)\textemdash negative edges within clusters and positive edges across clusters\textemdash can disrupt ideal clustering structures. %The \textbf{\textit{violations}} are illustrated in Figure \ref{fig:framework}. 
To address this, %Therefore, denoising is crucial for enhancing the signed graph clustering performance. 
we propose two graph rewiring methods to enhance clustering integrity: Violation Sign-Refine (VS-R), which corrects the signs of violated edges to align negative and positive edges with the expected inter-cluster and intra-cluster relationships; and Density-based Augmentation (DA), which adds new edges based on long-range interaction patterns to reinforce message passing. Both methods leverage Weak Balance and are used as preprocessing steps to denoise and augment the initial graph topology\textemdash specifically, the message-passing matrix. 
% The effectiveness of these rewiring techniques is subsequently evaluated based on the accuracy of node representations in downstream learning tasks.

\subsubsection{Violation Sign-Refine} To address noisy edges, we utilize high-order neighbor interactions to correct their signs. Based on Weak Balance, we first adapt the definitions of positive and negative walks for $K$-way clustering.
% \begin{figure}[t]
%     \centering
%     \includegraphics[width=0.35\textwidth]{figures/bridge_violations.pdf}
%     \caption{The illustration of effective friends and enemies. By Eq. \eqref{effectiveness score}. The relationship between nodes $v_1$ and $v_2$ can be identified as \textit{effective friends}, and so are $v_3$ and $v_4$. The relationship between $v_{5}$ and $v_6$ are identified as \textit{effective enemies}.}
%     \label{fig:bri_vio}
% \end{figure}
Following Social Balance Theory, \cite{diaz24signed} defines a positive walk as one containing an even number of negative edges and a negative walk as one containing an odd number of negative edges. However, they are not suitable for $K$-way clustering due to the uncertainty brought by the ``\textit{the enemy of my enemy might be my enemy or my friend}'' principle of Weak Balance. We formally redefine positive and negative walks as follows.
\begin{Definition}~\label{pos_neg_walk}
    A walk of length $l\in \mathbb{N}^{+}$ connecting nodes $v_i$ and $v_j$ is positive if all its edges are positive; it is negative when it contains exactly one negative edge and all other edges are positive. 
\end{Definition}

%Dfn.~\ref{pos_neg_walk} aligns with "\textit{FFF}", "\textit{FEE}""\textit{EFE}" while excluding "\textit{EEF}". 
Since violations are sparse in graphs, we assume that leveraging higher-order information from longer-range neighbors helps revise the signs of violated edges. Lemma~\ref{lem:dis_nodeij_length} specifies the non-noise score between $v_i$ and $v_j$ w.r.t. the $l$-length positive and negative walks.

\begin{lemma}~\label{lem:dis_nodeij_length}
    For $v_i$, $v_j\in \mathcal{V}$ in a signed graph $\mathcal{G}=\{\mathcal{V},\mathcal{E}, \mathbf{X}\}$, let $\mu_l^{+}(i,j)$ and $\mu_l^{-}(i,j)$ be the number of positive and negative walks with length $l$ connecting $v_i$ and $v_j$, respectively. Then, $\forall a\in \mathbb{N}$, 
    \begin{align}
        &\mu_l^{+}(i,j)-\mu_l^{-}(i,j)=(\mathbf{A}^{+})^{l}_{ij}-\sum_{a=0}^{l-1}((\mathbf{A}^{+})^{a}\mathbf{A}^{-}(\mathbf{A}^{+})^{l-1-a})_{ij}.
        % & \nonumber
    \end{align}
\end{lemma}
% Lemma~\ref{lem:dis_nodeij_length} formulates the difference between the number of positive and negative walks of $l$-length connecting node $v_i$ and $v_j$. 

If we consider all walks up to length $L'$, the non-noise score of the connection between $v_i$ and $v_j$ can be defined:
% If the number of positive walks connecting $v_i$ and $v_j$ exceeds the number of negative walks, the nodes $v_i$ and $v_j$ are \textit{effective friends}; otherwise, the two nodes are \textit{effective enemies}. Mathematically, it is as follows：
\begin{align}~\label{effectiveness score}
 &\Gamma_{ij}(L')=\sum_{l=1}^{L'}\alpha_{l}(\mu_l^{+}(i,j)-\mu_l^{-}(i,j)),\\
 &\text{where} \quad \alpha_l=\begin{cases} 1, & l=1 \\ 1/(l!), & 1 < l < L'\\ 1-\sum_{l=2}^{L'-1}1/(l!), & l = L'
 \end{cases}.
\end{align}
Here, $\alpha_l$ decreases with $l$, indicating that shorter walks have more influence. $\Gamma$ is utilized to correct violations as $\Gamma_{ij}$ extracts high-order information from neighbors of $v_i$ and $v_j$ within $L'$-hop. With $\Gamma_{ij}$, we obtain a refined adjacency matrix $\hat{A}$ via the following rules:
\begin{align}\label{new_Aij}
    \hat{\mathbf{A}}_{ij} =\begin{cases} 
     1, & \Gamma_{ij} > \delta^{+} \\
     \mathbf{A}_{ij}, & \delta^{-} \le \Gamma_{ij} \le \delta^{+}\\
     -1, & \Gamma_{ij} < \delta^{-}
     \end{cases},   
\end{align}
where $\delta^{+}>0$ and $\delta^{-}<0$ are two thresholds. $v_i$ and $v_j$ are considered \textit{effective friends} when $\Gamma_{ij}>\delta^{+}$, indicating a positive edge ($+$); $v_i$ and $v_j$ are considered \textit{effective enemies} when $\Gamma_{ij}<\delta^{-}$, indicating a negative edge ($-$); otherwise, the original adjacency entries in $\mathbf{A}$ retains. The magnitude $\left | \Gamma_{ij} \right | $ represents the confidence level of two nodes being \textit{effective friends} or \textit{enemies}. A larger (resp. smaller) $\left | \Gamma_{ij} \right | $ represents a stronger (resp. weaker) positive or negative relationship between $v_i$ and $v_j$. This method refines the adjacency matrix by reinforcing accurate relational signals and reducing the impact of noisy edges, thereby facilitating more effective clustering. 

\subsubsection{Density-based Augmentation} Following the noise corrections made by VS-R, the revised graph, denoted as $\hat{\mathcal{G}}=\{\mathcal{V}, \hat{\mathcal{E}},\mathbf{X}\}$, is processed through Density-based Augmentation to increase the density of positive edges within clusters and negative edges between clusters. The revised adjacency matrices for positive and negative edges, $\hat{\mathbf{A}}^{+}$ and $\hat{\mathbf{A}}^{-}$, are augmented as below:
\begin{equation}~\label{eq:clean_graph_pre}
    \mathbf{A}'^{+} = (\hat{\mathbf{A}}^{+})^{m^{+}};\quad \mathbf{A}'^{-} = \sum_{a=0}^{m^{-}}(\hat{\mathbf{A}}^{+})^{a}\hat{\mathbf{A}}^{-}(\hat{\mathbf{A}}^{+})^{m^{-}-a},
\end{equation}
where $m^{+}$ and $m^{-}$ are scalar hyper-parameters indicating the extent of augmentation. The augmented adjacency matrices are:
\begin{equation}
    \mathbf{A}''^{+}=\begin{cases}
      1, & \mathbf{A}'^{+}_{ij}>0, \ i\ne j\\
      0, & \mathbf{A}'^{+}_{ij}=0, \ i\ne j \\
      0, & \mathbf{A}'^{+}_{ij}, \ i=j \\
    \end{cases};\quad
\mathbf{A}''^{-}=\begin{cases}
      1, & \mathbf{A}'^{-}_{ij}>0, \ i\ne j\\
      0,  & \mathbf{A}'^{-}_{ij}=0, \ i\ne j \\
      0, & \mathbf{A}'^{-}_{ij}, \ i=j \\
    \end{cases}.
\end{equation}

If $m^{+}=1$ (resp. $m^{-}=0$), no augmentation is performed on $\hat{\mathbf{A}}^{+}$ (resp. $\hat{\mathbf{A}}^{-}$). For $m^{+}>1$, it adds a positive edge between any two nodes connected by a $m^{+}$-length positive walk (Dfn.~\ref{pos_neg_walk}). For $m^{-}>0$, it adds a negative edge between any two nodes connected by a $(m^{-}+1)$-length negative walk. 
% For a larger $m$ (resp. $s$), the method puts more emphasis on long-distance nodes, i.e., global information. For a small $m$ (resp. $s$), this means the method amplifies local information. 
%Therefore, both the positive-edge density within clusters and the negative-edge density across clusters can be increased, which is particularly effective for a signed graph with no or few violations.
This strategy effectively enhances the clustering potential by reinforcing intra-cluster connectivity with positive edges and inter-cluster separations with negative edges. It is particularly effective for a signed graph with few violations.
% Alternatively, we propose the diffusion-based rewiring method for the signed graph in a $K$-way clustering setting. More details and comparisons are in the Appendix. 

\subsection{Signed Clustering Encoder}
Signed Graph Convolution Network, our signed graph encoder in DSGC, is tailored for $K$-way clustering. %We first point out that the conventional principle of Social Balance, "\textit{the enemy of my enemy is my friend (EEF)}" reduces the representation discrimination among nodes linked negatively in Section~\ref{experiment_encoder}. 
It leverages Weak Balance Theory principles to learn discriminative node representations %, in particular nodes connected by negative edges. Since negative edges serve as critical cues for identifying cluster boundaries and dividing clusters. 
that signify greater separation between nodes connected by negative edges and closer proximity between those by positive edges~\footnote{This goal is often reflected in the loss function designs in previous work~\cite{derr2018signed,huang2019signed,li2023signed,li2020learning,liu2021signed} for link prediction. However, its importance has been overlooked in signed encoders.}. 

% So we only adopt the principles implied in Weak Balance to design our encoder for propagating and aggregating neighbors' embeddings.

% we aggregate neighbors' embeddings to obtain the positive and negative representations of the central node according to the principles implied in Weak Balance. Importantly,  abandoning the conventional principle of Social Balance, ''an enemy of one's enemy is a friend" can force the two-hop enemy neighbor and the central node to have dissimilar representations under the $K$-way clustering setting. 

% This signed graph encoder entails the following challenges: (1) How to design the convolutions allowing "the enemy of one's enemy may be still an enemy"? (2) How to enable the encoder to increase the distinctive between two positive nodes and decrease the distinctive between two negative nodes?

% In this work, we propose to address those specific challenges. 

Based on the rewired graph topology defined by $\mathbf{A}''^{+}$ and $\mathbf{A}''^{-}$, we first introduce self-loops to each node using $\Tilde{\mathbf{A}}^{+}=\mathbf{A}''^{+}+\epsilon^{+}\mathbf{I}$, $\Tilde{\mathbf{A}}^{-}=\mathbf{A}''^{-}+\epsilon^{-}\mathbf{I}$, where $\mathbf{I}$ is the identity matrix and $\epsilon^{+}$ and $\epsilon^{-}$ are the balance hyperparameters. The adjacency matrices are then normalized as follow: $\bar{\mathbf{A}}^{+}=(\Tilde{\mathbf{D}}^{+})^{-1}\Tilde{\mathbf{A}}^{+}$ and $\bar{\mathbf{A}}^{-}=(\Tilde{\mathbf{D}}^{-})^{-1}\Tilde{\mathbf{A}}^{-}$, where $\Tilde{\mathbf{D}}^{+}$ and $\Tilde{\mathbf{D}}^{-}$ are diagonal degree matrices with $\Tilde{\mathbf{D}}^{+}_{ii}=\sum_{j}\Tilde{\mathbf{A}}_{ij}^{+}$ and $\Tilde{\mathbf{D}}^{-}_{ii}=\sum_{j}\Tilde{\mathbf{A}}_{ij}^{-}$. We learn $d$-dimensional positive and negative embeddings, $\mathbf{Z}^{+}_i$ and $\mathbf{Z}^{-}_i$, for each node $v_i\in \mathcal{V}$, and concatenate them as the final node representation: $\mathbf{Z}_i=\text{CONCAT}(\mathbf{Z}^{+}_i, \mathbf{Z}^{-}_i)$, where $\mathbf{Z}_i^{+}\in \mathbb{R}^{ 1 \times d }$ and $\mathbf{Z}_i^{-}\in \mathbb{R}^{1 \times d }$ %can be obtained by adaptively combining the intermediate positive and negative embeddings of all layers:
are computed through layers of our signed graph convolution network:
\begin{equation}
    \mathbf{Z}^{+}_i = \sum_{l=0}^{L}\omega^{+(l)} \mathbf{Z}_i^{+(l)}, \ \mathbf{Z}^{-}_i = \sum_{l=0}^{L}\omega^{-(l)} \mathbf{Z}_i^{-(l)}.~\label{pos_neg_embeddings}
\end{equation}
$\omega^{+(l)}$ and $\omega^{-(l)}$, shared by all nodes, are layer-specific trainable weights that modulate the contribution of different convolution layers to the final node representation. $L$ is the number of layers in the neural network. %A Larger $L$ captures increasingly more global information. 
This design allows the encoder to leverage information from different neighborhood ranges. The intermediate representations of all nodes, $\mathbf{Z}^{+(l)} \in \mathbb{R}^{ |\mathcal{V}| \times d }$ and $\mathbf{Z}^{-(l)} \in \mathbb{R}^{ |\mathcal{V}| \times d }$, can be obtained as %the proposed non-parameters signed convolution layers:
\begin{align}
    & \mathbf{Z}^{+(l)} = (\bar{\mathbf{A}}^{+})^{l}\mathbf{Z}^{+(0)},~\label{pos_agg} \\ 
    & \mathbf{Z}^{-(l)} = \sum_{b=0}^{l-1}(\bar{\mathbf{A}}^{+})^{b}(-\bar{\mathbf{A}}^{-})(\bar{\mathbf{A}}^{+})^{l-1-b}\mathbf{Z}^{-(0)},~\label{neg_agg} 
\end{align}
where the superscript $(l)$ and $l$ denote the layer index and power number, respectively. The initial node embeddings, $\mathbf{Z}^{+(0)}\in \mathbb{R}^{\left |\mathcal{V} \right | \times d}$ and $\mathbf{Z}^{-(0)}\in \mathbb{R}^{\left |\mathcal{V} \right |\times d}$, are derived from the input feature matrix $\mathbf{X}\in \mathbb{R}^{\left | \mathcal{V} \right |\times d_0}$ by two graph-agnostic non-linear networks:
\begin{align}
    & \mathbf{Z}^{+(0)}=\sigma(\mathbf{X}\mathbf{W}_{0}^{+})\mathbf{W}_{1}^{+},\\
    & \mathbf{Z}^{-(0)}=\sigma(\mathbf{X}\mathbf{W}_{0}^{-})\mathbf{W}_{1}^{-},
\end{align}
where $\sigma$ is the $ReLU$ activation function. $\mathbf{W}_{0}^{+}\in \mathbb{R}^{d_0 \times d}$ and $\mathbf{W}_{1}^{+}\in \mathbb{R}^{d\times d}$ are the trainable parameters of the positive network; $\mathbf{W}_{0}^{-}\in \mathbb{R}^{d_0 \times d}$ and $\mathbf{W}_{1}^{-}\in \mathbb{R}^{d \times d}$ are that of the negative network. We claim that the positive aggregation function, Eq.~\eqref{pos_agg}, can pull the nodes linked by positive walks, thus reducing the intra-cluster variances. Meanwhile, the negative aggregation function, Eq.~\eqref{neg_agg}, can push nodes linked by negative walks, thus increasing the inter-cluster variances. We also investigate Weak Balance principles implied in Eq.~\eqref{pos_agg} and Eq.~\eqref{neg_agg}, as well as the effect of the minus sign ``-'' in the term $(-\bar{\mathbf{A}}^{-})$ to nodes representations linked negatively and the clustering boundary in App.~\ref{app:analyze_encoder}. 

\vspace{-3pt}
\subsection{$K$-way Signed Graph Clustering}
% Recall that the objective of signed clustering is to partition the network into $K$ clusters such that the majority of edges within clusters are positive and most edges across clusters are negative. 
With node representations $\mathbf{Z} \in \mathbb{R}^{ |\mathcal{V}| \times 2d}$ learned in our encoder, we propose a non-linear transformation to predict clusters.

\textbf{Clustering Assignment.} Considering a $K$-way clustering problem, where $K$ is the number of clusters ($K\ll d$), node $v_i$ is assigned a probabilities vector $\Pi_i=[\pi_i(1), \dots, \pi_i(K)]$, representing the likelihood of belonging to each cluster. $k\in \{1,\dots, K\}$ denotes the index of a cluster and $\sum_{k=1}^K\pi_i(k)=1$. This probability is computed using a learnable transformation followed by a softmax operation:
\begin{equation}
    \pi_i(k)=q_{\theta}(k|\mathbf{Z}_i)=\frac{\exp(\mathbf{Z}_i \cdot \theta_{k})}{\sum_{k'=1}^K\exp(\mathbf{Z}_i\cdot\theta_{k'})},
\end{equation}
where $\theta_k \in \mathbb{R}^{2d\times 1}$ is a parameter for cluster $k$ to be trained by minimizing the signed clustering loss. The assignment vectors of all nodes $\{\Pi_i\}_{i=1}^{|\mathcal{V}|}$ form an assignment matrix $\Pi\in \mathbb{R}^{|\mathcal{V}|\times K}$. 

\textbf{Differential Signed Clustering Loss.} Signed graph clustering aims to minimize violations, which was historically considered as an NP-Hard optimization problem~\cite{kunegis2010spectral} with designed discrete (non-differential) objectives in spectral methods. We transform it into a differentiable format by utilizing a soft assignment matrix $\Pi$ in place of a hard assignment matrix $\mathbf{C}$.
% $\{\mathbf{c}_1,\mathbf{c}_2,\dots, \mathbf{c}_{K}\}$
Specifically, given that the cluster number $K$ is known, let $\mathbf{C}\in \{0,1\}^{|\mathcal{V}|\times K}$
be a hard cluster assignment matrix, where $|\mathcal{V}|$ is the number of nodes. Each column $\mathbf{C}_{(:,k)}$ indicates membership in cluster $k$, with $\mathbf{C}_{(:,k)}(i)=1$ if node $v_i$ belong to the cluster $k$; otherwise $\mathbf{C}_{(:, k)}(i)=0$.

The number of positive edges between cluster $k$ and other clusters can be captured by $\mathbf{C}_{(:,k)}^{T}\mathbf{L}^{+}\mathbf{C}_{(:,k)}$ with the positive graph Laplacian $\mathbf{L}^{+}=\mathbf{D}^{+}-\mathbf{A}^{+}$. The number of negative edges within cluster $k$ can be measured by $\mathbf{C}_{(:,k)}^{T}\mathbf{A}^{-}\mathbf{C}_{(:,k)}$. So the violations w.r.t. cluster $k$ can be measured by $\mathbf{C}_{(:,k)}^{T}(\mathbf{L}^{+}+\mathbf{A}^{-})\mathbf{C}_{(:,k)}$. By replacing the hard assignment $\mathbf{C}_{(:,k)}$ with the soft assignment probability $\Pi_{(:,k)}$, the differential clustering loss is constructed as:
% \begin{equation}~\label{sc1} 
%     \mathcal{L}=\frac{1}{\left|\mathcal{V}\right|}\sum_{l=0}^{L}\sum_{k=1}^{K}\mathbf{\Pi}_{(:, k)}^{T}(\mathbf{L}^{+(l)}+\mathbf{A}^{-(l)})\mathbf{\Pi}_{(:, k)} +  \lambda\mathcal{L}_{\textit{regu}}
% \end{equation}
\begin{equation}~\label{sc1} 
    \mathcal{L}=\frac{1}{\left|\mathcal{V}\right|}\sum_{k=1}^{K}\mathbf{\Pi}_{(:,k)}^{T}(\mathbf{L}^{+}+\mathbf{A}^{-})\mathbf{\Pi}_{(:,k)} + \lambda\mathcal{L}_{\textit{regu}},
\end{equation}
% \begin{equation}
%     \text{where} \mathbf{L}^{+(l)} = \mathbf{D}^{+(l)} + (\mathbf{A}^{+})^{l};\quad \mathbf{A}^{-(l)} = \sum_{a=0}^{l}\theta_l(\mathbf{A}^{+})^{a}(\mathbf{A}^{-})(\mathbf{A}^{+})^{l-a}
% \end{equation}
where $\lambda$ is a hyperparameter, and $\mathcal{L}_{\textit{regu}}$ is a regularization term computing the degree volume in cluster to prevent model collapse:
% \begin{equation}
% \mathcal{L}_{\textit{regu}}=-\frac{1}{\left|\mathcal{V}\right|}\sum_{l=1}^l\sum_{k=1}^{K}\mathbf{\Pi}_{(:, k)}^{T}\overline{\mathbf{D}}^{(l)}\mathbf{\Pi}_{(:, k)}
% \end{equation}
\begin{equation}
\mathcal{L}_{\textit{regu}}=-\frac{1}{\left|\mathcal{V}\right|}\sum_{k=1}^{K}\mathbf{\Pi}_{(:, k)}^{T}\overline{\mathbf{D}}\mathbf{\Pi}_{(:, k)},
\end{equation}
where $\overline{\mathbf{D}}$ is the degree matrix of $\mathbf{A}$. Minimizing $\mathcal{L}$ equals finding a partition with minimal violations. We iteratively optimize the signed encoder and non-linear transformation by minimizing $\mathcal{L}$.%The signed encoder and non-linear assignment transformation are optimized by minimizing $\mathcal{L}$ iteratively. 

\textbf{Inference stage.} Each node $v_i\in \mathcal{V}$ is assigned to the cluster with the highest probability in its vector $\Pi_i$: $s_i = argmax_{k}\Pi_i$, where $s_i \in \{1,\dots,K\}$ is the cluster index for $v_i$. The set of all node cluster assignments, $\{s_i\}_{i=1}^{\left|\mathcal{V}\right|}$, is used to evaluate the performance of the clustering approach.

\section{Experiments}
This section evaluates our DSGC model with both synthetic and real-world graphs to address the following research questions. \textbf{RQ1:} Can DSGC achieve state-of-the-art clustering performance on signed graphs without any labels? \textbf{RQ2:} How does each component contribute to the effectiveness of DSGC? \textbf{RQ3:} How does the Violation Sign-Refine (VS-R) impact signed topology structures by correcting noisy edges? \textbf{RQ4:} How do the strategies in our signed encoder, specifically abandoning the ``\textit{EEF}'' principle and the minus sign in term $(-\bar{\mathbf{A}}^{-})$,  contribute to forming wider clustering boundaries? Our implementation is available in PyTorch\footnote{The code is provided in https://github.com/yaoyaohuanghuang/DSGC}.

%TODO: 检查一下所有 section, subsection, subsubsection使用相同的大写规则
% \vspace{-10pt}
\subsection{Experimental Settings}~\label{Experimental Settings}
\vspace{-15pt}
\subsubsection{Datasets.} Follow SPONGE~\cite{cucuringu2019sponge}, we evaluate DSGC with a variety of synthetic and real-world graphs: (i) \textbf{Synthetic SSBM graphs.} 
The Signed Stochastic Block Model (SSBM) is commonly used to generate labeled signed graphs~\cite{cucuringu2019sponge, Mercado2019Spectral}, parameterized by $N$ (number of nodes), $K$ (number of clusters), $p$ (edge probability or sparsity), and $\eta$ (sign flip probability). This model first sets edges within the same cluster as positive, and edges between clusters as negative. It then models noises by randomly flipping the sign of each edge with probability $\eta\in[0,1/2)$. Each generated graph can be represented as SSBM ($N$,$K$,$p$,$\eta$). 
% By variable-controlling approach, we respectively vary one parameter each time while fixing the other parameters to implement our comparison experiments. 
(ii) \textbf{Real-world graphs.}
% \textbf{Unlabeled graphs.}
S\&P is a stock correlation network from market excess returns during $2003-2005$, consisting of $1,193$ nodes, $1,069,319$ positive edges, and $353,930$ negative edges. Rainfall is a historical rainfall dataset from Australia, where edge weights are computed by the pairwise Pearson correlation. Rainfall is a complete signed graph with $306$ nodes, $64,408$ positive edges, and $29,228$ negative edges. Moreover, DSGC is compared against $15$ baselines demonstrated in App.~\ref{app:baselines}.
\vspace{-3pt}
\subsubsection{Evaluation Metrics} 
For \textit{Labeled graphs (SSBM)}, Accuracy (ACC), Adjusted Rand Index (ARI)~\cite{gates2017impact}, Normalized mutual information (NMI), and F1 score are used as the ground truths of nodes are available. % These measures indicate how well the predicted partition matches ground truth. With a value close to $1$, the predicted clusters almost perfectly match the ground truth partition; with a value close to $0$, the predicted clusters match an almost random assignment of the nodes into clusters. 
For \textit{Unlabeled graphs (S\&P and Rainfall)}, due to the lack of ground truths, clustering quality is visualized by plotting network adjacency matrices sorted by cluster membership. See more detailed settings in App.~\ref{app:setting}.

\begin{table*}[t]
\centering
\caption{Performance comparison of graph clustering methods on SSBM graphs with ACC (\%) and NMI (\%). Bold values indicate the best results; underlined values indicate the runner-up.}
\scalebox{0.64}{
\begin{tabular}{l|cc|cc|cc|cc|cc|cc|cc|cc|cc|cc}
\toprule[1.3pt]
\midrule
SSBM & \multicolumn{10}{c|}{($N=1000$, $K=5$, $p=0.01$, $\eta$)} & \multicolumn{10}{c}{($N=1000$, $K=10$, $p$, $\eta=0.02$)} \\
\midrule
SSBM  & \multicolumn{2}{c|}{$\eta=0$}  & \multicolumn{2}{c|}{$\eta=0.02$} & \multicolumn{2}{c|}{$\eta=0.04$} & \multicolumn{2}{c|}{$\eta=0.06$} & \multicolumn{2}{c|}{$\eta=0.08$} & \multicolumn{2}{c|}{$p=0.01$}  & \multicolumn{2}{c|}{$p=0.02$} & \multicolumn{2}{c|}{$p=0.03$} & \multicolumn{2}{c|}{$p=0.04$} & \multicolumn{2}{c}{$p=0.05$} \\
\midrule
\textbf{Metrics} & ACC & NMI & ACC & NMI & ACC & NMI & ACC & NMI & ACC & NMI & ACC & NMI & ACC & NMI & ACC & NMI & ACC & NMI & ACC & NMI \\
\midrule[0.3pt]
$\mathbf{A}^{*}$ & 71.60 & 41.06 & 68.10 & 35.95 & 62.20 & 27.86 & 41.60 & 15.25 & 43.80 & 12.65 & 16.70 & 3.90 & 2.13 & 7.89 & 42.80 & 24.43 & 79.70 & 62.52 & 93.30 & 86.60\\
$\overline{\mathbf{L}}_{sns}$ & 21.20 & 0.96 & 
21.20 & 0.85 & 21.10 & 1.23 & 20.50 & 0.83 & 20.60 & 1.05& 12.20 & 1.86 & 14.30 & 2.91 & 18.30 & 6.34 & 18.80 & 7.55 & 34.20 & 23.08\\
$\overline{\mathbf{L}}_{dns}$ & 41.70 & 16.39 & 38.30 & 12.47 & 31.70 & 6.45 & 31.20 & 6.46 & 29.10 & 3.75& 15.70 & 2.91 & 19.10 & 5.80 & 27.60 & 12.29 & 46.90 & 30.57 & 83.50 & 68.80\\
$\overline{\mathbf{L}}$ & 20.30 & 0.79 & 20.30 & 0.79 & 20.30 & 0.79 & 20.30 & 0.79 & 20.30 & 0.79 & 10.70 & 1.75 & 10.70 & 1.75 & 10.70 & 1.94 & 10.70 & 2.02 & 10.70 & 1.75\\
$\mathbf{L}_{sym}$ & 75.80 & 46.49 & 69.40 & 37.06 & 62.30 & 28.46 & 48.20 & 19.35 & 47.70 & 15.68 & 16.00 & 2.69 & \underline{19.60} & 6.47 & 40.00 & 21.55 & 78.60 & 61.12 & 93.70 & 86.03\\
BNC & 41.00 & 14.76 & 39.50 & 12.66 & 35.70 & 7.28 & 27.90 & 5.34 & 27.50 & 3.84 & 15.10 & 2.31 & 19.30 & 5.77 & 23.80 & 12.22 & 49.00 & 32.98 & 83.60 & 69.19\\
BRC & 20.30 & 0.79 & 20.30 & 0.79 & 20.40 & 0.78 & 20.30 & 0.79 & 20.30 & 0.79 & 11.10 & 1.95 & 10.70 & 1.82 & 12.10 & 3.20 & 13.80 & 4.73 & 10.70 & 1.75\\
SPONGE & 86.40 & 65.49 & \underline{81.40} & 55.73 & \underline{71.70} & 41.85 & \underline{55.00} & \underline{28.61} & \underline{49.90} & \underline{22.21} & 18.40 & 6.41 & 2.86 & 15.08 & 62.70 & 41.71 & 90.50 & 80.30 & 97.70 & 94.59\\
SPONGE$_{sym}$ & \underline{88.60} & \underline{77.89} & 67.60 & \underline{57.00} & 63.20 & \underline{46.35} & 35.00 & 20.90 & 32.20 & 12.12 & \underline{19.90} & 11.34 & 19.50 & \underline{20.53} & \underline{81.60} & \textbf{78.10} & \underline{95.90} & \underline{90.92} & \underline{98.90} & \underline{97.30}\\
SiNE & 22.70 & 0.55 & 23.30 & 0.52 & 24.10 & 0.53 & 25.00 & 0.77 & 22.40 & 0.36 & 14.20 & 1.55 & 14.80 & 2.59 & 15.30 & 2.27 & 15.30 & 2.86 & 15.10 & 2.66 \\
SNEA & 43.30 & 17.60 & 45.60 & 19.26 & 42.30 & 16.93 & 32.60 & 7.47 & 31.60 & 6.13 & 14.80 & 2.53 & 19.00 & 6.71 & 19.30 & 7.11 & 32.20 & 19.57 & 32.80 & 25.35\\
% sIR-LS & \underline{93.70} & \underline{81.34} & \textbf{91.00} & \textbf{74.53} & \underline{84.50} & \underline{61.53} & 20.30 & 0.79 & 20.30 & 0.79 & 10.40 & 1.75 & 30.90 & 16.26 & 97.20 & 93.43 & 99.50 & 98.78 & 99.90 & 99.76\\
% IR-LS & 93.40 & 80.52 & 90.70 & 73.23 & \textbf{85.80} & \textbf{63.29} & 20.30 & 0.79 & 20.30 & 0.79 & 10.40 & 1.75 & 10.60 & 1.75 & 87.80 & 80.22 & 99.40 & 98.54 & 99.90 & 99.76 \\

\midrule
DAEGC & 32.20 & 5.12 & 32.70 & 6.75 & 31.40 & 5.07 & 31.20 & 4.37 & 29.10 & 2.90 & 14.60 & \underline{14.60} & 15.90 & 3.22 & 17.10 & 4.38 & 18.50 & 7.43 & 21.80 & 11.70 \\
DFCN & 34.70 & 6.56 & 32.60 & 4.82 & 30.30 & 3.45 & 28.80 & 3.00 & 28.50 & 3.65 & 14.90 & 2.37 & 14.30 & 2.17 & 15.70 & 2.93 & 16.80 & 3.74 & 16.20 & 3.47 \\ 
DCRN & 48.40 & 23.18 & 44.40 & 18.80 & 43.70 & 21.07 & 37.10 & 11.40 & 33.30 & 10.37 & 16.70 & 3.61 & \underline{19.60} & 8.18 & 25.30 & 12.92 & 33.50 & 25.56 & 48.30 & 38.44\\
Dink-net & 27.20 & 2.21 & 27.00 & 1.73 & 27.70 & 1.87 & 26.50 & 1.59 & 26.40 & 2.03 & 14.60 & 1.80 & 15.10 & 2.00 & 15.20 & 25.20 & 16.90 & 3.47 & 18.80 & 4.57 \\
DGCLUSTER & 20.30 & 0.79 & 20.30 & 0.79 & 20.30 & 0.79 & 20.30 & 0.79 & 20.30 & 0.79 & 10.40 & 1.75 & 10.60 & 1.75 & 10.70 & 1.75 & 10.70 & 1.75 &10.60 & 1.75 \\ 
MAGI & 41.80 & 10.09 & 34.10 & 7.28 & 30.70 & 5.78 & 32.50 & 5.09 & 29.5 & 4.23 & 16.40  & 3.43 & 16.20 & 3.46 & 19.60 & 6.56 & 20.40 & 8.99 & 29.20 & 14.66 \\
% DAEGC & 82.40 & 61.72 & 79.80 & 54.96 & 73.40 & 43.23 & 55.40 & 26.41 & 53.30 & 22.13 & 21.50 & 9.25 & 46.70 & 30.02 & 77.60 & 61.22 & 95.10 & 89.06 & 98.60 & 96.59\\
% DFCN & 86.20 & 64.57 & 80.10 & 52.75 & 72.90 & 42.07 & 60.12 & 26.68 & 50.10 & 19.26 & 21.90 & 8.21 & 36.40 & 24.06 & 78.20 & 63.03 & 96.00 & 91.08 & 98.80 & 97.08\\
% DCRN & \underline{89.10} & 70.40 & \underline{83.30} & \underline{58.37} & \underline{75.50} & \underline{47.17} & \underline{64.00} & \underline{31.02} & \underline{54.80} & \underline{22.42} & \underline{20.40} & 7.26 & \underline{51.10} & \underline{31.68} & \underline{84.20} & 70.94 & \underline{96.20} & \underline{91.93} & \underline{99.10} & \underline{97.85}\\

\midrule
DSGC &\textbf{95.30} & \textbf{85.40} & \textbf{90.80}& \textbf{73.60} & \textbf{82.80} & \textbf{57.30} & \textbf{66.50} & \textbf{33.50} & \textbf{57.70} & \textbf{23.30} & \textbf{28.70} & \textbf{13.10} & \textbf{64.90} & \textbf{44.80} & \textbf{85.40} & \underline{71.90} & \textbf{96.90} & \textbf{92.80} & \textbf{99.20} & \textbf{98.10}\\ 
\midrule
SSBM & \multicolumn{10}{c|}{($N$, $K=5$, $p=0.01$, $\eta$=0)} & \multicolumn{10}{c}{($N=1000$, $K$, $p=0.01$, $\eta=0.02$)} \\
\midrule
SSBM  & \multicolumn{2}{c|}{$N=300$}  &\multicolumn{2}{c|}{$N=500$} & \multicolumn{2}{c|}{$N=800$} & \multicolumn{2}{c|}{$N=1000$} & \multicolumn{2}{c|}{$N=1200$} & \multicolumn{2}{c|}{$K=4$}  & \multicolumn{2}{c|}{$K=5$} & \multicolumn{2}{c|}{$K=6$} & \multicolumn{2}{c|}{$K=7$} & \multicolumn{2}{c}{$K=8$}\\
\midrule
\textbf{Metrics} & ACC & NMI & ACC & NMI & ACC & NMI & ACC & NMI & ACC & NMI & ACC & NMI & ACC & NMI & ACC & NMI & ACC & NMI & ACC & NMI\\
\midrule[0.3pt]
$\mathbf{A}$ & 28.67 & 4.62 & 35.80 & 11.40 & 50.62  & 20.98  & 71.60  & 41.06  & 83.08  & 58.26 & 90.70 & 71.11 & 68.10 & 35.95 & 32.80 & 9.20 & 24.70 & 6.62 & 20.70 & 3.74\\
$\overline{\mathbf{L}}_{sns}$ & 21.67 & 2.69 & 21.60 & 1.78 &  21.25 &  1.44 & 21.20  & 0.96 & 21.67  & 1.48 & 25.30 & 0.70 & 21.20 & 0.85 & 17.80 & 1.29 & 16.40 & 1.36 & 16.00 & 2.09\\
$\overline{\mathbf{L}}_{dns}$ & 21.33 & 3.00 & 23.60 & 2.50 & 24.38  & 2.16  & 41.70  &  16.39 & 50.42  & 18.22 & 60.90 & 33.87 & 38.30 & 12.47 & 24.20 & 3.56 & 18.60 & 2.17 & 19.30 & 2.75 \\
$\overline{\mathbf{L}}$ &25.20 & 0.59 & 20.30 & 0.79 & 17.10 & 0.98 & 15.00 & 1.55 & 12.90 & 1.37 & 21.33 & 2.17 & 21.20 & 0.94 & 20.37  & 0.98  & 20.30  & 0.79 & 20.25  & 0.66 \\
$\mathbf{L}_{sym}$ & 21.00 & 3.15 & 30.00 & 7.47 &  50.12 & 20.91  & 75.80  & 46.49 & 83.92  & 59.47 & 91.80 & 73.31 & 69.40 & 37.06 & 32.00 & 8.85 & 26.80 & 7.99 & 20.80 & 4.55\\
BNC &21.67 & 2.11& 25.20 & 3.08 & 24.88  & 1.89  & 41.00  & 14.76  & 54.50  & 21.56 & 60.80 & 31.74 & 39.50 & 12.66 &25.80 & 4.61 & 18.50 & 1.74 & 19.10 & 3.09\\
BRC & 21.00 & 2.54 & 20.60 & 1.45  & 24.88  & 1.89  & 20.30  & 0.79  & 20.25  & 0.66 & 25.40 & 0.71 & 20.30 & 0.79 & 17.10 & 1.02 & 14.80 & 1.17 &13.00 & 1.37\\
SPONGE & 21.33& 3.07 & 29.20 &  11.43 & 20.50  & 0.98  &  86.40 & 65.49  & \underline{94.75}  & \underline{83.37} & \underline{95.70} & \underline{84.24} & \underline{81.40} & 55.73 & 43.10 & 20.83 & \underline{45.30} & 19.93 & 23.10 & 8.97 \\
SPONGE$_{sym}$ &27.67 & \underline{7.24} & 35.20  & \underline{15.56}  & \underline{82.12} &  \underline{66.43} & \underline{88.60}  & \underline{77.89}  & 91.33  & 81.92 & 94.70 & 82.10 & 67.60 & \underline{57.00} & \underline{62.90} & \underline{44.58} & 32.70 & \underline{21.65} & \underline{25.00} & \underline{10.52}\\
SiNE & 25.00 & 1.76 & 26.20 & 1.57 & 24.25 & 0.96 & 22.70 & 0.55 & 24.08 & 1.14 & 27.50 & 0.29 & 23.30 & 0.52 & 20.60 & 0.68 & 17.80 & 1.03 & 16.50 & 1.64\\
SNEA & 30.00 & 4.54 & 32.80 & 7.12 & 39.00 & 9.41 & 43.30 & 17.60 & 51.25 & 29.90 & 65.30 & 48.55 & 45.70 & 19.24 & 27.80 & 5.48 & 26.30 & 6.50 & 20.10 & 3.35 \\
% sIR-LS & 20.67 & 2.55 & 20.80 & 1.55 & 20.37 & 0.98 & 93.7 & 81.34 & 98.5 & 94.35 & 97.50 & 90.41 & 91.00 & 74.53 & 17.00 & 0.98 & 14.60 & 1.18 & 12.80 & 1.37 \\
% IR-LS & 20.67 & 2.55 & 20.80 & 1.55 & 32.75 & 5.48 & 93.4 & 80.52 & 98.42 & 94.02 & 97.50 & 90.37 & 90.70 & 73.23 & 17.00 & 0.98 & 14.60 & 1.18 & 12.80 & 1.37 \\

\midrule
DAEGC & 26.67 & 3.56 & 27.00 & 2.74 & 27.88 & 3.47 & 32.20 & 5.12 & 35.42 & 9.17 & 39.30 & 8.00 & 32.70 & 6.75 & 24.30 & 2.66 & 20.60 & 2.24 & 18.20 & 2.74 \\
DFCN & 28.33 & 4.95 & \underline{36.80} & 6.00 & 31.50 & 4.47 & 34.70 & 6.56 & 30.25 & 3.79 & 43.70 & 8.47 & 32.60 & 4.82 & 23.70 & 1.59 & 19.40 & 1.60 & 17.40 & 1.95 \\
DCRN & 28.00 & 4.53 & 33.40 & 8.69 & 32.00 & 10.69 & 48.40 & 23.18 & 49.83 & 29.91 & 67.30 & 39.13 & 44.40 & 18.80 & 33.60 & 12.89 & 24.20 & 6.66 & 19.80 & 4.47 \\
Dink-net & 29.33 & 2.74 & 28.60 & 2.41 & 27.38 & 1.75 & 27.20 & 2.21 & 29.33 & 2.46 & 35.00 & 2.76 & 27.00 & 1.73 & 21.90 & 1.61 & 20.70 & 1.55 & 18.00 & 1.74 \\
DGCLUSTER & 20.67 & 2.55 & 20.80 & 1.55 & 20.37 & 0.98 & 20.30 & 0.79 & 20.25 & 0.66 & 25.20 & 0.59 & 20.30 & 0.79 & 17.00 & 0.98 & 14.60 & 1.18 & 12.80 & 1.37 \\
MAGI & \underline{33.00} & 6.11 & 35.00 & 5.46 & 32.38 & 5.79 & 41.80 & 10.09 & 45.42 & 14.34 & 46.70 & 13.49 & 34.10 & 7.28 & 24.00 & 2.76 & 20.60 & 2.64 & 17.30 & 2.16 \\
% DAEGC & 31.33 & 6.26 & 40.00 & 12.56 & 57.25 & 36.91 & 82.40 & 61.72 & 93.80 & 83.52 & 92.70 & 76.36 & 79.80 & 54.96 & 47.50 & 23.48 & 44.30 & 19.96 & 31.90 & \underline{11.91} \\
% DFCN &\underline{35.00} &11.44& 45.20 &16.17 & 69.87 & 40.34 & 86.20 & 64.57 & 95.92 & 85.62 & 93.90 & 80.03 & 80.10 & 52.75 & 55.70 & 25.69 & 35.50 & 11.41 & 31.70 & 11.50\\
% DCRN & 33.00 & \underline{10.13} & \underline{45.40} & \underline{16.64} & 73.12 & 44.39 & \underline{89.10} & 70.40 & \underline{96.42} & \underline{87.93} & 95.40 & \underline{84.35} & \underline{83.30} & \underline{58.37} & 57.80 & 29.74 & 42.30 & 16.94 & \underline{33.30} & 11.56\\

\midrule

DSGC &\textbf{37.70} & \textbf{10.30} & \textbf{54.60} & \textbf{30.80} & \textbf{89.40} & \textbf{71.40} & \textbf{95.30} & \textbf{85.40} & \textbf{98.40} & \textbf{94.40} & \textbf{97.40} & \textbf{90.10} & \textbf{90.80} & \textbf{73.60} & \textbf{70.90} & \textbf{45.10} & \textbf{51.30} & \textbf{23.90} & \textbf{35.90} & \textbf{14.40}\\ 
\midrule[0.8pt]
\bottomrule[1.3pt]
\end{tabular}}
\label{tab:overall_perf_acc_nmi}
\end{table*}

\vspace{-3pt}
\subsection{Overall Performance}
To address \textbf{RQ1}, we evaluated our DSGC and baselines on a variety of labeled signed graphs generated from four SSBM configurations, including SSBM ($N=1000$, $K=5$, $p=0.01$, $\eta$) with $\eta \in \{0, 0.02, 0.04, 0.06, 0.08\}$, SSBM ($N=1000$, $K=10$, $p$, $\eta=0.02$) with $p \in \{0.01, 0.02, 0.03, 0.04, 0.05\}$, SSBM ($N$, $K=5$, $p=0.01$, $\eta=0$) with $N \in \{300, 500, 800, 1000, 1200\}$, and SSBM ($N=1000$, $K$, $p=0.01$, $\eta=0.02$) with $K \in \{4, 5, 6, 7, 8\}$. The performance of each experiment was measured by taking the average of 5 repeated executions. Table~\ref{tab:overall_perf_acc_nmi} reports the results in ACC and NMI. Appendix~\ref{sec:over_per_ari_f1} provides the results in ARI and F1 score. 

Table~\ref{tab:overall_perf_acc_nmi} shows: (i) \textit{Superior performance:} Our DSGC significantly outperforms all baseline models across all metrics, even though SPONGE and SPONGE$_{sym}$ are known for their effectiveness on such datasets. (ii) \textit{Robustness:} Regardless of whether the graph is dense or sparse ($p$), large or small ($N$), noisy or clean ($\eta$), and the number of clusters is few or many ($K$), DSGC maintains notably superior performance on all $20$ labeled signed graphs.
%As $\eta$ and $K$ increase, the performance of all methods decreases gradually with the performance gap remaining pronounced. As the $N$ and $p$ increase, the performance of all methods increases gradually and the extent of the performance gap becomes less pronounced. Regardless of whether the graph is dense or sparse ($p$), large or small ($N$), noisy or clean ($\eta$), and the number of clusters is few or many ($K$), DSGC exhibits notably superior performance on all $20$ labeled signed graphs. This observation highlights the effectiveness and robustness of our approach to four parameters $\eta$, $p$, $N$, and $K$. 
% Especially for SSBM ($1000$, $10$, $0.02$, $0.05$), our method DSGC achieved an ACC gain of $45.3\%$ than the runner-up ACC of $\mathbf{L}_{sym}$. 
(iii) \textit{Comparative analysis:} While deep unsigned clustering methods (DAEGC, DFCN, DCRN, Dink-net, DGCLUSTER MAGI) consistently underperform  our DSGC due to the limitation of their capabilities to only handle non-negative edges. DSGC still has a clear advantage, highlighting the effectiveness of its design specifically tailored for signed graph clustering. 

% generally outperform non-deep spectral methods due to their advanced representation learning capabilities, 

\begin{figure}[ht!]
	\centering
  		\subfigure[Edge probability ($p$)]{
			\includegraphics[width=0.2\textwidth, trim=0 0 0 73]{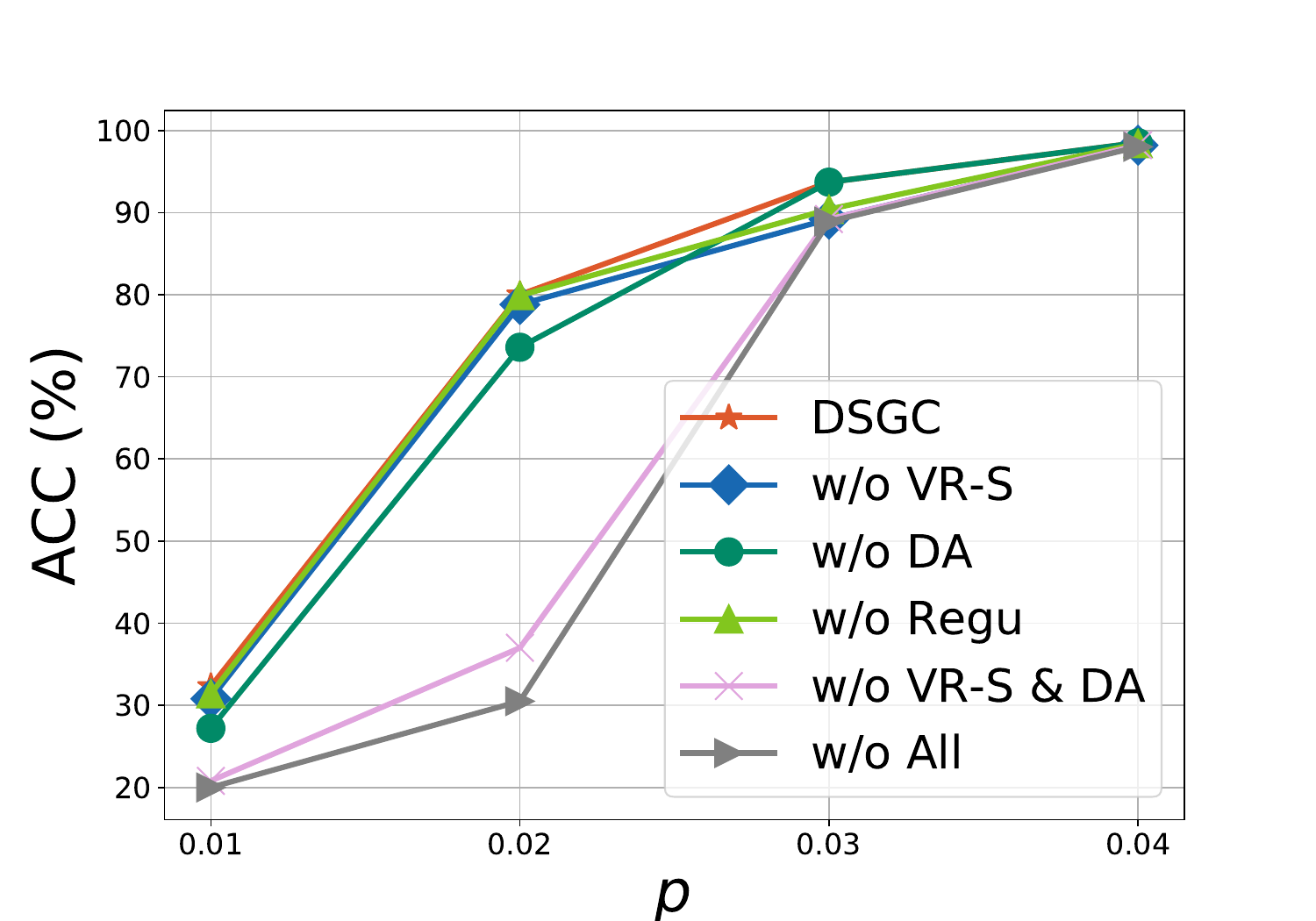}
	}
		\subfigure[Flip probability ($\eta$)]{
			\includegraphics[width=0.2\textwidth, trim=0 0 0 73]{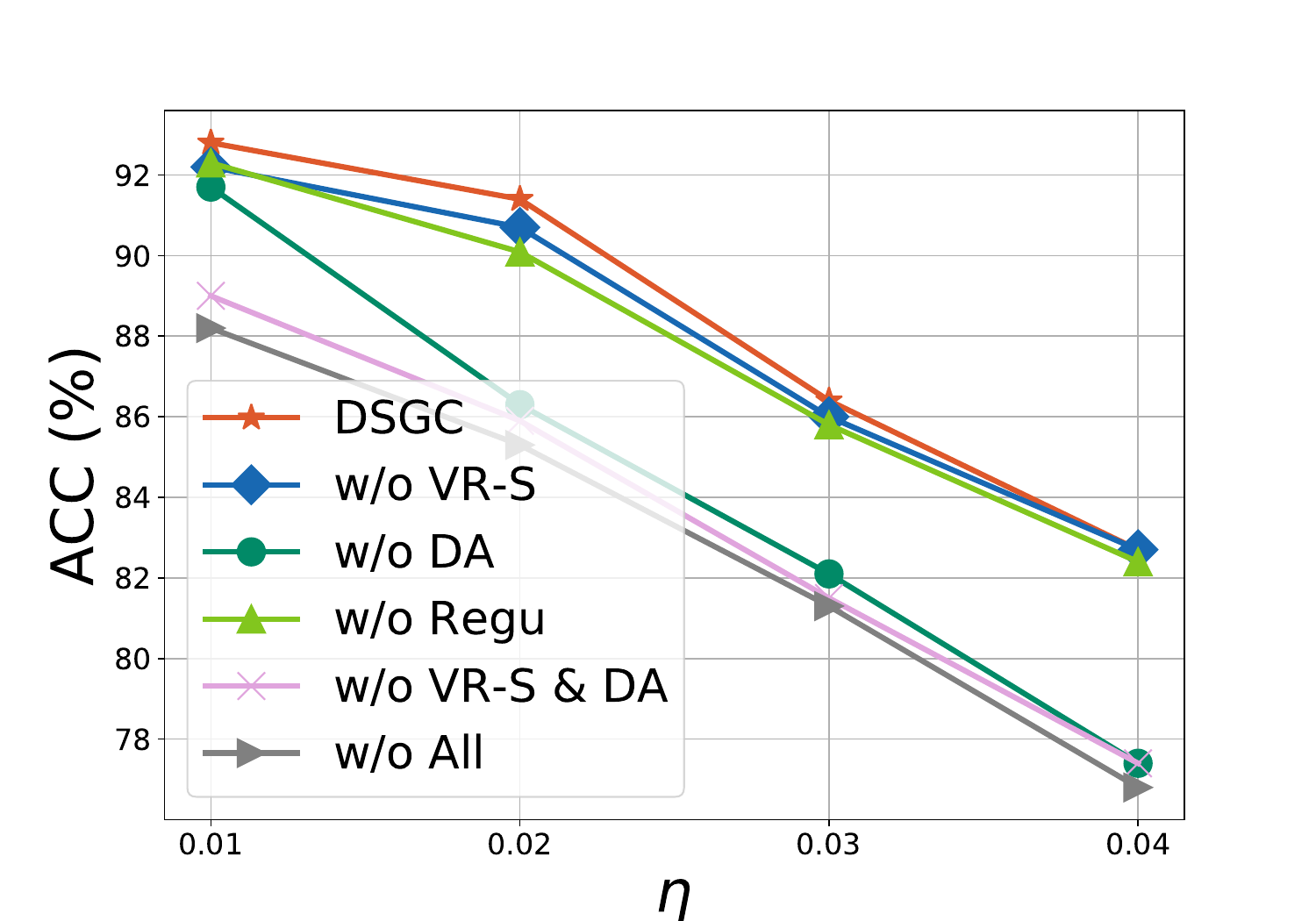}
	}
\\ 
            \subfigure[Node number ($N$)]{
			\includegraphics[width=0.2\textwidth, trim=0 0 0 73]{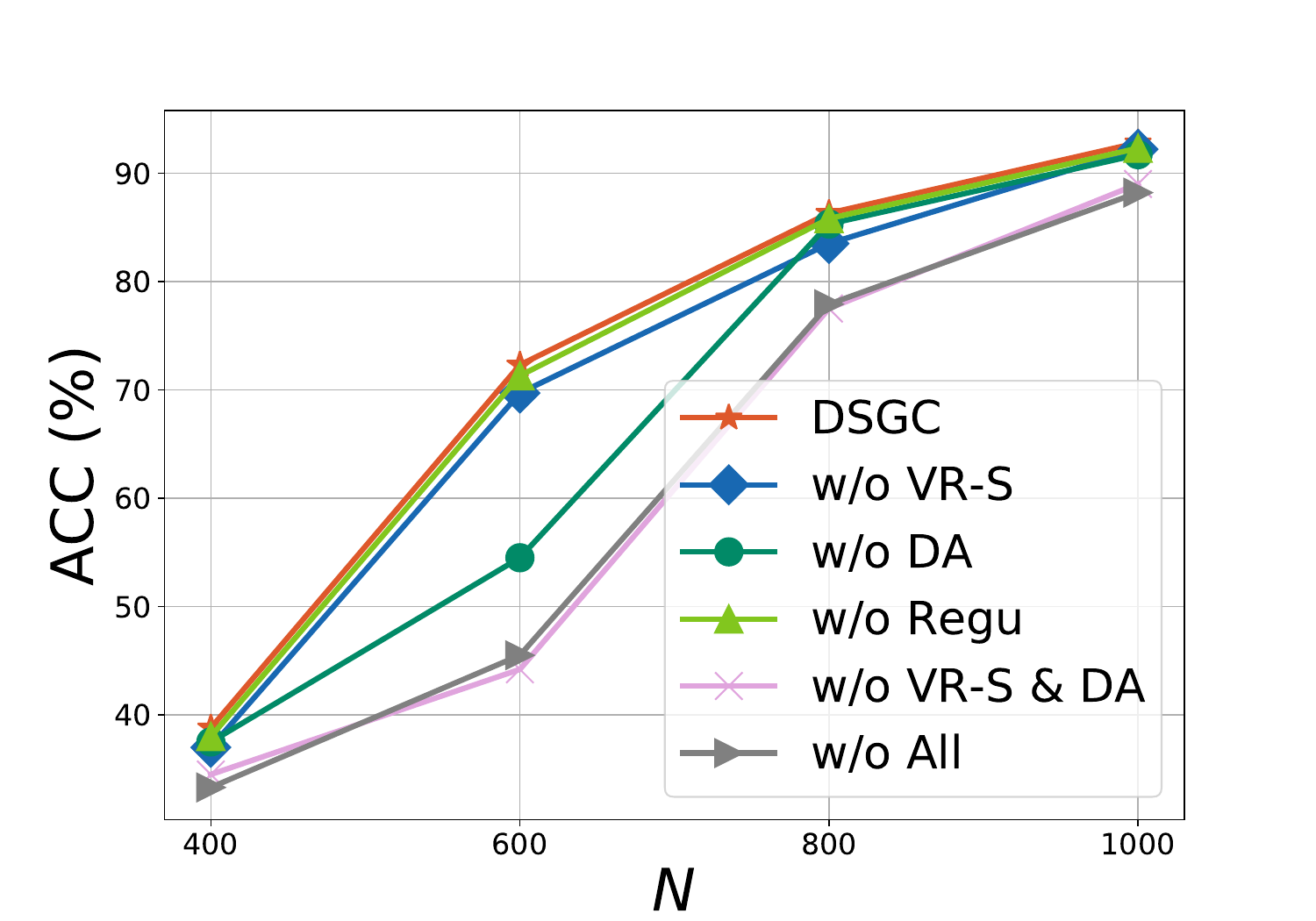}
	}
            \subfigure[Cluster number ($K$)]{
			\includegraphics[width=0.2\textwidth, trim=0 0 0 73]{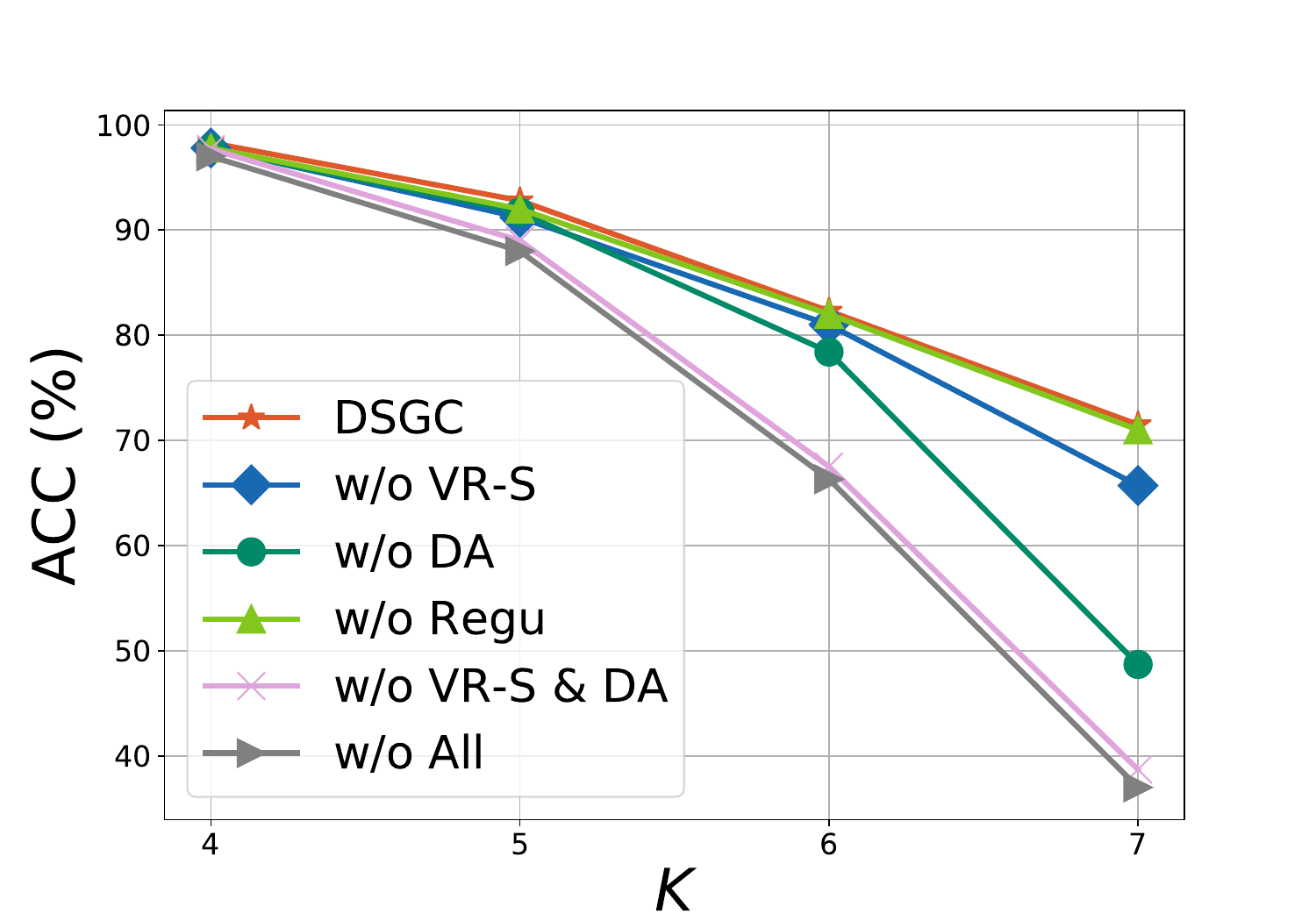}
	}
	\caption{Ablation study. (a)$\sim$(d) ACC($\%$) vs. edge probability $p$, flip probability $\eta$, node number $N$ and cluster number $K$.}
        \label{fig:ablation_study}
\end{figure}
\vspace{-3pt}
\subsection{Ablation Study}
To address \textbf{RQ2} and evaluate the contributions of key components of DSGC, we performed an ablation study using labeled signed graphs, including SSBM ($1000$, $10$, $p$, $0.01$), SSBM ($1000$, $5$, $0.01$, $\eta$), SSBM ($N$, $5$, $0.01$, $0.01$), and SSBM ($1000$, $K$, $0.01$, $0.01$). The variants of DSGC tested are: \textbf{w/o VS-R} is DSGC without Violation Sign-Refine; \textbf{w/o DA} is DSGC without Density-based Augmentation; \textbf{w/o Regu} is DSGC without Regularization term; \textbf{w/o VS-R \& DA} is DSGC without VS-R and DA; \textbf{w/o DA \& Regu} is DSGC without DA and Regu; \textbf{w/o VS-R \& Regu} is DSGC without VS-R and Regu; and \textbf{w/o All} is DSGC without VS-R, DA, and Regu.

From the results depicted in Fig.~\ref{fig:ablation_study}, it is evident that: (i) \textit{Performance trends}: As the edge probability ($p$) and the number of nodes ($N$) increase, the accuracy (ACC) of DSGC and its variants consistently improves. Conversely, increases in the sign flip probability ($\eta$) and the number of clusters ($K$) lead to a decline in ACC across all models. (ii) \textit{Component impact}: DSGC outperforms all variants on all labeled signed graphs, demonstrating the significant role each component plays in enhancing clustering performance. Specifically, DA emerges as the most influential component, affirming its effectiveness in reinforcing the graph structure and improving node representations by adding strategically placed new edges. % Overall, these individual components, i.e., VS-R, DA, and Regu, have well contributed to RSGNN.

\begin{table}[ht!]
\centering
\caption{Improvements in ACC (\%) ($\uparrow$) and NMI (\%) ($\uparrow$) and reductions in violation ratio ($\downarrow$) with VS-R.}
\scalebox{0.63}{
	\begin{tabular}{l|cc|cc|cc|cc|cc}
		\toprule[1.3pt]
		\midrule[0.3pt]
		SSBM  & \multicolumn{4}{c|}{($1000,K,0.01,0.02$)} & \multicolumn{2}{c|}{($1000,5,0.01,\eta$)} & \multicolumn{4}{c}{($1000,10,p,0.02$)}   \\
		\midrule
		SSBM & \multicolumn{2}{c|}{$K=5$} & \multicolumn{2}{c|}{$K=6$} & \multicolumn{2}{c|}{$\eta=0.04$} & \multicolumn{2}{c|}{$p=0.01$}& \multicolumn{2}{c}{$p=0.02$}\\
		\midrule
		\textbf{Metrics} & ACC  & NMI  & ACC  & NMI  & ACC  & NMI  & ACC  & NMI & ACC  & NMI \\
		\midrule[0.3pt]
		$\mathbf{A}$ & -12.10 & 1.49 & 15.00 & 15.89 & 13.70 & 0.74 & 9.00 & 3.42 & 19.80 & 27.24\\
		$\overline{\mathbf{L}}_{sns}$ & 61.11 & 55.75 & 39.10 & 25.88 & 56.10 & 47.02 & 5.10 & 2.54 & 51.10 & 40.57\\
		$\overline{\mathbf{L}}_{dns}$ & 44.20 & 44.28 & 33.10 & 24.69 & 45.50 & 41.77 & 2.50 & 1.90 & 46.10 & 38.18\\
		$\overline{\mathbf{L}}$ & 0.10 & -0.01 & -0.1 & 0.00 & 0.10 & -0.01 & 0.00 & 0.00 & 0.00 & 0.00 \\
		$\mathbf{L}_{sym}$ & 12.80 & 19.74 & 25.10 & 19.78 & 16.40 & 22.15 & 3.80 & 3.84 & 45.70 & 36.68\\
		BNC & 42.70 & 43.87 & 32.10 & 24.72 & 41.70 & 41.18 & 2.00 & 26.80 & 47.20 & 39.31\\
		BRC & 0.1 & -0.01 & -0.1 & -0.04 & 0.00 & 0.00 & -0.7 & -0.2 & -0.3 & -0.06\\
		SPONGE & 2.70 & 4.35 & 14.70 & 8.30 & 5.60 & 6.38 & 0.30 & 0.50 & 29.40 & 24.15\\
		SPONGE$_{sym}$ & 17.10 & 9.34 & 9.70 & 8.99 & 12.40 & 3.64 & 5.10 & 3.71 & 25.40 & 22.50\\
		\midrule
		violation ratio  & \multicolumn{2}{c|}{2.72} & \multicolumn{2}{c|}{1.18} & \multicolumn{2}{c|}{3.48} & \multicolumn{2}{c|}{0.46} & \multicolumn{2}{c}{0.3} \\
		\midrule[0.8pt]
		\bottomrule[1.3pt]
	\end{tabular}}\label{tab:perfgain_spec_signed_rewiring}
%\vspace*{-\baselineskip}
\end{table}

\vspace{-3pt}
\subsection{Analysis of Violation Sign-Refine}
% The ablation study above has demonstrated that the Signed Graph Rewiring (SGR) module significantly benefits our model DSGC.
To investigate \textbf{RQ3}, we analyzed the impact of applying Violation Sign-Refine (VS-R) on the performance of spectral clustering methods. VS-R was first used to pre-process and denoise signed graphs to generate new graphs. Then we compared the performance of all spectral methods before and after applying VS-R. Signed graphs were generated by fixing $N=1000$ and varying ($K, \eta, p$), including SSBM ($1000, K, 0.01, 0.02$) with $K \in \{5,6\}$, SSBM ($1000, 5, 0.01, 0.04$), and SSBM ($1000,10,p,0.02$) with $p \in \{0.01, 0.02\}$. 

Table~\ref{tab:perfgain_spec_signed_rewiring} shows that VS-R significantly improves the clustering performance across all tested spectral methods w.r.t. ACC and NMI by generating cleaner graphs with better clustering structure. Specifically, the performance increments vary inversely with the strength of the baseline methods—stronger baselines show smaller gains, whereas weaker baselines benefit more substantially from the VS-R preprocessing. VS-R also consistently reduces the \textit{violation ratio}, defined as the ratio of the number of violated edges to the number of non-violated edges, across various graph configurations. 
% Besides, we utilize $t$-SNE to visualize the original and new embeddings by applying strong spectral methods on the original and denoised graphs in App.~\ref{App:vis_VS-R}.
In addition to numerical analysis, Fig.~\ref{fig:visua_specemb_wsigncorrection} in App.~\ref{app:vis_VS-R} provides visual evidence of the impact of VS-R. the embeddings of new graphs, displayed in the bottom row, exhibit clearer clustering boundaries than those of the original graphs in the top row. 
% Utilizing $t$-SNE, we compared the embeddings of original and denoised graphs learned by strong spectral methods on SSBM ($1000$, $5$, $0.01$, $0.04$). We can observe that the embeddings of new graphs, displayed in the bottom row, exhibit clearer clustering boundaries than those of the original graphs in the top row. That is, spectral methods, including BNC, SPONGE, and SPONGE$_{sym}$, achieve enhanced clustering performance on cleaner graph structures after employing VS-R.  
\begin{figure}[ht!]
    \centering
    \includegraphics[width=0.44\textwidth, trim=0 0 0 20, clip]{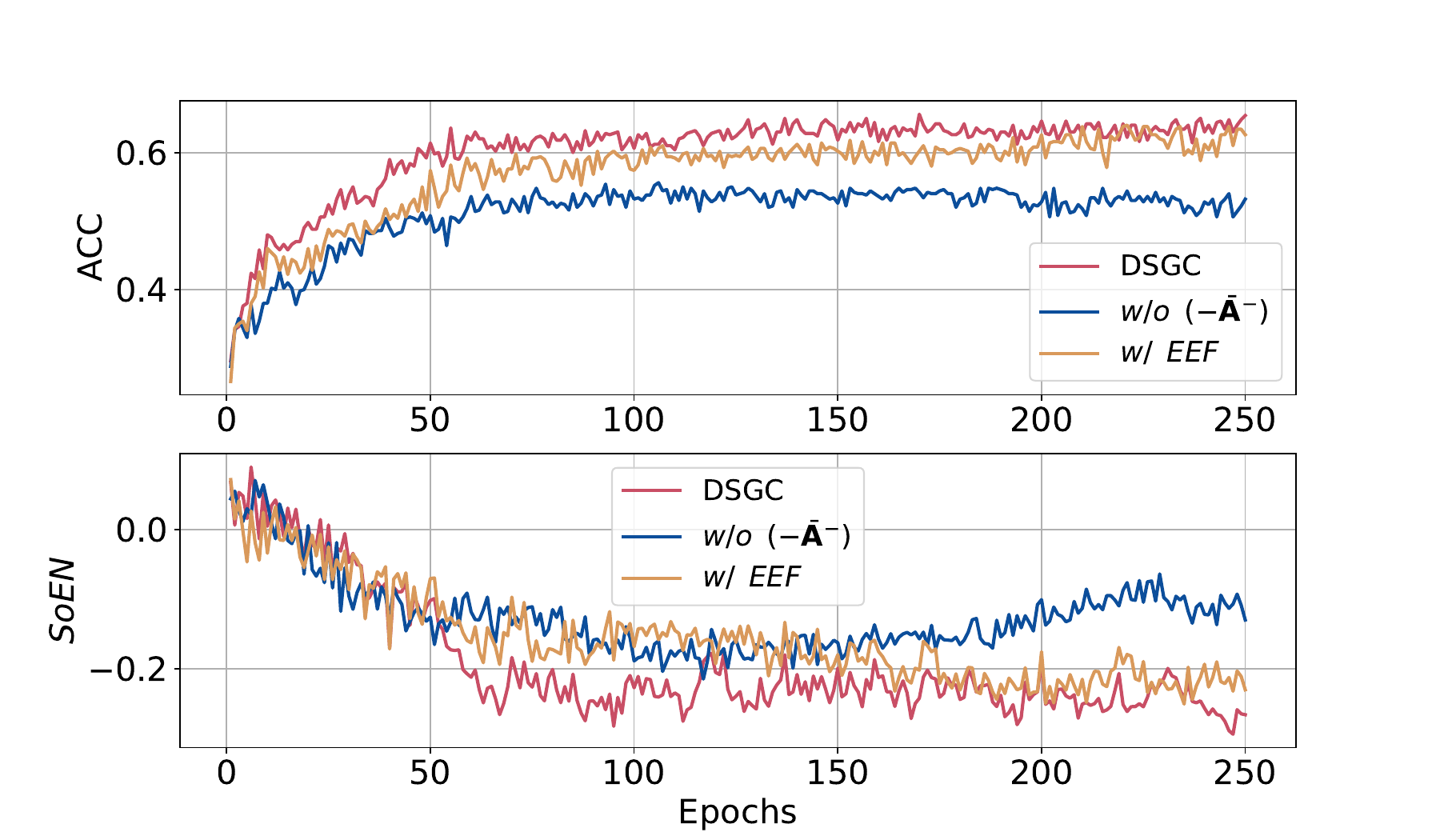}
    \caption{The impact of the term $(-\bar{\mathbf{A}}^{-})$ and ``\textit{EEF}'' principle on ACC (\%) (Top) and \textit{SoEN} (Bottom).}
    \label{fig:analyze_encoder}
\end{figure}
% \vspace{-1pt}
\subsection{Impact of Signed Encoder to Clustering}~\label{experiment_encoder}
% In the design of DSGC, we claim that the term $(-\bar{\mathbf{A}}^{-})$ and abandoning the conventional principle induced in Balance Theory, "the \textit{E}nemy of one's \textit{E}nemy is a \textit{F}riend (\textit{EEF})", can push the nodes connected negatively away from each other, leading to clearer and wider boundaries between clusters, since negative edges exist across clusters. 
To address \textbf{RQ4}, we developed two variants of DSGC encoder, including DSGC w/o $(-\bar{\mathbf{A}}^{-})$ that replaces $(-\bar{\mathbf{A}}^{-})$ with $(\bar{\mathbf{A}}^{-})$,  and DSGC w/ \textit{EEF} that incorporates the ``\textit{the enemy of my enemy is my friend (EEF)}'' principle from Balance Theory to DSGC. Both variants and DSGC used the layer number $L=2$. 
The positive and negative representations of DSGC w/ \textit{EEF} are $\mathbf{Z}_{\textit{eef}}^{+} = \mathbf{Z}^{+}+(\bar{\mathbf{A}}^{-})^{2}\mathbf{Z}^{+(0)};\ \mathbf{Z}_{\textit{eef}}^{-} = \mathbf{Z}^{-}$
where $\mathbf{Z}^{+}$ and $\mathbf{Z}^{-}$ are the positive and negative representations computed by Eq.~\ref{pos_neg_embeddings} in DSGC. Similarly, the positive and negative embeddings of DSGC \textit{w/o} $(-\bar{\mathbf{A}}^{-})$ are $\mathbf{Z}^{+}_{-a} = \mathbf{Z}^{+}; \ \mathbf{Z}^{-}_{-a} = -\mathbf{Z}^{-}~\label{-A_emb}$. 
We define a metric, \textit{SoEN}, to measure the distance between nodes linked by negative edges:
% the average \textit{S}imilarity \textit{o}f two node \textit{E}mbeddings linked along \textit{N}egative edges (\textit{SoEN}) normalized by that of two embeddings linked along the positive edge as a metric to measure the distance of negative edges:
\begin{equation}
    \textit{SoEN} = \frac{|\mathcal{E}^{+}|}{|\mathcal{E}^{-}|} \cdot \frac{\sum_{e_{ij}\in\mathcal{E}^{-}} s(\mathbf{z}_{i}, \mathbf{z}_{j})}{\sum_{e_{i'j'}\in\mathcal{E}^{+}} s(\mathbf{z}_{i'}, \mathbf{z}_{j'})}, \nonumber
\end{equation} 
where $s(\cdot, \cdot)$ is the inner product, indicating the similarity between two nodes. Ideally, \textit{SoEN} is a negative value and a lower \textit{SoEN} indicates a greater distance between nodes connected by negative edges and a clearer clustering boundary. 

Fig.~\ref{fig:analyze_encoder} illustrates the ACC and \textit{SoEN} of DSGC and its variants. The results show that: (i) DSGC consistently outperforms its variants. Incorporating the ``\textit{EEF}'' principle or altering the sign of $(-\bar{\mathbf{A}}^{-})$ significantly impacts clustering performance because DSGC achieves lower \textit{SoEN} along with epochs than its variants. This demonstrates its advantage in separating nodes linked by negative edges, leading to clearer clustering boundaries and larger inter-cluster variances. (ii) The term $(-\bar{\mathbf{A}}^{-})$ has higher impact than the inclusion of \textit{EEF}, suggesting the original negative edge handling in DSGC is critical for maintaining clear cluster separations. %contributes more (replacing $(-\bar{\mathbf{A}}^{-})$ with $(\bar{\mathbf{A}}^{-})$ causes a significant decrease in accuracy) than the abandoning \textit{EEF}. That is because DSGC w/o $(-\bar{\mathbf{A}}^{-})$ gets a larger \textit{SoEN} than DSGC w/ \textit{EEF}, thus yielding better clustering structure.

\begin{figure}[ht!]
	\centering
		\subfigure[BNC]{
			\includegraphics[width=0.11\textwidth, trim=0 0 0 80]{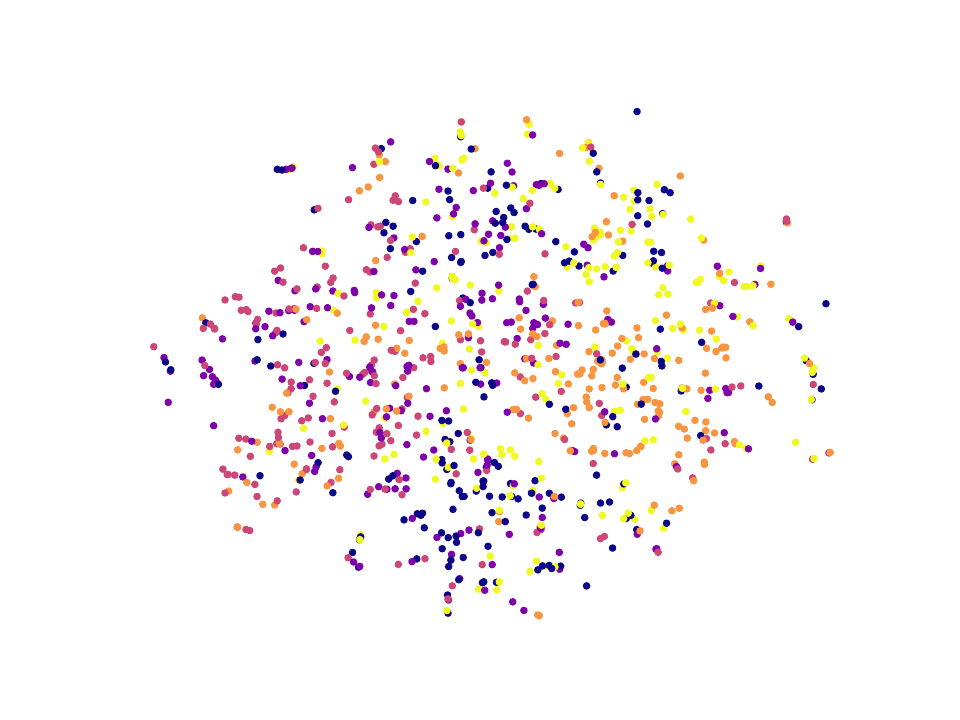}
	}\hspace{-2mm}
 % 		\subfigure[BRC]{
	% 		\includegraphics[width=0.14\textwidth]{figures/1000BRC_5_0.01_0.02_visu.pdf}
	% }
            \subfigure[SPONGE]{
			\includegraphics[width=0.11\textwidth, trim=0 0 0 80]{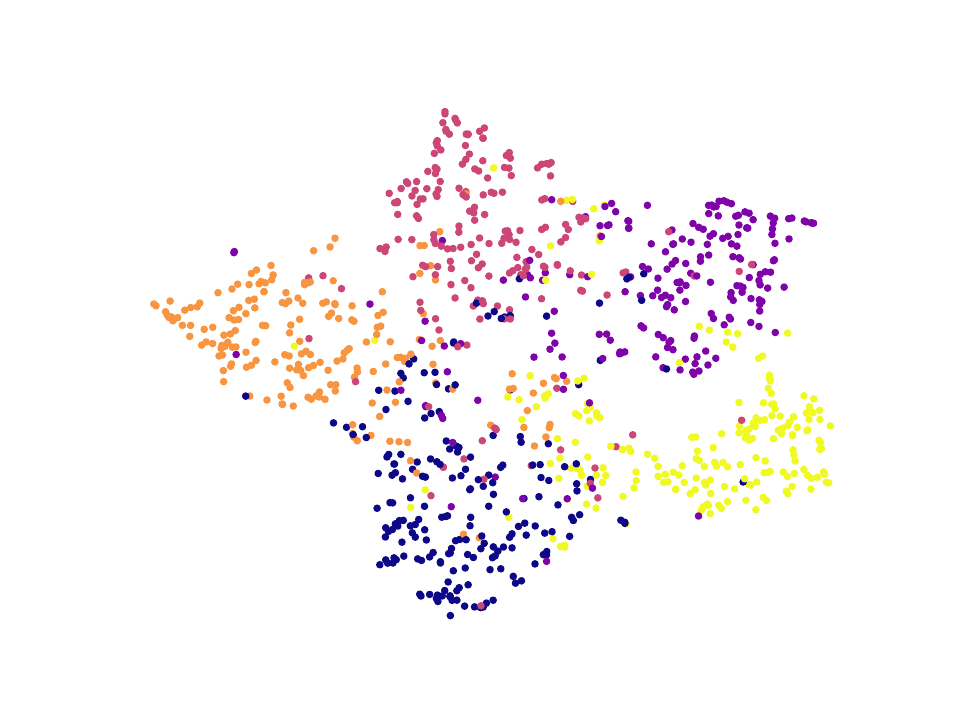}
	}\hspace{-2mm}
            \subfigure[SPONGE$_{sym}$]{
			\includegraphics[width=0.11\textwidth, trim=0 0 0 80]{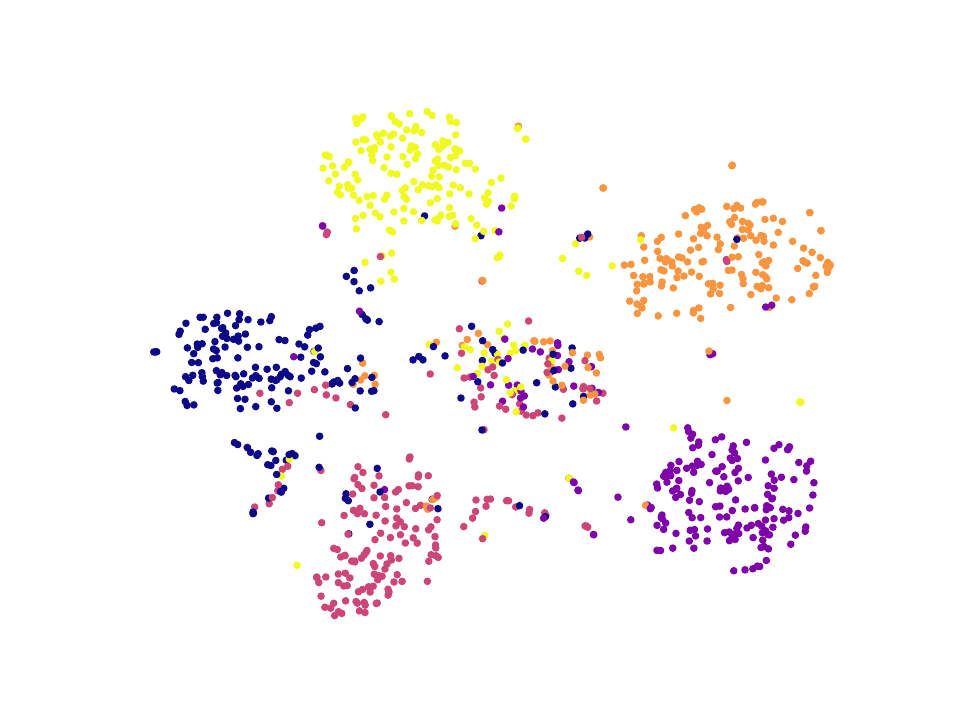}
	}\hspace{-2mm}
             \subfigure[Ours]{
			\includegraphics[width=0.11\textwidth, trim=0 0 0 80]{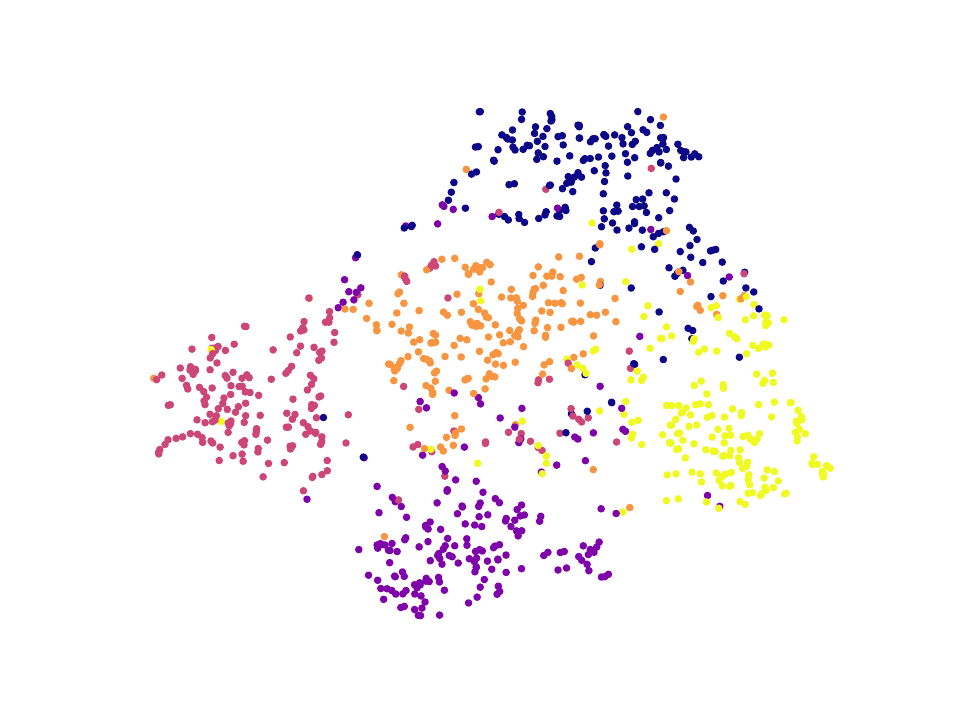}
	}

	\caption{Visualization of clustering results from different algorithms. The ground truth class number is $5$. }
        \label{visua_all_methods}
\end{figure}
\vspace{-10pt}
\subsection{Visualization}
We utilized t-SNE to visualize the embeddings produced by DSGC and several strong baselines, including BNC~\cite{chiang2012scalable}, BRC~\cite{chiang2012scalable}, SPONGE~\cite{cucuringu2019sponge}, and SPONGE$_{sym}$~\cite{cucuringu2019sponge}, on SSBM ($N=1000$, $K=5$, $p=0.01$, $\eta=0.02$) in Fig.~\ref{visua_all_methods}. Both BNC and BRC exhibit mode collapse, where most nodes are grouped into one or a few clusters. SPONGE and SPONGE$_{sym}$ show improved clustering structures. However, SPONGE lacks a clear boundary between clusters while SPONGE$_{sym}$ appears to form $6$ clusters with a central cluster where nodes from different true clusters are mixed. This indicates its potential issue with handling nodes connected by negative edges, which are typically located at cluster boundaries. In contrast, DSGC successfully pushes nodes linked by negative edges apart, effectively eliminating the central cluster phenomenon in SPONGE$_{sym}$. This result is attributed to the exclusion of the ``\textit{EEF}'' principle and the incorporation of the term $(-\textbf{A}^{-})$ in the graph encoder.
% We conjecture that those nodes in that center cluster are likely to be connected by noisy edges or located at the boundary, leading to low clustering assignment confidence. on the contrary, our DSGC successfully generates $5$ clusters, consistent with the ground truth number of clusters, shown in Fig.~\ref{visua_all_methods}(e). That is because the term $(-\textbf{A}^{-})$ and the abandonment of the traditional principle "the Enemy of one's Enemy is a Friend (EEF)" in our encoder can push nodes with negative links, located at the boundary, away from each other, so the center cluster has disappeared.

% \begin{figure}[t]
% 	\centering
% 		\subfigure[BNC]{
% 			\includegraphics[width=0.12\textwidth]{figures/1000BNC_5_0.01_0.04_visu.pdf}
% 	}
%  		\subfigure[BRC]{
% 			\includegraphics[width=0.12\textwidth]{figures/1000BRC_5_0.01_0.04_visu.pdf}
% 	}
%             \subfigure[SPONGE]{
% 			\includegraphics[width=0.12\textwidth]{figures/1000_5_0.01_0.04_SPONGE_visu.pdf}
% 	}
%             \subfigure[SPONGE$_{sym}$]{
% 			\includegraphics[width=0.12\textwidth]{figures/1000_5_0.01_0.04_SPONGE_sym_visu.pdf}
% 	}
%              \subfigure[ours]{
% 			\includegraphics[width=0.12\textwidth]{figures/ours_1000_5_0.01_0.04.pdf}
% 	}

% 	\caption{Comparison of clustering results}
%         \label{balance theory}
% \end{figure}

\subsection{Unlabeled Graphs}
We also evaluated DSGC on unlabeled real-world signed graphs,  S\&P1500~\cite{cucuringu2019sponge} and Rainfall~\cite{cucuringu2019sponge}, comparing it  against three baselines, BRC~\cite{chiang2012scalable}, BNC~\cite{chiang2012scalable}, and SPONGE\_{sym}~\cite{cucuringu2019sponge}. The adjacency matrices of these graphs were sorted by predicted cluster membership to visually assess clustering outcomes. More evaluations are provided in App.~\ref{app:unlabel_data}.
% We compare the clustering results of our DSGC against three algorithms, i.e., BRC~\cite{chiang2012scalable}, BNC~\cite{chiang2012scalable}, SPONGE~\cite{cucuringu2019sponge}. 

\begin{figure}[ht!]
	\centering
		\subfigure[BRC]{
			\includegraphics[width=0.07\textwidth, trim=0 20 0 20]{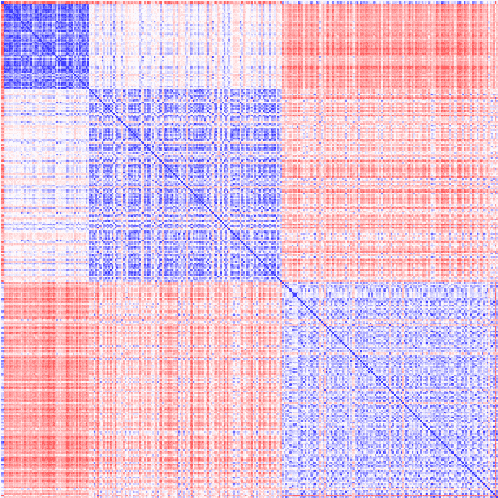}
	}\hspace{4mm} 
 		\subfigure[BNC]{
			\includegraphics[width=0.07\textwidth, trim=0 20 0 20]{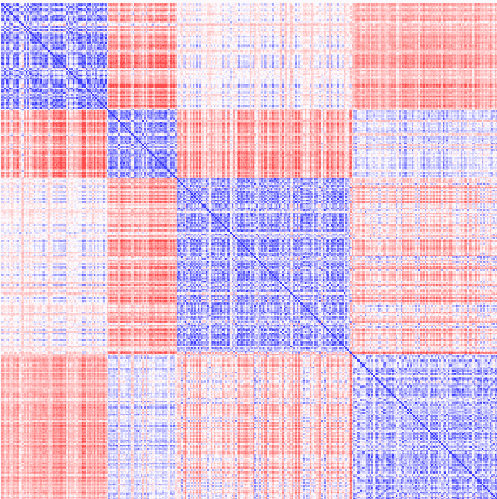}
	}\hspace{4mm}
            \subfigure[SPONGE]{
			\includegraphics[width=0.07\textwidth, trim=0 20 0 20]{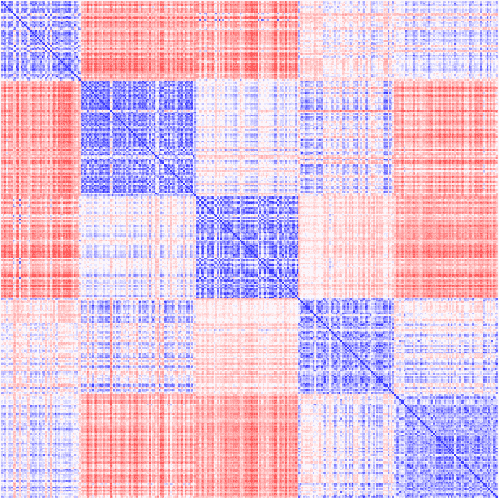}
	}\hspace{4mm}
             \subfigure[Ours]{
			\includegraphics[width=0.07\textwidth, trim=0 20 0 20]{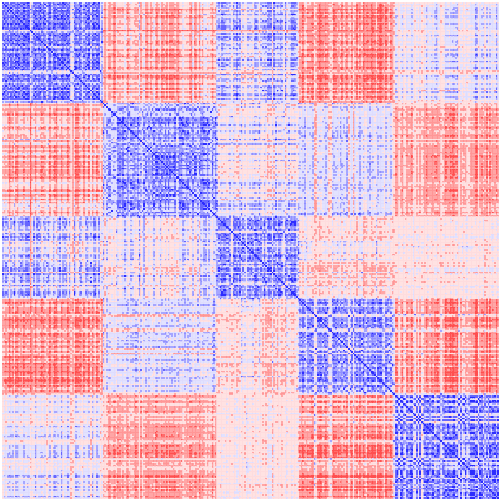}
	}\\
        		\subfigure[BRC]{
			\includegraphics[width=0.07\textwidth, trim=0 20 0 20]{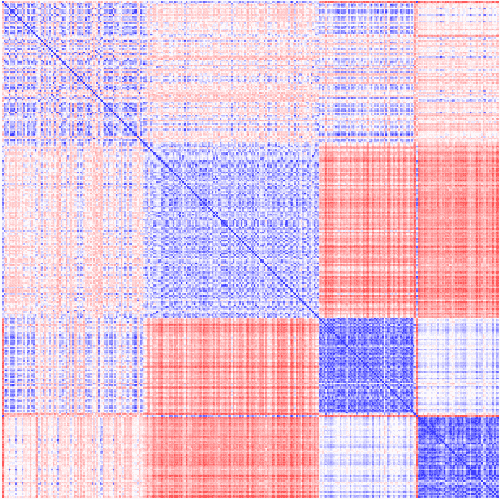}
	}\hspace{4mm}
 		\subfigure[BNC]{
			\includegraphics[width=0.07\textwidth, trim=0 20 0 20]{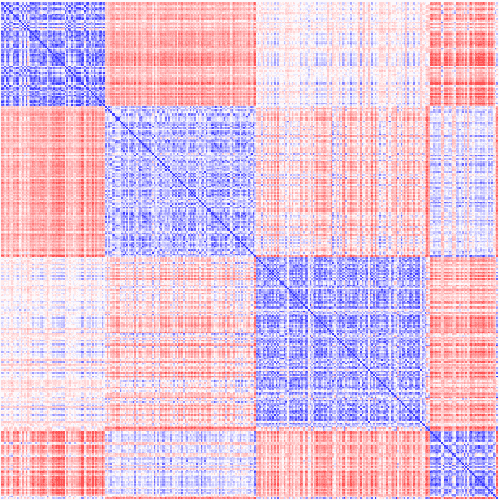}
	}\hspace{4mm}
            \subfigure[SPONGE]{
			\includegraphics[width=0.07\textwidth, trim=0 20 0 20]{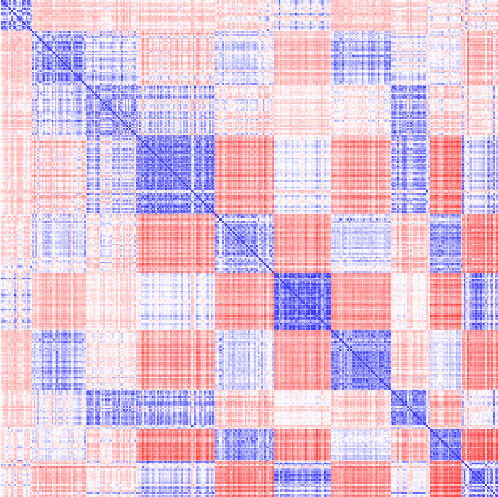}
	}\hspace{4mm}
             \subfigure[Ours]{
			\includegraphics[width=0.07\textwidth, trim=0 20 0 20]{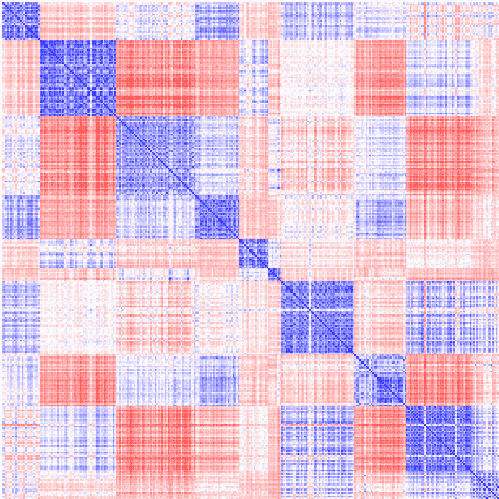}
	}
	\caption{Sorted adjacency matrix for the Rainfall dataset with $K=5$ (top row) and $K=10$ (bottom row).}
        \label{fig:rainfall}
\end{figure}

\textbf{Rainfall.} Following \cite{cucuringu2019sponge}, we analyzed the clustering structures for $K=\{5, 10\}$ in Fig.~\ref{fig:rainfall}, where blue and red denote positive and negative edges, respectively\footnote{Darker blue diagonal blocks indicate more cohesive clusters, while darker pink non-diagonal blocks signify stronger negative relationships between clusters, enhancing the clarity of the clustering semantics.}. Both BRC and BNC fail to identify the expected number of clusters, resulting in model collapse. In contrast, DSGC successfully identifies the specified clusters ($5$ or $10$) and exhibits higher ratios of positive internal edges and stronger negative inter-cluster edges compared to SPONGE, indicating more cohesive and well-defined clusters.

\begin{figure}[ht!]
	\centering
 		\subfigure[BRC]{
 			\includegraphics[width=0.07\textwidth, trim=0 0 0 20]{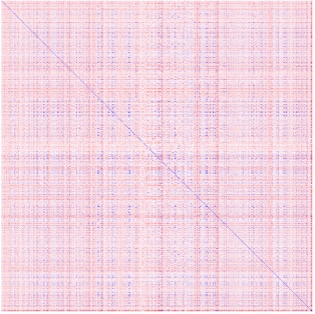}
	}\hspace{4mm}
  		\subfigure[BNC]{
 			\includegraphics[width=0.07\textwidth, trim=0 0 0 20]{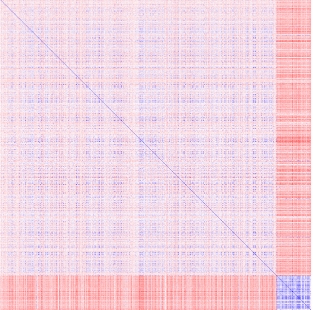}
	}\hspace{4mm}
             \subfigure[SPONGE]{
 			\includegraphics[width=0.07\textwidth, trim=0 0 0 20]{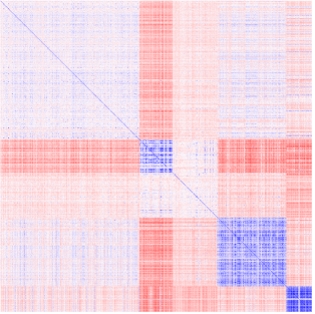}
	}\hspace{4mm}
              \subfigure[Ours]{
 			\includegraphics[width=0.07\textwidth, trim=0 0 0 20]{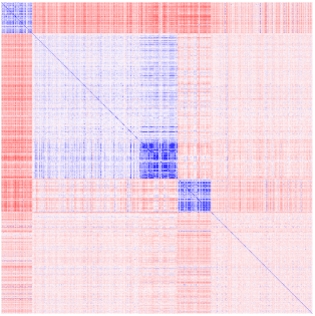}
	}\\
         		\subfigure[BRC]{
 			\includegraphics[width=0.07\textwidth, trim=0 0 0 20]{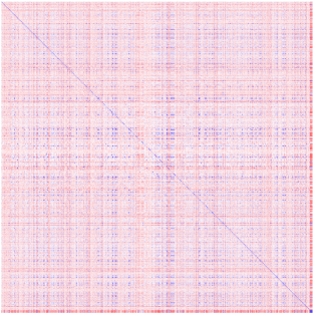}
	}\hspace{4mm}
  		\subfigure[BNC]{
 			\includegraphics[width=0.07\textwidth, trim=0 0 0 20]{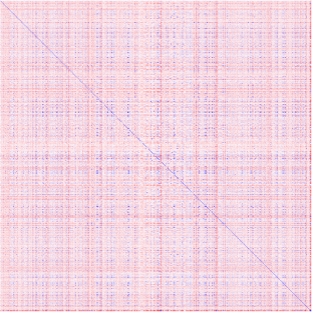}
	}\hspace{4mm}
             \subfigure[SPONGE]{
 			\includegraphics[width=0.07\textwidth, trim=0 0 0 20]{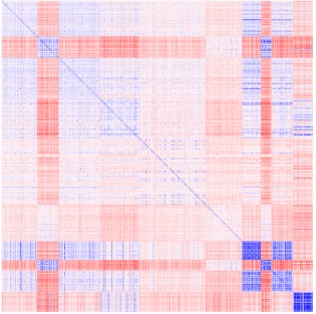}
	}\hspace{4mm}
              \subfigure[Ours]{
 			\includegraphics[width=0.07\textwidth, trim=0 0 0 20]{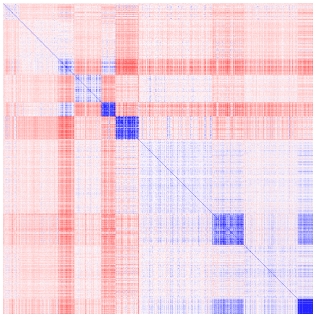}
	}

	\caption{Sorted adjacency matrix for S\&P$1500$ with $K=5$ (top row) and $K=10$ (bottom row).}
         \label{fig:sp1500}
\end{figure}

\textbf{S\&P1500.} Fig.~\ref{fig:sp1500} shows the clustering structures for $K=\{5, 10\}$. BRC and BNC suffer model collapse, placing most nodes into a single large, sparse cluster. In contrast, DSGC produces clear, compact clusters with significantly higher ratios of positive to negative internal edges than the entire graph, indicating more effective clustering that even surpasses SPONGE in identifying relevant groupings.
% BNC only identifies $2$ clusters. Our DSGC succeeded in finding $5$ clusters (top row), and exhibit discernible associations. By our DSGC, all clusters demonstrate a significantly higher ratio of positive to negative internal edges, compared to that of the graph as a whole.

\vspace{-15pt}
\section{CONCLUSION}
In this paper, we introduce DSGC, a novel deep signed graph clustering method, to enhance the clarity of cluster boundaries. Existing approaches generally rely on the Social Balance Theory, primarily suitable for $2$-way clustering. In contrast, DSGC leverages the Weak Balance Theory to address more general $K$-way clustering without the need for explicit labels. DSGC first introduces two pre-processing techniques, VS-R and DA, to denoise and structurally enhance signed graphs before clustering. Then, DSGC constructs a clustering-oriented signed neural network that produces more discriminative representations, specifically for nodes linked negatively. By optimizing a non-linear transformation for node clustering assignments, DSGC significantly outperforms existing methods, establishing clearer and more meaningful cluster distinctions in complex multi-cluster scenarios.

\begin{acks}
% The authors would also like to thank the anonymous reviewers for their constructive feedback, which greatly helped us to improve the quality and clarity of the paper. 
This work is partially supported by National Natural Science Foundation of China (NSFC) under the grant No. 92270125 and No. 62276024.

\end{acks}

\bibliographystyle{last_ACM-Reference-Format}
\balance
\bibliography{final_references}


\begin{thebibliography}{45}

%%% the Web address.

\ifx \showCODEN    \undefined \def \showCODEN     #1{\unskip}     \fi
\ifx \showISBNx    \undefined \def \showISBNx     #1{\unskip}     \fi
\ifx \showISBNxiii \undefined \def \showISBNxiii  #1{\unskip}     \fi
\ifx \showISSN     \undefined \def \showISSN      #1{\unskip}     \fi
\ifx \showLCCN     \undefined \def \showLCCN      #1{\unskip}     \fi
\ifx \shownote     \undefined \def \shownote      #1{#1}          \fi
\ifx \showarticletitle \undefined \def \showarticletitle #1{#1}   \fi
\ifx \showURL      \undefined \def \showURL       {\relax}        \fi
% invisible to TeX
\providecommand\bibfield[2]{#2}
\providecommand\bibinfo[2]{#2}
\providecommand\natexlab[1]{#1}
\providecommand\showeprint[2][]{arXiv:#2}

\bibitem[Bhowmick et~al\mbox{.}(2024)]%
        {BhowmickKHSM24}
\bibfield{author}{\bibinfo{person}{Aritra Bhowmick}, \bibinfo{person}{Mert
  Kosan}, \bibinfo{person}{Zexi Huang}, \bibinfo{person}{Ambuj~K. Singh}, {and}
  \bibinfo{person}{Sourav Medya}.} \bibinfo{year}{2024}\natexlab{}.
\newblock \showarticletitle{{DGCLUSTER:} {A} Neural Framework for Attributed
  Graph Clustering via Modularity Maximization}. In
  \bibinfo{booktitle}{\emph{Thirty-Eighth {AAAI} Conference on Artificial
  Intelligence}}. \bibinfo{pages}{11069--11077}.
\newblock


\bibitem[Bo et~al\mbox{.}(2020)]%
        {bo2020structural}
\bibfield{author}{\bibinfo{person}{Deyu Bo}, \bibinfo{person}{Xiao Wang},
  \bibinfo{person}{Chuan Shi}, \bibinfo{person}{Meiqi Zhu},
  \bibinfo{person}{Emiao Lu}, {and} \bibinfo{person}{Peng Cui}.}
  \bibinfo{year}{2020}\natexlab{}.
\newblock \showarticletitle{Structural Deep Clustering Network}. In
  \bibinfo{booktitle}{\emph{The Web Conference 2020}}.
  \bibinfo{pages}{1400--1410}.
\newblock


\bibitem[Cartwright and Harary(1956)]%
        {cartwright1956structural}
\bibfield{author}{\bibinfo{person}{Dorwin Cartwright} {and}
  \bibinfo{person}{Frank Harary}.} \bibinfo{year}{1956}\natexlab{}.
\newblock \showarticletitle{Structural balance: a generalization of Heider's
  theory.}
\newblock \bibinfo{journal}{\emph{Psychological review}} \bibinfo{volume}{63},
  \bibinfo{number}{5} (\bibinfo{year}{1956}), \bibinfo{pages}{277}.
\newblock


\bibitem[Chen et~al\mbox{.}(2018)]%
        {chen2018bridge}
\bibfield{author}{\bibinfo{person}{Yiqi Chen}, \bibinfo{person}{Tieyun Qian},
  \bibinfo{person}{Huan Liu}, {and} \bibinfo{person}{Ke Sun}.}
  \bibinfo{year}{2018}\natexlab{}.
\newblock \showarticletitle{"Bridge": Enhanced Signed Directed Network
  Embedding}. In \bibinfo{booktitle}{\emph{Proceedings of the 27th {ACM}
  International Conference on Information and Knowledge Management, {CIKM}}}.
  \bibinfo{pages}{773--782}.
\newblock


\bibitem[Chiang et~al\mbox{.}(2012)]%
        {chiang2012scalable}
\bibfield{author}{\bibinfo{person}{Kai{-}Yang Chiang},
  \bibinfo{person}{Joyce~Jiyoung Whang}, {and} \bibinfo{person}{Inderjit~S.
  Dhillon}.} \bibinfo{year}{2012}\natexlab{}.
\newblock \showarticletitle{Scalable clustering of signed networks using
  balance normalized cut}. In \bibinfo{booktitle}{\emph{21st {ACM}
  International Conference on Information and Knowledge Management, {CIKM}}}.
  \bibinfo{pages}{615--624}.
\newblock


\bibitem[Cucuringu et~al\mbox{.}(2019)]%
        {cucuringu2019sponge}
\bibfield{author}{\bibinfo{person}{Mihai Cucuringu}, \bibinfo{person}{Peter
  Davies}, \bibinfo{person}{Aldo Glielmo}, {and} \bibinfo{person}{Hemant
  Tyagi}.} \bibinfo{year}{2019}\natexlab{}.
\newblock \showarticletitle{{SPONGE:} {A} generalized eigenproblem for
  clustering signed networks}. In \bibinfo{booktitle}{\emph{The 22nd
  International Conference on Artificial Intelligence and Statistics,
  {AISTATS}}}, Vol.~\bibinfo{volume}{89}. \bibinfo{pages}{1088--1098}.
\newblock


\bibitem[Davis(1967)]%
        {davis1967clustering}
\bibfield{author}{\bibinfo{person}{James~A Davis}.}
  \bibinfo{year}{1967}\natexlab{}.
\newblock \showarticletitle{Clustering and structural balance in graphs}.
\newblock \bibinfo{journal}{\emph{Human relations}} \bibinfo{volume}{20},
  \bibinfo{number}{2} (\bibinfo{year}{1967}), \bibinfo{pages}{181--187}.
\newblock


\bibitem[Derr et~al\mbox{.}(2018a)]%
        {DerrAT18}
\bibfield{author}{\bibinfo{person}{Tyler Derr}, \bibinfo{person}{Charu~C.
  Aggarwal}, {and} \bibinfo{person}{Jiliang Tang}.}
  \bibinfo{year}{2018}\natexlab{a}.
\newblock \showarticletitle{Signed Network Modeling Based on Structural Balance
  Theory}. In \bibinfo{booktitle}{\emph{Proceedings of the 27th {ACM}
  International Conference on Information and Knowledge Management, {CIKM}}}.
  \bibinfo{pages}{557--566}.
\newblock


\bibitem[Derr et~al\mbox{.}(2018b)]%
        {derr2018signed}
\bibfield{author}{\bibinfo{person}{Tyler Derr}, \bibinfo{person}{Yao Ma}, {and}
  \bibinfo{person}{Jiliang Tang}.} \bibinfo{year}{2018}\natexlab{b}.
\newblock \showarticletitle{Signed Graph Convolutional Networks}. In
  \bibinfo{booktitle}{\emph{{IEEE} International Conference on Data Mining,
  {ICDM}}}. \bibinfo{pages}{929--934}.
\newblock


\bibitem[Diaz{-}Diaz and Estrada(2024)]%
        {diaz24signed}
\bibfield{author}{\bibinfo{person}{Fernando Diaz{-}Diaz} {and}
  \bibinfo{person}{Ernesto Estrada}.} \bibinfo{year}{2024}\natexlab{}.
\newblock \showarticletitle{Signed graphs in data sciences via communicability
  geometry}.
\newblock \bibinfo{journal}{\emph{arXiv preprint arXiv:2403.07493}}
  (\bibinfo{year}{2024}).
\newblock


\bibitem[Fujita et~al\mbox{.}(2012)]%
        {FujitaSKSPM12}
\bibfield{author}{\bibinfo{person}{Andr{\'{e}} Fujita},
  \bibinfo{person}{Patricia Severino}, \bibinfo{person}{Kaname Kojima},
  \bibinfo{person}{Jo{\~{a}}o~Ricardo Sato},
  \bibinfo{person}{Alexandre~Galv{\~{a}}o Patriota}, {and}
  \bibinfo{person}{Satoru Miyano}.} \bibinfo{year}{2012}\natexlab{}.
\newblock \showarticletitle{Functional clustering of time series gene
  expression data by Granger causality}.
\newblock \bibinfo{journal}{\emph{{BMC} Syst. Biol.}}  \bibinfo{volume}{6}
  (\bibinfo{year}{2012}), \bibinfo{pages}{137}.
\newblock


\bibitem[Gates and Ahn(2017)]%
        {gates2017impact}
\bibfield{author}{\bibinfo{person}{Alexander~J. Gates} {and}
  \bibinfo{person}{Yong{-}Yeol Ahn}.} \bibinfo{year}{2017}\natexlab{}.
\newblock \showarticletitle{The Impact of Random Models on Clustering
  Similarity}.
\newblock \bibinfo{journal}{\emph{J. Mach. Learn. Res.}}  \bibinfo{volume}{18}
  (\bibinfo{year}{2017}), \bibinfo{pages}{87:1--87:28}.
\newblock


\bibitem[Ghasemian et~al\mbox{.}(2020)]%
        {GhasemianHC20}
\bibfield{author}{\bibinfo{person}{Amir Ghasemian}, \bibinfo{person}{Homa
  Hosseinmardi}, {and} \bibinfo{person}{Aaron Clauset}.}
  \bibinfo{year}{2020}\natexlab{}.
\newblock \showarticletitle{Evaluating Overfit and Underfit in Models of
  Network Community Structure}.
\newblock \bibinfo{journal}{\emph{{IEEE} Trans. Knowl. Data Eng.}}
  \bibinfo{volume}{32}, \bibinfo{number}{9} (\bibinfo{year}{2020}),
  \bibinfo{pages}{1722--1735}.
\newblock


\bibitem[Graham et~al\mbox{.}(2013)]%
        {chung2013dirichlet}
\bibfield{author}{\bibinfo{person}{Fan~Chung Graham},
  \bibinfo{person}{Alexander Tsiatas}, {and} \bibinfo{person}{Wensong Xu}.}
  \bibinfo{year}{2013}\natexlab{}.
\newblock \showarticletitle{Dirichlet PageRank and Ranking Algorithms Based on
  Trust and Distrust}.
\newblock \bibinfo{journal}{\emph{Internet Math.}} \bibinfo{volume}{9},
  \bibinfo{number}{1} (\bibinfo{year}{2013}), \bibinfo{pages}{113--134}.
\newblock


\bibitem[Harary(1953)]%
        {harary1953notion}
\bibfield{author}{\bibinfo{person}{Frank Harary}.}
  \bibinfo{year}{1953}\natexlab{}.
\newblock \showarticletitle{On the notion of balance of a signed graph.}
\newblock \bibinfo{journal}{\emph{Michigan Mathematical Journal}}
  \bibinfo{volume}{2}, \bibinfo{number}{2} (\bibinfo{year}{1953}),
  \bibinfo{pages}{143--146}.
\newblock


\bibitem[He et~al\mbox{.}(2022)]%
        {he2022sssnet}
\bibfield{author}{\bibinfo{person}{Yixuan He}, \bibinfo{person}{Gesine
  Reinert}, \bibinfo{person}{Songchao Wang}, {and} \bibinfo{person}{Mihai
  Cucuringu}.} \bibinfo{year}{2022}\natexlab{}.
\newblock \showarticletitle{{SSSNET:} Semi-Supervised Signed Network
  Clustering}. In \bibinfo{booktitle}{\emph{Proceedings of the 2022 {SIAM}
  International Conference on Data Mining, {SDM}}}. \bibinfo{pages}{244--252}.
\newblock


\bibitem[Huang et~al\mbox{.}(2019)]%
        {huang2019signed}
\bibfield{author}{\bibinfo{person}{Junjie Huang}, \bibinfo{person}{Huawei
  Shen}, \bibinfo{person}{Liang Hou}, {and} \bibinfo{person}{Xueqi Cheng}.}
  \bibinfo{year}{2019}\natexlab{}.
\newblock \showarticletitle{Signed Graph Attention Networks}. In
  \bibinfo{booktitle}{\emph{Artificial Neural Networks and Machine Learning}},
  Vol.~\bibinfo{volume}{11731}. \bibinfo{publisher}{Springer},
  \bibinfo{pages}{566--577}.
\newblock


\bibitem[Huang et~al\mbox{.}(2021)]%
        {HuangSHC21}
\bibfield{author}{\bibinfo{person}{Junjie Huang}, \bibinfo{person}{Huawei
  Shen}, \bibinfo{person}{Liang Hou}, {and} \bibinfo{person}{Xueqi Cheng}.}
  \bibinfo{year}{2021}\natexlab{}.
\newblock \showarticletitle{{SDGNN:} Learning Node Representation for Signed
  Directed Networks}. In \bibinfo{booktitle}{\emph{Thirty-Fifth {AAAI}
  Conference on Artificial Intelligence}}. \bibinfo{pages}{196--203}.
\newblock


\bibitem[Islam et~al\mbox{.}(2018)]%
        {islam2018signet}
\bibfield{author}{\bibinfo{person}{Mohammad~Raihanul Islam},
  \bibinfo{person}{B.~Aditya Prakash}, {and} \bibinfo{person}{Naren
  Ramakrishnan}.} \bibinfo{year}{2018}\natexlab{}.
\newblock \showarticletitle{SIGNet: Scalable Embeddings for Signed Networks}.
  In \bibinfo{booktitle}{\emph{Advances in Knowledge Discovery and Data
  Mining}}, Vol.~\bibinfo{volume}{10938}. \bibinfo{pages}{157--169}.
\newblock


\bibitem[Jung et~al\mbox{.}(2020)]%
        {jung2020signed}
\bibfield{author}{\bibinfo{person}{Jinhong Jung}, \bibinfo{person}{Jaemin Yoo},
  {and} \bibinfo{person}{U Kang}.} \bibinfo{year}{2020}\natexlab{}.
\newblock \showarticletitle{Signed Graph Diffusion Network}.
\newblock  (\bibinfo{year}{2020}).
\newblock
\showeprint[arXiv]{2012.14191}


\bibitem[Knyazev(2001)]%
        {knyazev2001toward}
\bibfield{author}{\bibinfo{person}{Andrew~V. Knyazev}.}
  \bibinfo{year}{2001}\natexlab{}.
\newblock \showarticletitle{Toward the Optimal Preconditioned Eigensolver:
  Locally Optimal Block Preconditioned Conjugate Gradient Method}.
\newblock \bibinfo{journal}{\emph{{SIAM} J. Sci. Comput.}}
  \bibinfo{volume}{23}, \bibinfo{number}{2} (\bibinfo{year}{2001}),
  \bibinfo{pages}{517--541}.
\newblock


\bibitem[Kunegis et~al\mbox{.}(2010)]%
        {kunegis2010spectral}
\bibfield{author}{\bibinfo{person}{J{\'{e}}r{\^{o}}me Kunegis},
  \bibinfo{person}{Stephan Schmidt}, \bibinfo{person}{Andreas Lommatzsch},
  \bibinfo{person}{J{\"{u}}rgen Lerner}, \bibinfo{person}{Ernesto William~De
  Luca}, {and} \bibinfo{person}{Sahin Albayrak}.}
  \bibinfo{year}{2010}\natexlab{}.
\newblock \showarticletitle{Spectral Analysis of Signed Graphs for Clustering,
  Prediction and Visualization}. In \bibinfo{booktitle}{\emph{Proceedings of
  the {SIAM} International Conference on Data Mining, {SDM}}}.
  \bibinfo{pages}{559--570}.
\newblock


\bibitem[Lee et~al\mbox{.}(2020)]%
        {LeeSK20}
\bibfield{author}{\bibinfo{person}{Yeon{-}Chang Lee}, \bibinfo{person}{Nayoun
  Seo}, {and} \bibinfo{person}{Sang{-}Wook Kim}.}
  \bibinfo{year}{2020}\natexlab{}.
\newblock \showarticletitle{Are Negative Links Really Beneficial to Network
  Embedding?: In-Depth Analysis and Interesting Results}. In
  \bibinfo{booktitle}{\emph{The 29th {ACM} International Conference on
  Information and Knowledge Management}}. \bibinfo{pages}{2113--2116}.
\newblock


\bibitem[Li et~al\mbox{.}(2023)]%
        {li2023signed}
\bibfield{author}{\bibinfo{person}{Yu Li}, \bibinfo{person}{Meng Qu},
  \bibinfo{person}{Jian Tang}, {and} \bibinfo{person}{Yi Chang}.}
  \bibinfo{year}{2023}\natexlab{}.
\newblock \showarticletitle{Signed Laplacian Graph Neural Networks}. In
  \bibinfo{booktitle}{\emph{Thirty-Seventh {AAAI} Conference on Artificial
  Intelligence, {AAAI} 2023, Thirty-Fifth Conference on Innovative Applications
  of Artificial Intelligence}}. \bibinfo{publisher}{{AAAI} Press},
  \bibinfo{pages}{4444--4452}.
\newblock


\bibitem[Li et~al\mbox{.}(2020)]%
        {li2020learning}
\bibfield{author}{\bibinfo{person}{Yu Li}, \bibinfo{person}{Yuan Tian},
  \bibinfo{person}{Jiawei Zhang}, {and} \bibinfo{person}{Yi Chang}.}
  \bibinfo{year}{2020}\natexlab{}.
\newblock \showarticletitle{Learning Signed Network Embedding via Graph
  Attention}. In \bibinfo{booktitle}{\emph{The Thirty-Fourth {AAAI} Conference
  on Artificial Intelligence, {AAAI}}}. \bibinfo{pages}{4772--4779}.
\newblock


\bibitem[Liu et~al\mbox{.}(2021)]%
        {liu2021signed}
\bibfield{author}{\bibinfo{person}{Haoxin Liu}, \bibinfo{person}{Ziwei Zhang},
  \bibinfo{person}{Peng Cui}, \bibinfo{person}{Yafeng Zhang},
  \bibinfo{person}{Qiang Cui}, \bibinfo{person}{Jiashuo Liu}, {and}
  \bibinfo{person}{Wenwu Zhu}.} \bibinfo{year}{2021}\natexlab{}.
\newblock \showarticletitle{Signed Graph Neural Network with Latent Groups}. In
  \bibinfo{booktitle}{\emph{The 27th {ACM} {SIGKDD} Conference on Knowledge
  Discovery and Data Mining}}. \bibinfo{pages}{1066--1075}.
\newblock


\bibitem[Liu et~al\mbox{.}(2024)]%
        {LiuLCWWZTSFMWC24}
\bibfield{author}{\bibinfo{person}{Yunfei Liu}, \bibinfo{person}{Jintang Li},
  \bibinfo{person}{Yuehe Chen}, \bibinfo{person}{Ruofan Wu},
  \bibinfo{person}{Ericbk Wang}, \bibinfo{person}{Jing Zhou},
  \bibinfo{person}{Sheng Tian}, \bibinfo{person}{Shuheng Shen},
  \bibinfo{person}{Xing Fu}, \bibinfo{person}{Changhua Meng},
  \bibinfo{person}{Weiqiang Wang}, {and} \bibinfo{person}{Liang Chen}.}
  \bibinfo{year}{2024}\natexlab{}.
\newblock \showarticletitle{Revisiting Modularity Maximization for Graph
  Clustering: {A} Contrastive Learning Perspective}. In
  \bibinfo{booktitle}{\emph{Proceedings of the 30th {ACM} {SIGKDD} Conference
  on Knowledge Discovery and Data Mining}}. \bibinfo{pages}{1968--1979}.
\newblock


\bibitem[Liu et~al\mbox{.}(2023)]%
        {00080XZYLL23}
\bibfield{author}{\bibinfo{person}{Yue Liu}, \bibinfo{person}{Ke Liang},
  \bibinfo{person}{Jun Xia}, \bibinfo{person}{Sihang Zhou},
  \bibinfo{person}{Xihong Yang}, \bibinfo{person}{Xinwang Liu}, {and}
  \bibinfo{person}{Stan~Z. Li}.} \bibinfo{year}{2023}\natexlab{}.
\newblock \showarticletitle{Dink-Net: Neural Clustering on Large Graphs}. In
  \bibinfo{booktitle}{\emph{International Conference on Machine Learning,
  {ICML}}}, Vol.~\bibinfo{volume}{202}. \bibinfo{pages}{21794--21812}.
\newblock


\bibitem[Liu et~al\mbox{.}(2022)]%
        {LiuTZLSYZ22}
\bibfield{author}{\bibinfo{person}{Yue Liu}, \bibinfo{person}{Wenxuan Tu},
  \bibinfo{person}{Sihang Zhou}, \bibinfo{person}{Xinwang Liu},
  \bibinfo{person}{Linxuan Song}, \bibinfo{person}{Xihong Yang}, {and}
  \bibinfo{person}{En Zhu}.} \bibinfo{year}{2022}\natexlab{}.
\newblock \showarticletitle{Deep Graph Clustering via Dual Correlation
  Reduction}. In \bibinfo{booktitle}{\emph{Thirty-Sixth {AAAI} Conference on
  Artificial Intelligence, {AAAI}}}. \bibinfo{pages}{7603--7611}.
\newblock


\bibitem[Mercado et~al\mbox{.}(2019a)]%
        {bosch2018node}
\bibfield{author}{\bibinfo{person}{Pedro Mercado}, \bibinfo{person}{Jessica
  Bosch}, {and} \bibinfo{person}{Martin Stoll}.}
  \bibinfo{year}{2019}\natexlab{a}.
\newblock \showarticletitle{Node Classification for Signed Social Networks
  Using Diffuse Interface Methods}. In \bibinfo{booktitle}{\emph{Machine
  Learning and Knowledge Discovery in Databases}},
  Vol.~\bibinfo{volume}{11906}. \bibinfo{pages}{524--540}.
\newblock


\bibitem[Mercado et~al\mbox{.}(2016)]%
        {Mercado2016clustering}
\bibfield{author}{\bibinfo{person}{Pedro Mercado}, \bibinfo{person}{Francesco
  Tudisco}, {and} \bibinfo{person}{Matthias Hein}.}
  \bibinfo{year}{2016}\natexlab{}.
\newblock \showarticletitle{Clustering Signed Networks with the Geometric Mean
  of Laplacians}. In \bibinfo{booktitle}{\emph{Advances in Neural Information
  Processing Systems 29}}. \bibinfo{pages}{4421--4429}.
\newblock


\bibitem[Mercado et~al\mbox{.}(2019b)]%
        {Mercado2019Spectral}
\bibfield{author}{\bibinfo{person}{Pedro Mercado}, \bibinfo{person}{Francesco
  Tudisco}, {and} \bibinfo{person}{Matthias Hein}.}
  \bibinfo{year}{2019}\natexlab{b}.
\newblock \showarticletitle{Spectral Clustering of Signed Graphs via Matrix
  Power Means}. In \bibinfo{booktitle}{\emph{Proceedings of the 36th
  International Conference on Machine Learning, {ICML}}},
  Vol.~\bibinfo{volume}{97}. \bibinfo{pages}{4526--4536}.
\newblock


\bibitem[Pan et~al\mbox{.}(2018)]%
        {PanHLJYZ18}
\bibfield{author}{\bibinfo{person}{Shirui Pan}, \bibinfo{person}{Ruiqi Hu},
  \bibinfo{person}{Guodong Long}, \bibinfo{person}{Jing Jiang},
  \bibinfo{person}{Lina Yao}, {and} \bibinfo{person}{Chengqi Zhang}.}
  \bibinfo{year}{2018}\natexlab{}.
\newblock \showarticletitle{Adversarially Regularized Graph Autoencoder for
  Graph Embedding}. In \bibinfo{booktitle}{\emph{Proceedings of the
  Twenty-Seventh International Joint Conference on Artificial Intelligence,
  {IJCAI}}}. \bibinfo{pages}{2609--2615}.
\newblock


\bibitem[Pan et~al\mbox{.}(2024)]%
        {pan2024csgdn}
\bibfield{author}{\bibinfo{person}{Yiru Pan}, \bibinfo{person}{Xingyu Ji},
  \bibinfo{person}{Jiaqi You}, \bibinfo{person}{Lu Li},
  \bibinfo{person}{Zhenping Liu}, \bibinfo{person}{Xianlong Zhang},
  \bibinfo{person}{Zeyu Zhang}, {and} \bibinfo{person}{Maojun Wang}.}
  \bibinfo{year}{2024}\natexlab{}.
\newblock \showarticletitle{CSGDN: Contrastive Signed Graph Diffusion Network
  for Predicting Crop Gene-Trait Associations}.
\newblock \bibinfo{journal}{\emph{arXiv:2410.07511}} (\bibinfo{year}{2024}).
\newblock


\bibitem[Shahriari and Jalili(2014)]%
        {shahriari2014ranking}
\bibfield{author}{\bibinfo{person}{Moshen Shahriari} {and}
  \bibinfo{person}{Mahdi Jalili}.} \bibinfo{year}{2014}\natexlab{}.
\newblock \showarticletitle{Ranking Nodes in Signed Social Networks}.
\newblock \bibinfo{journal}{\emph{Soc. Netw. Anal. Min.}} \bibinfo{volume}{4},
  \bibinfo{number}{1} (\bibinfo{year}{2014}), \bibinfo{pages}{172}.
\newblock


\bibitem[Tu et~al\mbox{.}(2021)]%
        {tu2021deep}
\bibfield{author}{\bibinfo{person}{Wenxuan Tu}, \bibinfo{person}{Sihang Zhou},
  \bibinfo{person}{Xinwang Liu}, \bibinfo{person}{Xifeng Guo},
  \bibinfo{person}{Zhiping Cai}, \bibinfo{person}{En Zhu}, {and}
  \bibinfo{person}{Jieren Cheng}.} \bibinfo{year}{2021}\natexlab{}.
\newblock \showarticletitle{Deep Fusion Clustering Network}. In
  \bibinfo{booktitle}{\emph{Thirty-Fifth {AAAI} Conference on Artificial
  Intelligence, {AAAI}}}. \bibinfo{pages}{9978--9987}.
\newblock


\bibitem[Tzeng et~al\mbox{.}(2020)]%
        {tzeng2020discovering}
\bibfield{author}{\bibinfo{person}{Ruo{-}Chun Tzeng}, \bibinfo{person}{Bruno
  Ordozgoiti}, {and} \bibinfo{person}{Aristides Gionis}.}
  \bibinfo{year}{2020}\natexlab{}.
\newblock \showarticletitle{Discovering conflicting groups in signed networks}.
  In \bibinfo{booktitle}{\emph{Advances in Neural Information Processing
  Systems 33}}.
\newblock


\bibitem[Wang et~al\mbox{.}(2019)]%
        {wang2019attributed}
\bibfield{author}{\bibinfo{person}{Chun Wang}, \bibinfo{person}{Shirui Pan},
  \bibinfo{person}{Ruiqi Hu}, \bibinfo{person}{Guodong Long},
  \bibinfo{person}{Jing Jiang}, {and} \bibinfo{person}{Chengqi Zhang}.}
  \bibinfo{year}{2019}\natexlab{}.
\newblock \showarticletitle{Attributed Graph Clustering: {A} Deep Attentional
  Embedding Approach}. In \bibinfo{booktitle}{\emph{Proceedings of the
  Twenty-Eighth International Joint Conference on Artificial Intelligence,
  {IJCAI}}}. \bibinfo{pages}{3670--3676}.
\newblock


\bibitem[Wang et~al\mbox{.}(2017a)]%
        {WangATL17}
\bibfield{author}{\bibinfo{person}{Suhang Wang}, \bibinfo{person}{Charu~C.
  Aggarwal}, \bibinfo{person}{Jiliang Tang}, {and} \bibinfo{person}{Huan Liu}.}
  \bibinfo{year}{2017}\natexlab{a}.
\newblock \showarticletitle{Attributed Signed Network Embedding}. In
  \bibinfo{booktitle}{\emph{Proceedings of the 2017 {ACM} on Conference on
  Information and Knowledge Management, {CIKM}}}. \bibinfo{publisher}{{ACM}},
  \bibinfo{pages}{137--146}.
\newblock


\bibitem[Wang et~al\mbox{.}(2017b)]%
        {wang2017signed}
\bibfield{author}{\bibinfo{person}{Suhang Wang}, \bibinfo{person}{Jiliang
  Tang}, \bibinfo{person}{Charu~C. Aggarwal}, \bibinfo{person}{Yi Chang}, {and}
  \bibinfo{person}{Huan Liu}.} \bibinfo{year}{2017}\natexlab{b}.
\newblock \showarticletitle{Signed Network Embedding in Social Media}. In
  \bibinfo{booktitle}{\emph{Proceedings of the 2017 {SIAM} International
  Conference on Data Mining, Houston}}. \bibinfo{publisher}{{SIAM}},
  \bibinfo{pages}{327--335}.
\newblock


\bibitem[Xu et~al\mbox{.}(2019)]%
        {xu2019link}
\bibfield{author}{\bibinfo{person}{Pinghua Xu}, \bibinfo{person}{Wenbin Hu},
  \bibinfo{person}{Jia Wu}, {and} \bibinfo{person}{Bo Du}.}
  \bibinfo{year}{2019}\natexlab{}.
\newblock \showarticletitle{Link Prediction with Signed Latent Factors in
  Signed Social Networks}. In \bibinfo{booktitle}{\emph{Proceedings of the 25th
  {ACM} {SIGKDD} International Conference on Knowledge Discovery {\&} Data
  Mining}}. \bibinfo{pages}{1046--1054}.
\newblock


\bibitem[Yang et~al\mbox{.}(2007)]%
        {yang2007community}
\bibfield{author}{\bibinfo{person}{Bo Yang}, \bibinfo{person}{William~K.
  Cheung}, {and} \bibinfo{person}{Jiming Liu}.}
  \bibinfo{year}{2007}\natexlab{}.
\newblock \showarticletitle{Community Mining from Signed Social Networks}.
\newblock \bibinfo{journal}{\emph{{IEEE} Trans. Knowl. Data Eng.}}
  \bibinfo{volume}{19}, \bibinfo{number}{10} (\bibinfo{year}{2007}),
  \bibinfo{pages}{1333--1348}.
\newblock


\bibitem[Zhang et~al\mbox{.}(2024)]%
        {zhang2024dropedge}
\bibfield{author}{\bibinfo{person}{Zeyu Zhang}, \bibinfo{person}{Lu Li},
  \bibinfo{person}{Shuyan Wan}, \bibinfo{person}{Sijie Wang},
  \bibinfo{person}{Zhiyi Wang}, \bibinfo{person}{Zhiyuan Lu},
  \bibinfo{person}{Dong Hao}, {and} \bibinfo{person}{Wanli Li}.}
  \bibinfo{year}{2024}\natexlab{}.
\newblock \showarticletitle{DropEdge not Foolproof: Effective Augmentation
  Method for Signed Graph Neural Networks}.
\newblock \bibinfo{journal}{\emph{NeurIPS}} (\bibinfo{year}{2024}).
\newblock


\bibitem[Zhang et~al\mbox{.}(2023a)]%
        {zhang2023contrastive}
\bibfield{author}{\bibinfo{person}{Zeyu Zhang}, \bibinfo{person}{Jiamou Liu},
  \bibinfo{person}{Kaiqi Zhao}, \bibinfo{person}{Song Yang},
  \bibinfo{person}{Xianda Zheng}, {and} \bibinfo{person}{Yifei Wang}.}
  \bibinfo{year}{2023}\natexlab{a}.
\newblock \showarticletitle{Contrastive learning for signed bipartite graphs}.
  In \bibinfo{booktitle}{\emph{Proceedings of the 46th International ACM SIGIR
  Conference on Research and Development in Information Retrieval}}.
  \bibinfo{pages}{1629--1638}.
\newblock


\bibitem[Zhang et~al\mbox{.}(2023b)]%
        {ZhangLZWHW2023RSGNN}
\bibfield{author}{\bibinfo{person}{Zeyu Zhang}, \bibinfo{person}{Jiamou Liu},
  \bibinfo{person}{Xianda Zheng}, \bibinfo{person}{Yifei Wang},
  \bibinfo{person}{Pengqian Han}, \bibinfo{person}{Yupan Wang},
  \bibinfo{person}{Kaiqi Zhao}, {and} \bibinfo{person}{Zijian Zhang}.}
  \bibinfo{year}{2023}\natexlab{b}.
\newblock \showarticletitle{{RSGNN:} {A} Model-agnostic Approach for Enhancing
  the Robustness of Signed Graph Neural Networks}. In
  \bibinfo{booktitle}{\emph{Proceedings of the {ACM} Web Conference 2023,
  {WWW}}}. \bibinfo{pages}{60--70}.
\newblock


\end{thebibliography}

\newpage
\appendix
\section{Sociological Theories }~\label{app:sociological_theories}
\begin{figure}[ht!]
	\centering
		\subfigure[$2$-way clustering]{
			\includegraphics[width=0.16\textwidth]{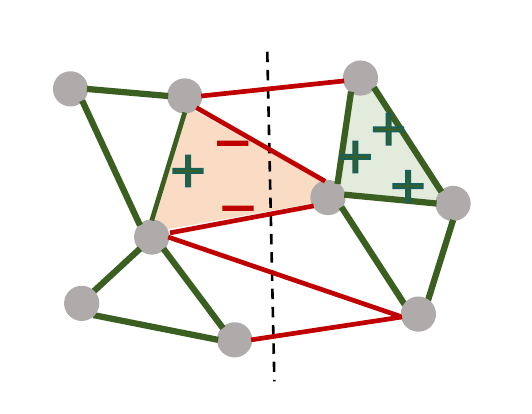}
	}
		\subfigure[$3$-way clustering]{
			\includegraphics[width=0.17\textwidth]{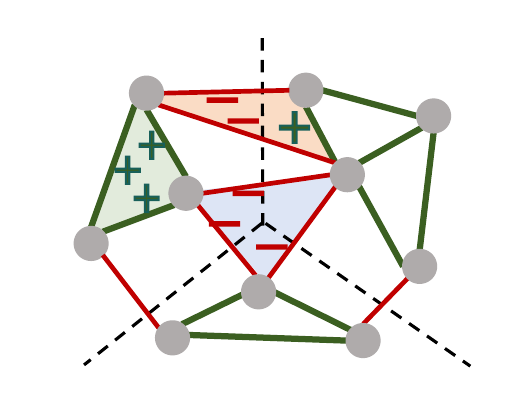}
	}
	\caption{Illustrative comparison of Balance and Weak Balance Theory. The orange and blue triangles prove ``\textit{EEF}'' and ``\textit{EEE}'' principles.}
        \label{app:balance_theory}
\end{figure}

In this section, we specifically demonstrate two sociological theories: Social Balance and Weak Balance Theory.
\begin{figure}[ht!]
	\centering
		\subfigure[Balance Theory]{
			\includegraphics[width=0.22\textwidth, trim=0 1.5 0 0]{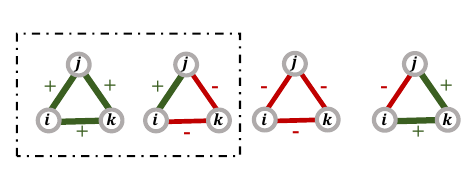}
	}
		\subfigure[Weak Balance Theory]{
			\includegraphics[width=0.22\textwidth, trim=0 0 0 0]{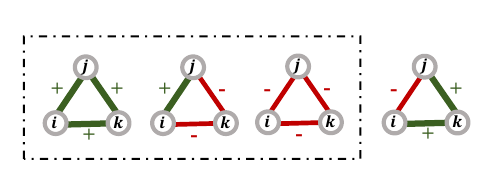}
	}
	\caption{Illustrations of sociological theories. Triads in boxes are considered as balanced and weakly balanced, respectively.}
        \label{app_fig:triads_theories}
\end{figure}
% The study of clusterability in signed graphs can be traced back to the foundational \textit{Social Balance Theory}~, stating that a signed undirected network without attributes can naturally exhibit a global structure conducive to $2$-way clustering.
\begin{theorem}[\textbf{Social Balance Theory}]~\label{Balance Theory}
    The signed network is balanced if and only if (i) all the edges are positive, or (ii) the node set can be partitioned into two mutually exclusive subsets, such that all edges within the same subset are positive and all edges between the two subsets are negative.
\end{theorem}
\begin{theorem}[\textbf{Weak Balance Theory}]~\label{Weak Balance Theory}
    The signed network is weakly balanced if and only if (i) all the edges are positive, or (ii) all nodes can be partitioned into $K\in \mathbb{N}^{+}$ disjoint sets, such that positive edges exist only within clusters, and negative edges exist only between clusters.
\end{theorem}
Social Balance Theory induces four fundamental principles: ``\textit{the friend of my friend is my friend (FFF)}'', ``\textit{the enemy of my friend is my enemy (EFE)}'', ``\textit{the friend of my enemy is my enemy (FEE)}'', and ``\textit{the enemy of my enemy is my friend (\textit{EEF})}''. A signed network is balanced if it does not violate these principles. For example, triads with an even number of negative edges are balanced, as shown by the first two triads in Fig.~\ref{app_fig:triads_theories}(a), which have 0 and 2 negative edges, respectively. These principles are traditionally applied to $2$-way clustering. 
% To accommodate $K$-way clustering,  proposes Weak Balance Theory, a relaxed version of Social Balance Theory. 
Conceptually, Weak Balance Theory replaces the ``\textit{EEF}'' principle in Social Balance Theory with ``\textit{the enemy of my enemy might be my enemy} (\textit{EEE})''. Accordingly, the first three triads in Fig.~\ref{app_fig:triads_theories}(b) are considered weakly balanced. 
% Importantly, the ``\textit{EEE}'' principle, applicable for $K$-way ($K>2$) clustering, allows nodes in a triangle to belong to three different clusters (e.g., the blue triangle in Fig.~\ref{app:balance_theory}(b)), illustrating a relaxation of the stricter Social Balance Theory.
% \textbf{Comparison of Social Balance Theory and Weak Balance Theory.} The partition $\{\mathcal{C}_1,\dots,\mathcal{C}_K\}$ of a signed graph $\mathcal{G}$ satisfying both theories can be uniformly defined such that the following conditions hold: 
% \begin{equation}
%     \begin{cases}
%        \textbf{A}_{ij}>0 & (e_{ij}\in\mathcal{E}) \cap(v_i \in \mathcal{C}_k) \cap (v_j \in \mathcal{C}_k)\\
%        \textbf{A}_{ij}<0 & (e_{ij}\in\mathcal{E}) \cap(v_i \in \mathcal{C}_k) \cap (v_j \in \mathcal{C}_l) (k\ne l)\\
%     \end{cases}
% \end{equation}
%TODO: k, l是否可以等于 K？可以
% where $\mathbf{A}_{ij}$ is the weight of edge $e_{ij}$ and $0<k,l<K$.
% Weak Balance Theory in Theorem \ref{Weak Balance Theory} generalized \textit{Social Balance} to $K$-way ($K>2$) clustering, relaxing the node relationship by \textit{allowing} "the enemy of my enemy is still my enemy". 
However, the ``\textit{EEE}'' principle is specific to $K$-clusterable ($K>2$) networks (e.g., Fig.~\ref{app:balance_theory}(b)) and does not appear in $2$-clusterable systems (Fig.~\ref{app:balance_theory}(a)). 
% Recent literature~has primarily leveraged Social Balance Theory principles to improve node representations for signed graphs, potentially overlooking the broader applicability of Weak Balance Theory in datasets with more than $2$ antagonistic groups, especially when explicit labels are lacking. Our work aims to fully explore Weak Balance Theory and its principles in the design of a signed graph encoder for $K$-way clustering.
% For example, the positive representation is partially obtained by aggregating the enemies ($v_k$) of the central node ($v_i$)'s enemy ($v_j$), which will decrease their discrimination in the latent space.
\section{Baselines}~\label{app:baselines}
DSGC is compared against $9$ representative signed spectral clustering methods. Five are basic signed spectral methods utilizing various forms of Laplacian matrix: (1) symmetric adjacency matrix $\mathbf{A}^{*}=\frac{1}{2}(\mathbf{A}+\mathbf{A}^{T})$; (2) simple normalized signed Laplacian $\overline{\mathbf{L}}_{sns}=\overline{\mathbf{D}}^{-1}(\mathbf{D}^{+}-\mathbf{D}^{-}\mathbf{A}^{*})$; (3) balanced normalized signed Laplacian $\overline{\mathbf{L}}_{bns}=\overline{\mathbf{D}}^{-1}(\mathbf{D}^{+}-\mathbf{A}^{*})$; (4) signed Laplacian graph $\overline{\mathbf{L}}=\overline{\mathbf{D}}-\mathbf{A}$ with a diagonal matrix $\overline{\mathbf{D}}$; and (5) its symmetrically normalized version $\mathbf{L}_{sym}$~\cite{kunegis2010spectral}. Two are $K$-way spectral clustering methods: (6) Balanced Normalized Cut (BNC) and (7) Balanced Ratio Cut (BRC)~\cite{chiang2012scalable}. The last two~ \cite{cucuringu2019sponge} are two generalized eigenproblem formulations: (8) SPONGE and (9) SPONGE$_{sym}$. Moreover, we compare with $6$ state-of-the-art deep unsigned graph clustering methods: (10) DAEGC~\cite{wang2019attributed}, (11) DFCN~\cite{tu2021deep}, (12) DCRN~\cite{LiuTZLSYZ22}, (13) Dink-net~\cite{00080XZYLL23}, (14) DGCLUSTER~\cite{BhowmickKHSM24} (15) MAGI~\cite{LiuLCWWZTSFMWC24}. Please refer to App.~\ref{app:setting} for hyperparameters settings and experiment details. Two deep signed graph works: (15) SiNE~\cite{wang2017signed} and (16) SNEA~\cite{li2020learning} (only their signed neural networks are adapted to DSGC). 

\section{Impact of Signed Encoder to Cluster Boundary}~\label{app:analyze_encoder}
In this section, we specifically analyze the principles of Weak Balance Theory implied in positive and negative aggregation functions (Eq.~\eqref{pos_agg} and Eq.~\eqref{neg_agg}) and the term $(-\bar{\mathbf{A}}^{-})$ in Eq.~\eqref{neg_agg} for the perspective of their impact to the node representations and the clustering boundaries.

\begin{figure}[ht!]
	\centering
		\subfigure[Eq.~\eqref{pos_agg}]{
			\includegraphics[width=0.13\textwidth]{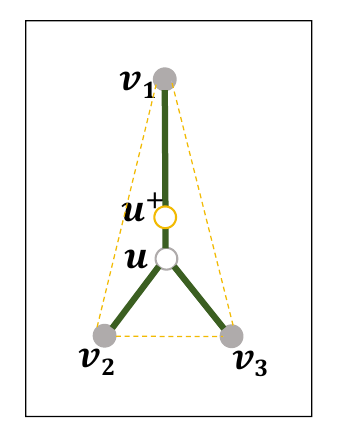}
	}
		\subfigure[Eq.~\eqref{neg_agg}]{
			\includegraphics[width=0.20\textwidth]{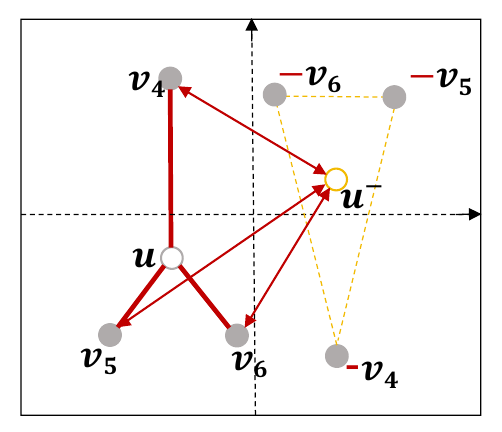}
	}
	\caption{Impact of the term $(-\bar{\mathbf{A}}^{-})$ in Eq.~\eqref{neg_agg} when both $\tau^{+}$ and $\tau^{-}$ are $0$. (a) The positive embedding $u^{+}$ is placed at the mean of its ``\textit{friend neighbors}'', including $v_1$, $v_2$, and $v_3$. (b) Due to "-" in the term, the negative embedding $u^{-}$ is placed at the mean of its ``\textit{enemy neighbors'}'' antipodal points, including $-v_4$, $-v_5$, and $-v_6$, resulting in $u^{-}$ further away from $v_4$, $v_5$, and $v_6$.}
        \label{posA_neg-A_smooth}
\end{figure}

\textbf{The term $(-\bar{\mathbf{A}}^{-})$.} The minus sign "$-$" helps push nodes linked by negative edges further apart in the latent space. For example, in Fig.~\ref{posA_neg-A_smooth} (a), the node $u$ has three ``\textit{friend neighbors}'', $v_1$, $v_2$, and $v_3$. The positive embedding $u^{+}$ of $u$ is placed at the mean of these three ``\textit{friend neighbors}'' according to Eq.~\eqref{pos_agg}, thus narrowing the distance between them and its central node $u$. In Fig.~\ref{posA_neg-A_smooth} (b), the node $u$ has three ``\textit{enemy neighbors}'', $v_4$, $v_5$, and $v_6$. "$-$" in the term ($-\bar{\mathbf{A}}^{-}$) indicates that two vertices with a negative edge should be placed on opposite sides.
Then, the negative embedding $u^{-}$ is placed at the mean of $-v_4$, $-v_5$, and $-v_6$, which are the opposite coordinates of  $v_4$, $v_5$, and $v_6$, respectively. This leads to further distance between $u$ and its ``\textit{enemy neighbors}''. As nodes linked negatively are likely to be located at the cluster boundary, pushing them away from each other will create clearer cluster boundaries, thus effectively increasing inter-cluster variances. Section~\ref{experiment_encoder} quantitatively analyzes the effect of $(-\bar{\mathbf{A}}^{-})$ on node embeddings.
\begin{figure}[t]
    \centering
    \includegraphics[width=0.78\linewidth]{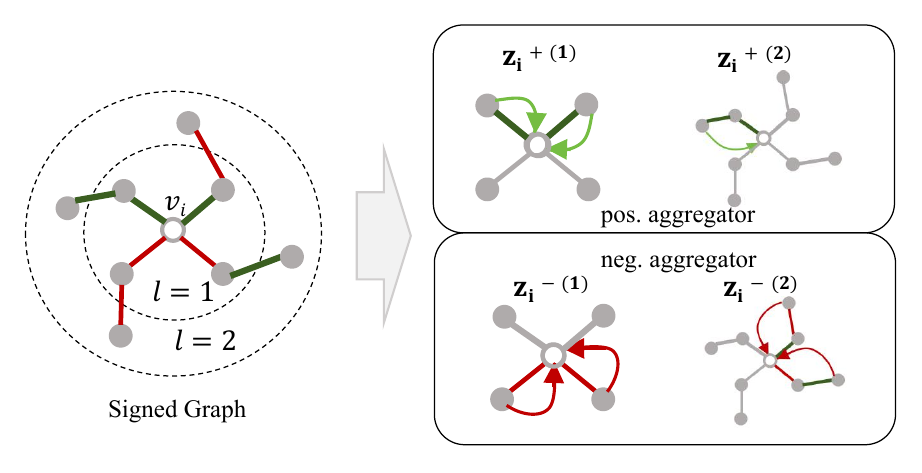}
    \caption{Illustration of positive and negative aggregations in Eq.~\eqref{pos_agg} and Eq.~\eqref{neg_agg} on a signed graph with the central node $v_i$ and its $2$-hop neighbors.}
    \label{fig:pos_neg_agg}
\end{figure}

%TODO: 字数太多下面这段和图也可以放到 Appendix
\textbf{Positive and Negative aggregation.} 
%We explain the effect of the principles implied in positive (Pos.) and negative (Neg.) aggregation functions. 
Eq.~\eqref{pos_agg} aggregates the node embeddings of all \textit{$l$-hop "friend neighbors"} along the $l$-length positive walk (Dfn~\ref{pos_neg_walk}), implying the principle ``\textit{the friend of my friend is my friend (FFF)}'' and its transitivity. It pulls \textit{``friend neighbors''} within $L$-hop toward the central node, thus reducing intra-cluster variances. Eq.~\eqref{neg_agg} aggregates the node embeddings of all \textit{$l$-hop ``enemy neighbors''} along the $l$-length negative walk (Dfn~\ref{pos_neg_walk}), implying the principles of ``\textit{the enemy of my friend is my enemy (EFE)}'',  ``\textit{the friend of my enemy is my enemy (FEE)}'', and the transitivity of ``\textit{FFF}''. Importantly, we no longer consider the specific principle ``\textit{the Enemy of my Enemy is my Friend (EEF)}'' of Social Balance so that the distance between nodes linked negatively can effectively increase, which can be verified by quantitatively comparing our DSGC and its variant that incorporates ``\textit{EEF}''. Taking a signed graph with $L=2$ as an example, Fig.~\ref{fig:pos_neg_agg} illustrates our positive and negative aggregation rules in Eq.~\eqref{pos_agg} and ~\eqref{neg_agg}.

\begin{figure}[ht!]
	\centering
  		\subfigure[ACC vs. $\delta$]{
			\includegraphics[width=0.22\textwidth]{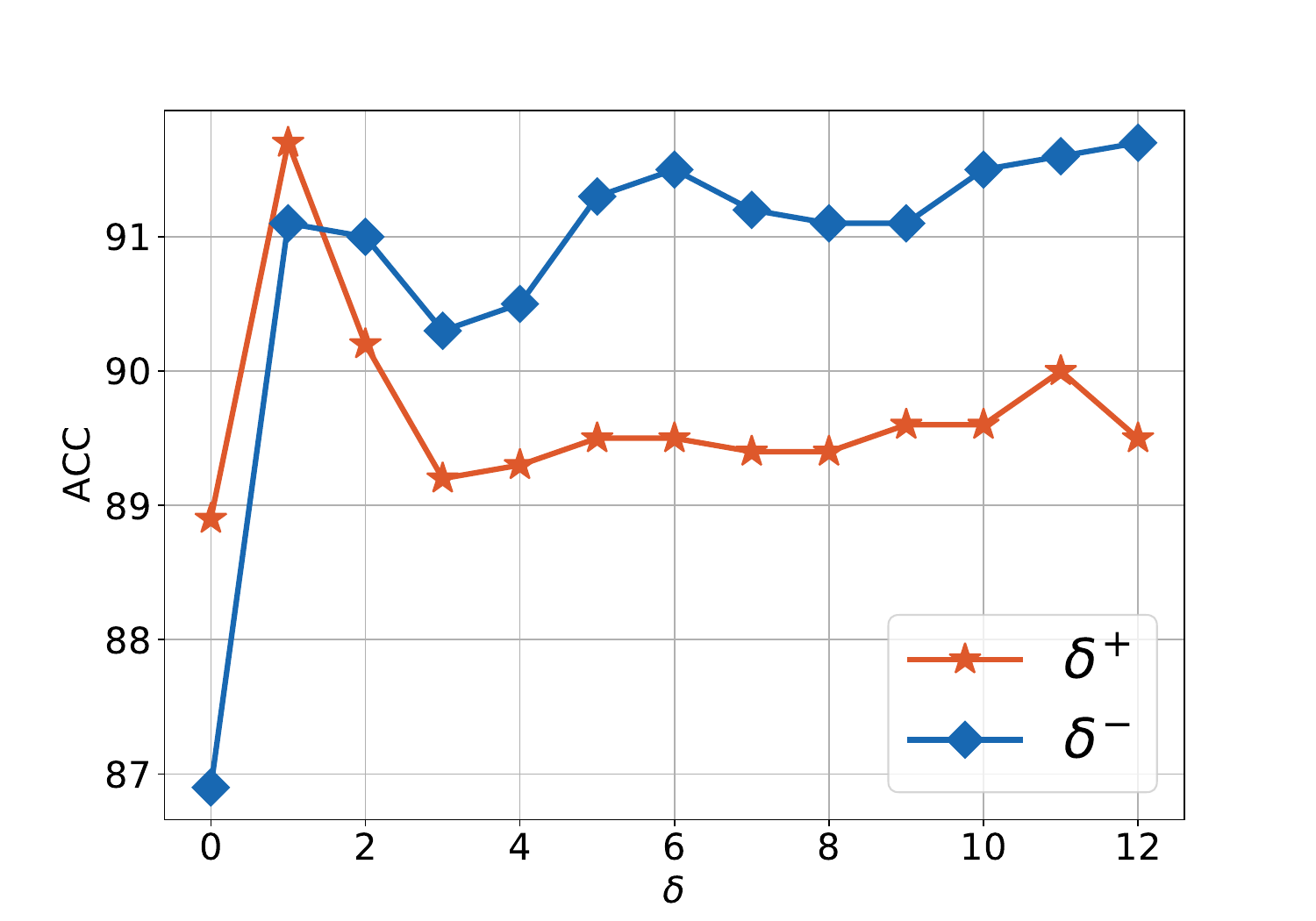}

	}
		\subfigure[ACC vs. $m$]{
			\includegraphics[width=0.22\textwidth]{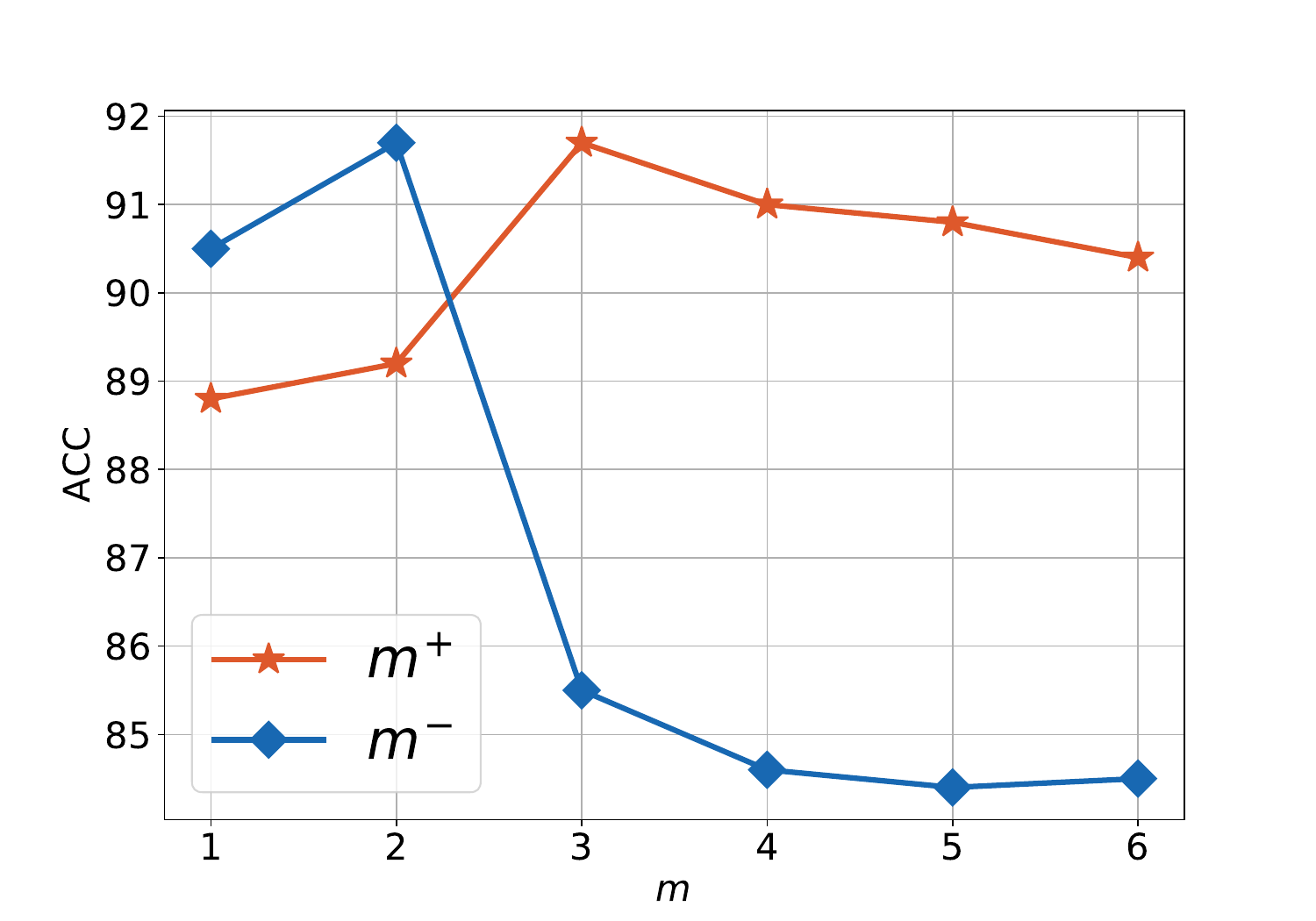}
	}
	\caption{Hyperparameter analysis on the signed graph.}
        \label{fig:hyperparameters}
\end{figure}

\begin{table*}
	\centering
	\caption{Performance comparison of signed graph clustering on SSBM with ARI (\%) and F1 (\%). Bold values indicate the best results; underlined values indicate the runner-up.}
	\scalebox{0.60}{
		\begin{tabular}{l|cc|cc|cc|cc|cc|cc|cc|cc|cc|cc}
			\toprule[1.3pt]
			\midrule[0.3pt]
			SSBM  & \multicolumn{10}{c|}{(N=1000, K=5, p=0.01, $\eta$)} & \multicolumn{10}{c}{(N=1000, K=10, p, $\eta=0.02$)}\\
			\midrule
			SSBM  & \multicolumn{2}{c|}{$\eta=0$}  & \multicolumn{2}{c|}{$\eta=0.02$} & \multicolumn{2}{c|}{$\eta=0.04$} & \multicolumn{2}{c|}{$\eta=0.06$} & \multicolumn{2}{c|}{$\eta=0.08$}  & \multicolumn{2}{c|}{p=0.01}  & \multicolumn{2}{c|}{p=0.02} & \multicolumn{2}{c|}{p=0.03} & \multicolumn{2}{c|}{p=0.04} & \multicolumn{2}{c}{p=0.05}  \\
			\midrule
			\textbf{Metrics} & ARI & F1 & ARI & F1 & ARI & F1 & ARI & F1 & ARI & F1 & ARI & F1 & ARI & F1 & ARI & F1 & ARI & F1 & ARI & F1\\
			\midrule[0.3pt]
			$\mathbf{A}$ & 41.66 & 71.58 & 35.90 & 68.26 & 28.09 & 62.14 & 12.03 & 41.74 & 9.94 & 43.98 & 0.77 & 16.13 & 3.01 & 21.32 & 16.85 & 43.04 & 60.21 & 79.80 & 86.91 & 93.91\\
			$\overline{\mathbf{L}}_{sns}$ & 0.01 & 10.17 & 0.01 & 10.36 & 0.01 & 9.01 & 0.00 & 8.36 & 0.00 & 8.08 & 0.00 & 8.17 & 0.11 & 11.34 & 1.68 & 13.48 & 3.40 & 11.20 & 15.78 & 25.01 \\
			$\overline{\mathbf{L}}_{dns}$ & 15.11 & 31.92 & 11.42 & 30.52 & 5.32 & 24.17 & 6.11 & 24.47 & 3.05 & 21.12 & 0.62 & 12.63 & 2.14 & 18.09 & 6.26 & 27.56 & 23.36 & 45.17 & 67.10 & 83.40\\
			$\overline{\mathbf{L}}$ & 0.00 & 7.29 & 0.00 & 7.29 & 0.00 & 7.29 & 0.00 & 7.29 & 0.00 & 7.29 & 0.00 & 3.22 & 0.00 & 3.22 & 0.00 & 3.22 & 0.00 & 3.21 & 0.00 & 3.22\\
			$\mathbf{L}_{sym}$ & 48.56 & 75.91 & 38.62 & 69.37 & 28.83 & 62.26 & 16.16 & 48.49 & 14.01 & 47.88 & 0.49 & 15.91 & 2.41 & 19.64 & 14.66 & 39.91 & 58.49 & 78.69 & 86.48 & 93.69\\
			BNC & 14.55 & 35.93 & 12.32 & 32.50 & 6.80 & 29.86 & 4.56 & 23.78 & 1.67 & 22.18 & 0.17 & 13.06 & 2.31 & 18.78 & 6.09 & 23.47 & 24.89 & 48.76 & 67.28 & 83.51\\
			BRC & 0.00 & 7.29 & 0.00 & 7.29 & 0.00 & 7.48 & 0.00 & 7.29 & 0.00 & 7.29 & 0.00 & 3.88 & 0.00 & 3.22 & 0.04 & 5.56 & 0.50 & 6.40 & 0.00 & 3.22\\
			SPONGE & 68.90 & 86.39 & \underline{58.65} & \underline{81.42} & \underline{41.36} & \underline{71.73} & \underline{24.28} & \underline{50.20} & \underline{17.22} & \underline{45.82} & 1.61 & 15.17 & \underline{8.25} & \underline{28.01} & 35.52 & 6.31 & 80.11 & 90.48 & 94.94 & 97.69\\
			SPONGE$_{sym}$ & \underline{71.34} & \underline{89.38} & 46.42 & 63.15 & 38.31 & 58.64 & 14.79 & 21.07 & 9.02 & 18.91 & \underline{1.64} & \underline{17.44} & 7.94 & 9.90 & \underline{67.53} & \underline{79.43} & \underline{90.97} & \underline{95.93} & \underline{97.56} & \underline{98.90}\\
                % sIR-LS & 84.80 & 93.71 & 78.65 & 91.04 & 65.03 & 84.52 & 0.00 & 7.29 & 0.00 & 7.29 & 0.00 & 2.63 & 9.02 & 31.19 & 93.79 & 97.21 & 98.89 & 99.50 & 99.78 & 99.90 \\
                % IR-LS & 84.12 & 93.39 & 78.09 & 90.70 & 67.62 & 85.81 & 0.00 & 7.29 & 0.00 & 7.29 & 0.00 & 2.63 & 0.00 & 3.02 & 78.62 & 85.87 & 98.69 & 99.40 & 99.78 & 99.90 \\
                SiNE & 0.05 & 22.62 & 0.00 & 23.25 & 0.03 & 24.08 & 0.22 & 24.94 & 0.00 & 22.24 & 0.00 & 14.02 & 0.43 & 14.54 & 0.23 & 15.16 & 0.50 & 15.16 & 0.36 & 14.92\\
                SNEA & 15.13 & 38.10 & 15.95 & 40.47 & 15.05 & 37.14 &5.67 & 28.19 & 4.38 & 27.69 & 0.65 & 9.09 & 3.11 & 12.17 & 3.32 & 13.18 & 13.01 & 21.82 & 15.80 & 22.71\\
			\midrule
                DAEGC & 3.66 & 27.98 & 5.27 & 26.81 & 3.65 & 26.57 & 3.09 & 26.55 & 2.03 & 24.29 & 0.26 & 11.99 & 0.73 & 12.78 & 1.05 & 15.40 & 2.51 & 16.14 & 4.93 & 19.97 \\
                DFCN & 4.26 & 35.07 & 3.10 & 32.73 & 2.09 & 30.42 & 1.71 & 27.97 & 1.92 & 28.51 & 0.23 & 14.74 & 0.19 & 13.85 & 0.39 & 14.39 & 0.83 & 16.48 & 0.68 & 15.83\\
                DCRN & 11.20 & 48.95 & 9.10 & 44.13 & 11.34 & 40.30 & 4.05 & 36.09 & 4.05 & 34.72 & 0.84 & 13.95 & 2.59 & 18.99 & 5.45 & 25.27 & 13.70 & 32.84 & 24.72 & 49.26 \\
                Dink-net & 1.37 & 26.40 & 0.93 & 26.12 & 1.09 & 27.03 & 0.90 & 26.43 & 1.15 & 26.22 & 0.00 & 14.37 & 0.06 & 15.08 & 0.22 & 13.80 & 0.83 & 15.85 & 1.54 & 15.56 \\
                DGCLUSTER & 0.00 & 7.29 & 0.00 & 7.29 & 0.00 & 7.29 & 0.00 & 7.29 & 0.00 & 7.29 & 0.00 & 2.63 & 0.00 & 3.02 & 0.00 & 3.22 & 0.00 & 3.22 & 0.00 & 3.02 \\
                MAGI & 9.47 & 39.24 & 6.69 & 28.48 & 4.42 & 28.25 & 4.03 & 30.40 & 3.09 & 26.75 & 0.79 & 13.83 & 1.03 & 14.45 & 2.59 & 19.62 & 4.06 & 20.18 & 7.98 & 29.16 \\
			% DAEGC & 58.40 & 82.16 & 51.72& 79.12 & 43.69 & 73.32 & 25.98 & 55.12 & 19.77 & 52.32 &\underline{2.76} & \underline{20.91} & 22.10 & 46.78 &56.59 & 77.23 & 89.33 & 95.11 & 96.68 & 98.50\\
			% DFCN & 67.40 & 86.11 & 55.48 & 80.14 & 43.25 & 72.81 & 23.43 & 57.19  & 16.91 & 49.34 & 1.73 & 22.08 & 11.08 & 34.31 & 57.57 & 78.83 & 91.18 & 96.02 & 97.33 & 98.80\\
			% DCRN & \underline{74.62} & 89.09 & \underline{60.38} & \underline{82.19} & \underline{49.48} & \underline{76.63} & \underline{30.91} & \underline{64.11} & \underline{21.02} & \underline{55.04} & 1.38 & 20.47 & \underline{22.29} & \underline{51.91} & \underline{68.21} & \underline{85.06} & \underline{93.17} & \underline{96.21} & \underline{98.22} & \underline{99.20}\\
			\midrule
			DSGC &\textbf{88.50} & \textbf{94.30} & \textbf{78.30} & \textbf{90.80} & \textbf{61.60} & \textbf{82.80} & \textbf{34.40} & \textbf{66.40} & \textbf{23.50} & \textbf{57.60} & \textbf{6.90} & \textbf{29.10} & \textbf{38.40} & \textbf{65.30} & \textbf{70.40} & \textbf{85.30} & \textbf{93.20} & \textbf{96.90} & \textbf{98.40} & \textbf{99.40}\\ 
			\midrule[0.8pt]
			SSBM & \multicolumn{10}{c|}{($N$, $K=5$, $p=0.01$, $\eta=0$)} & \multicolumn{10}{c}{($N=1000$, $K$, $p=0.01$, $\eta=0.02$)}\\
			\midrule
			SSBM  & \multicolumn{2}{c|}{$N=300$}  &\multicolumn{2}{c|}{$N=500$} & \multicolumn{2}{c|}{$N=800$} & \multicolumn{2}{c|}{$N=1000$} & \multicolumn{2}{c|}{$N=1200$} & \multicolumn{2}{c|}{K=4}  & \multicolumn{2}{c|}{K=5} & \multicolumn{2}{c|}{K=6} & \multicolumn{2}{c|}{K=7} & \multicolumn{2}{c}{K=8} \\
			\midrule
			\textbf{Metrics} & ARI & F1 & ARI & F1 & ARI & F1 & ARI & F1 & ARI & F1 & ARI & F1 & ARI & F1 & ARI & F1 & ARI & F1 & ARI & F1\\
			\midrule[0.3pt]
			$\mathbf{A}$ & 0.64  & 25.13  & \underline{6.24}  & 35.70 &  15.11  & 51.42 & 41.66 & 71.58 & 62.24 & 83.10 & 76.81 & 90.69 & 35.90 & 68.26 & 6.44 & 31.68 & 3.45 & 24.20 & 1.48 & 19.77 \\
			$\overline{\mathbf{L}}_{sns}$ & 0.00  & 11.42  & 0.01  & 11.00  & 0.02 & 9.15 & 0.01 & 10.17  & 0.04 & 10.54  & 0.00 & 10.81 & 0.01 & 10.36 & 0.00 & 7.98 & 0.01 & 9.90 & 0.07 & 11.16  \\
			$\overline{\mathbf{L}}_{dns}$ & 0.01  & 9.36  &  0.27 & 17.36  &  0.46 & 15.22   & 15.11 & 31.92 & 16.88 & 49.98  & 32.63 & 53.01 & 11.42 & 30.52 & 2.10 & 18.55& 0.77 & 15.11 & 0.96 & 16.26\\
			$\overline{\mathbf{L}}$ & 0.00  & 9.95  & 0.00  & 12.35  & 0.00 & 7.44 & 0.00   & 7.29  & 0.00 & 7.18 & 0.00 & 10.42 & 0.00 & 7.29 & 0.00 & 5.59 & 0.00 & 4.98 & 0.00 & 3.59 \\
			$\mathbf{L}_{sym}$ & 0.02  & 8.71  & 4.86  & 29.81  & 19.16 & 50.35   & 48.56   & 75.91  & 63.80 & 83.91 & 79.31 & 91.79 & 38.62 & 69.37 & 6.17 & 31.87 & 4.93 & 26.51 & 1.96 & \underline{20.80}\\
			BNC & 0.00 & 11.06 & 0.62  & 18.09  & 0.60 & 15.88   & 14.55   & 35.93 & 21.13 & 53.83 & 30.89 & 53.48 & 12.32 & 32.50 & 3.51 & 20.71 & 0.65 & 15.29 & 1.34 & 16.76 \\
			BRC & 0.00 &  8.71 & 0.00  & 8.92   & 0.00   & 7.69 & 0.00 & 7.29 & 0.00 & 7.18 & 0.00 & 10.83 & 0.00 & 7.29 & 0.00 & 5.76 & 0.00 & 4.59 & 0.00 & 3.79 \\
			SPONGE & 0.01 & 9.38  & 3.72  & 20.06   & 25.22   & 58.50   & 68.90 & 86.39 & \underline{87.28}  & \underline{94.75} & \underline{88.84} & \underline{95.71} & \underline{58.65} & \underline{81.42} & 16.74 & 35.71 & \underline{15.30} & \underline{42.37} & 3.86 & 19.64\\
			SPONGE$_{sym}$ & 0.85 & 22.87  &  4.65  & 30.32   & \underline{58.48}  & \underline{83.23} & \underline{71.34} & \underline{89.38} & 78.18 & 91.82 & 86.32 & 94.70 & 46.42 & 63.15 & \underline{31.01} & \underline{65.55} & 9.08 & 30.05 & \underline{3.87} & 19.21\\
            SiNE & 0.03 & 21.54 & 0.46 & 26.24 & 0.29 & 24.12 & 0.05 & 22.62 & 0.62 & 23.81 & 0.00 & 27.51 & 0.00 & 23.25 & 0.00 & 20.59 & 0.07 & 17.49 & 0.22 & 16.26\\
            SNEA & 2.35 & 22.23 & 5.32 & 27.82 & 7.58 & 34.76 & 15.13 & 38.10 & 27.66 & 44.02 & 47.35 & 56.51 & 15.92 & 40.66 & 4.34 & 20.46 & 4.55 & 19.66 & 1.79 & 13.15\\
                % sIR-LS & 0.01 & 8.05 & 0.00 & 8.03 & 0.00 & 7.44 & 84.80 & 93.71 & 96.92 & 98.50 & 93.40 & 97.51 & 78.65 & 91.04 & 0.00 & 5.39 & 0.00 & 4.19 & 0.00 & 3.39\\
                % IR-LS & 0.01 & 8.05 & 0.00 & 8.03 & 4.78 & 27.80 & 84.12 & 93.39 & 96.07 & 98.42 & 93.40 & 97.51 & 78.09 & 90.70 & 0.00 & 5.39 & 0.00 & 4.19 & 0.00 & 3.39\\
			\midrule
                DAEGC & 0.63 & 21.81 & 0.68 & 22.89 & 1.72 & 22.66 & 3.66 & 27.98 & 6.74 & 29.28 & 6.47 & 36.34 & 5.27 & 26.81 & 1.27 & 19.82 & 0.85 & 15.16 & 0.80 & 15.89\\ 
                DFCN & 1.23 & 26.14 & 3.94 & \underline{36.00} & 2.75 & 31.70 & 4.14 & 34.61 & 2.34 & 30.45 & 6.61 & 44.18 & 3.10 & 32.73 & 0.72 & 23.27  & 0.40 & 19.14 & 0.47 & 17.23\\
                DCRN & 0.67 & 26.21 & 3.59 & 33.40 & 3.50 & 30.88 & 11.20 & 48.95 & 18.67 & 50.84 & 29.98 & 68.75 & 9.10 & 44.13 & 6.09 & 31.83 & 2.61 & 24.23 & 1.10 & 17.04\\
                Dink-net & 0.70 & 28.18 & 1.33 & 25.38 & 0.97 & 27.12 & 1.37 & 26.40 & 1.72 & 29.36 & 2.68 & 32.11 & 0.93 & 26.12 & 0.60 & 21.79 & 0.40 & 18.95 & 0.28 & 16.55 \\
                DGCLUSTER & 0.01 & 8.05 & 0.00 & 8.30 & 0.00 & 7.44 & 0.00 & 7.29 & 0.00 & 7.18 & 0.00 & 10.42 & 0.00 & 7.29 & 0.00 & 5.39 & 0.00 & 4.19 & 0.00 & 3.39 \\
                MAGI & \underline{3.99} & \underline{31.39} & 3.74 & 35.09 & \underline{4.55} & 30.30 & 9.32 & 38.42 & 13.19 & 42.20 & 13.64 & 42.45 & 6.74 & 27.46 & 1.52 & 22.54 & 1.49 & 16.92 & 0.46 & 15.46 \\
			% DAEGC & 2.16 & 31.29 & 7.99 & 39.47 & 32.62 & 57.21 & 58.40 & 82.16 & 81.26 & 93.43 & 78.64 & 92.54 & 54.21 & 79.72 & 18.65 & 47.14 & 16.27 & 44.18 & 6.78 & 30.94\\
			% DFCN & 6.51 & 30.82 & \underline{13.74} & 43.20 & 37.70 & 69.88 & 67.40 & 86.11 & 88.63 & 95.93 & 84.53 & 93.88 & 55.48 & 80.14 & 22.06 & 55.15 & 8.03 & 34.55 & 5.08 & 31.48\\
			% DCRN & \underline{5.36} & \underline{32.80} & 12.87 & \underline{44.89} & 45.55 & 72.91 & \underline{74.62} & \underline{89.09} & \underline{91.20} & \underline{96.42} & 88.20 & 95.39 & \underline{62.41} & \underline{83.34} & 26.10 & 58.29 & 12.17 & \underline{42.65} & \underline{6.77} & \underline{32.84}\\
			\midrule
			DSGC &\textbf{6.80} & \textbf{37.40} & \textbf{34.10} & \textbf{66.60} & \textbf{75.40} & \textbf{89.30} & \textbf{88.50} & \textbf{95.30} & \textbf{96.10} & \textbf{98.40} & \textbf{93.20} & \textbf{97.40} & \textbf{78.30} & \textbf{90.80} & \textbf{44.40} & \textbf{70.50} & \textbf{21.10} & \textbf{50.90} & \textbf{10.00} & \text{35.90}\\ 
			\bottomrule[1.3pt]
	\end{tabular}}
	\label{tab:eta_ari_f1}
\end{table*}
\section{Hyperparameter sensitivity analysis}~\label{hyperparam_sen_analy}
This section explores the sensitivity of DSGC's performance to variations in its hyperparameters, specifically focusing on $\delta^{+}$, $\delta^{-}$, $m^{+}$, and $m^{-}$. $\delta^{+}$, $\delta^{-}$ are thresholds that determine the confidence levels for nodes being classified as \textit{effective friends} or \textit{effective enemies}, respectively. $m^{+}$ and $m^{-}$ control the augmentation of positive and negative edge densities within and across clusters, respectively. We used synthetic signed graphs from SSBM ($1000$, $5$, $0.01$, $0.02$) for this analysis. The results illustrated in Fig.~\ref{fig:hyperparameters} show that: (i) Optimal performance is achieved when both $\delta^{+}$ and $\delta^{-}$ are set to $1$. Increasing $\delta^{+}$ generally worsens accuracy (ACC) as less noisy edges are effectively refined. (ii) Both excessively high and low values of $m^{+}$ degrade clustering performance due to imbalances in capturing local versus more extended neighborhood information. Setting $m^{+}$ to $3$ and $m^{-}$ to $2$ achieve optimal performance.

\section{Visualization of VS-R on Spectral Methods}~\label{app:vis_VS-R}
\begin{figure}[ht!]
	\centering
		\subfigure[BNC]{
			\includegraphics[width=0.14\textwidth]{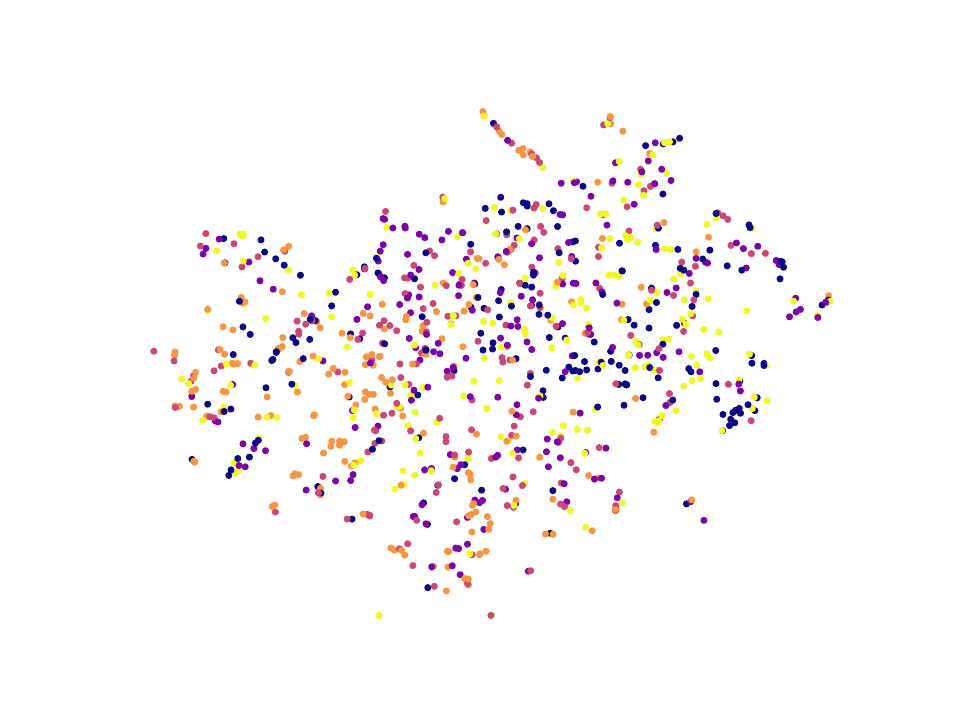}
	}
 		\subfigure[SPONGE]{
			\includegraphics[width=0.14\textwidth]{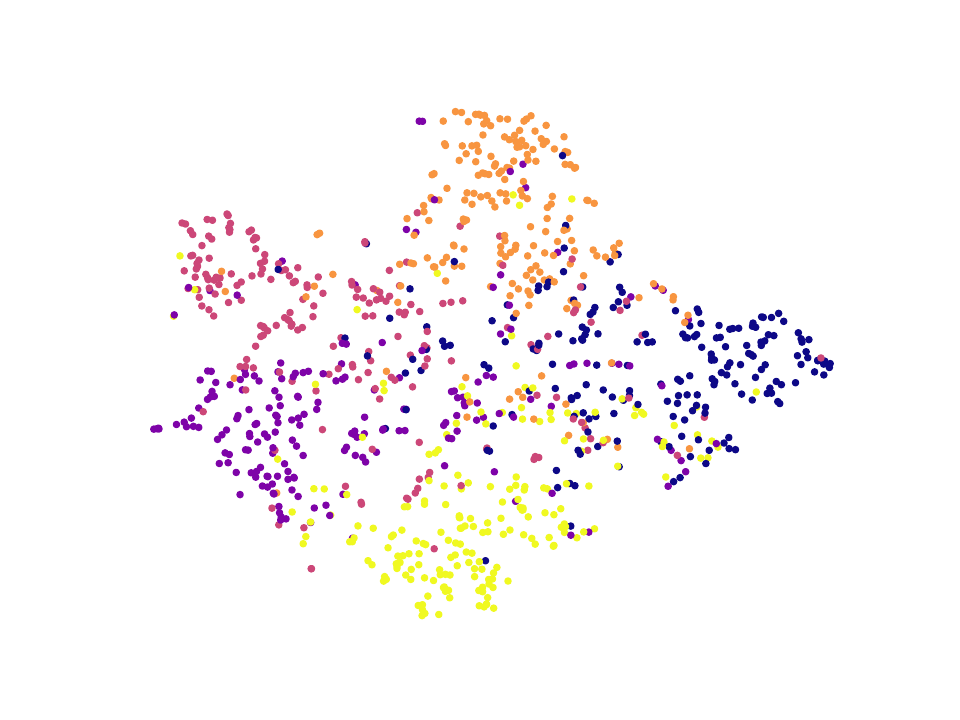}
   	}
            \subfigure[SPONGE$_{sym}$]{
			\includegraphics[width=0.14\textwidth]{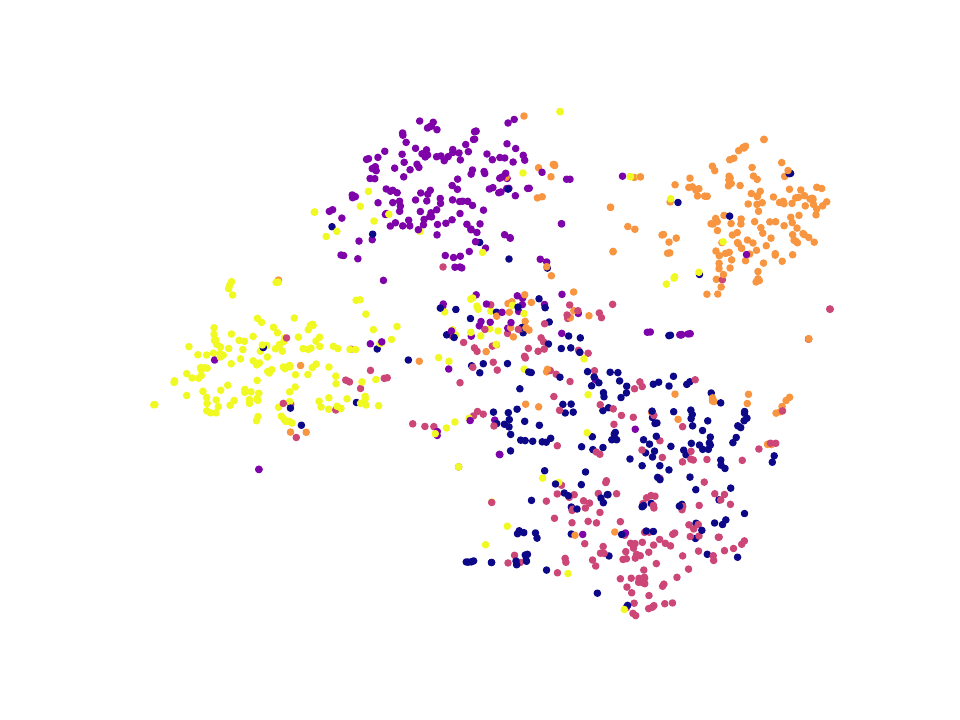}
	}\\
            \subfigure[BNC+VS-R]{
			\includegraphics[width=0.14\textwidth]{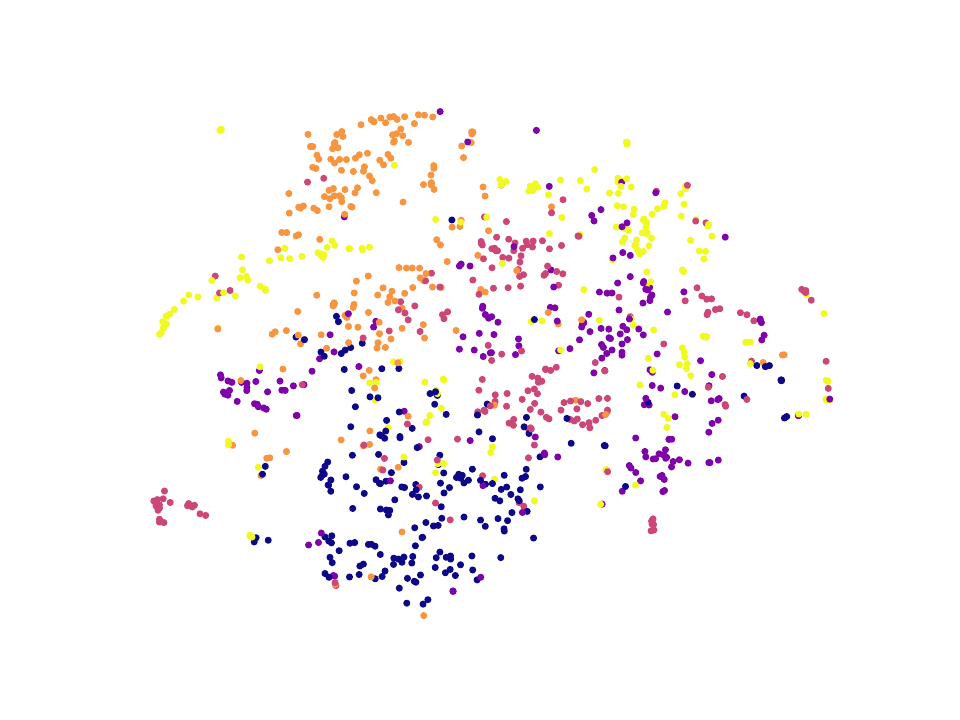}
	}
            \subfigure[SPONGE+VS-R]{
			\includegraphics[width=0.14\textwidth]{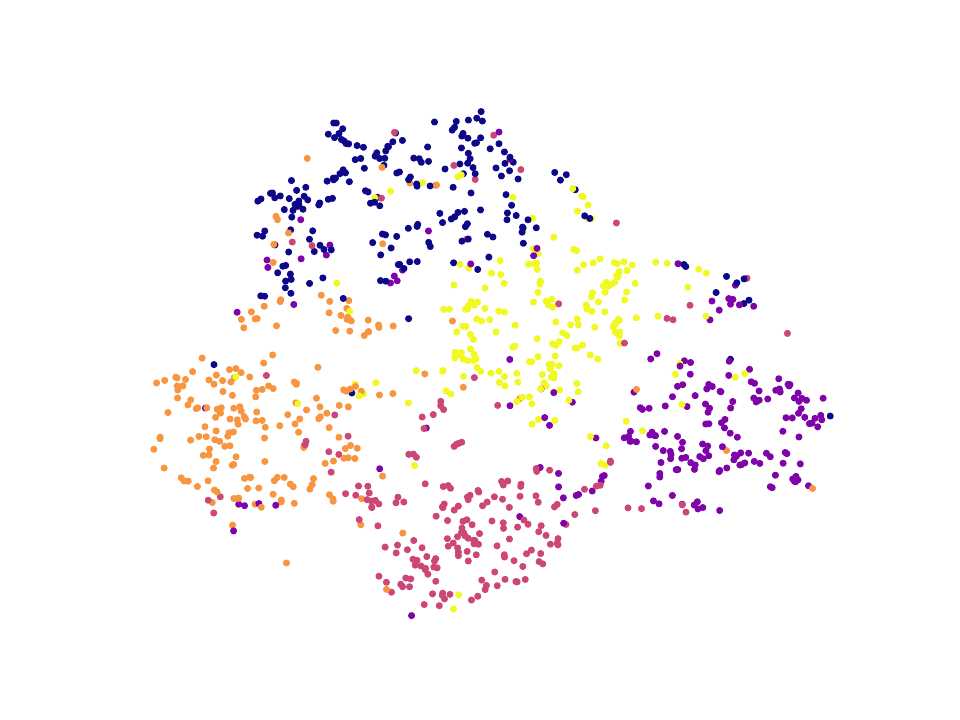}
	}
		\subfigure[SPONGE$_{sym}$+VS-R]{
			\includegraphics[width=0.14\textwidth]{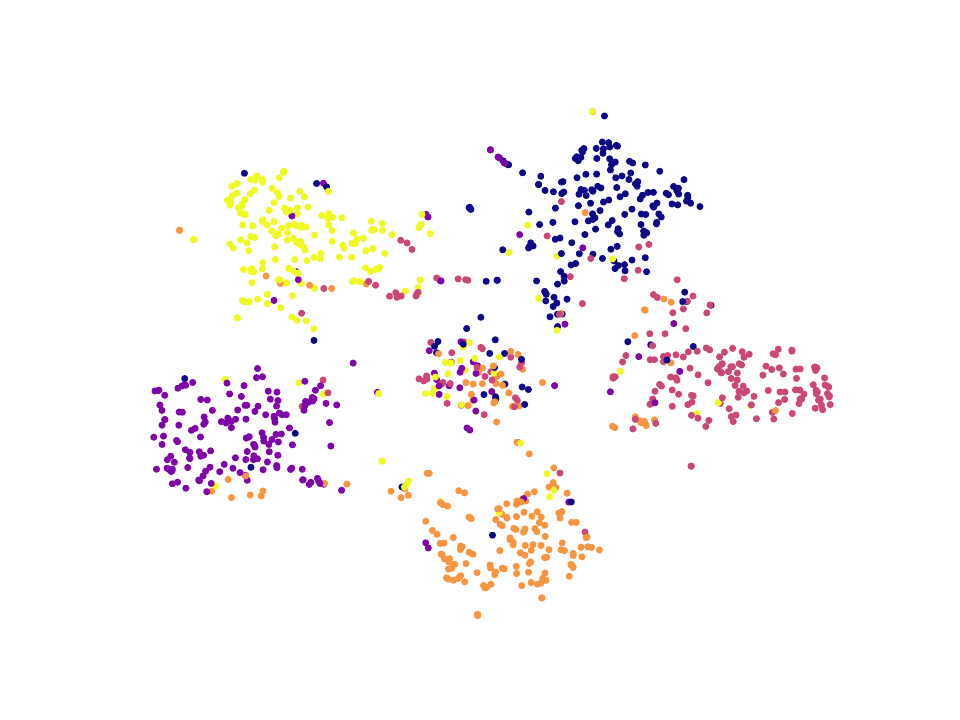}
	}

	\caption{Node embeddings of three signed spectral methods before and after applying VS-R. Different colors represent different clusters.}
        \label{fig:visua_specemb_wsigncorrection}
\end{figure}

In addition to numerical analysis, Fig.~\ref{fig:visua_specemb_wsigncorrection} provides visual evidence of the impact of VS-R. Utilizing $t$-SNE, we compared the embeddings of original and denoised graphs learned by strong spectral methods on SSBM ($1000$, $5$, $0.01$, $0.04$). We can observe that the embeddings of new graphs, displayed in the bottom row, exhibit clearer clustering boundaries than those of the original graphs in the top row. That is, spectral methods, including BNC, SPONGE, and SPONGE$_{sym}$, achieve enhanced clustering performance on cleaner graph structures after employing VS-R.  

\section{Implementation Setting}~\label{app:setting}
All experiments are implemented on PYtorch. The length of positive and negative walks $L'$ in VS-R is set to $3$. The balance parameter $\lambda$ in loss $\mathcal{L}$ is set to $0.03$. The layer number $L$ in our graph encoder is set to $2$. Node features $\mathbf{X}$ are derived from the $K$-dimmensional embeddings corresponding to the largest $K$ eigenvalues of the symmetrized adjacency matrix. The hidden dimension $d$ is $32$ in the our cluster-specific signed graph clustering. Besides, the hyperparameters sensitivity analysisi of $\delta^{+}$, $\delta^{-}$, $m^{+}$, and $m^{-}$ in Eq.~\eqref{new_Aij} and Eq.~\eqref{eq:clean_graph_pre} can be found in App.~\ref{hyperparam_sen_analy}. Following \cite{LeeSK20}, we change a signed graph to an unsigned graph by revising all negative edges to positive edges, which are inputted to above unsigned graph clustering methods (DAEGC, DFCN, DCRN, Dink-net, DGCLUSTER MAGI).

\section{ARI and F1 of labeled datasets}~\label{sec:over_per_ari_f1}
This results in ARI and F1 score for DSGC and all baselines are reported in Table~\ref{tab:eta_ari_f1}. They are generally consistent with the ACC and NMI results presented in the main text, reinforcing the conclusions drawn from those analyses. We can observe: (i) \textit{Superior performance:} DSGC still significantly outperforms all baseline models in ARI and F1 score. (ii) \textit{Robustness:} DSGC exhibits notably superior performance on all $20$ labeled signed graphs in ARI and F1 score. This observation highlights the effectiveness and robustness of our approach regarding $\eta$, $p$, $N$, and $K$. 
% Regardless of whether the graph is dense or sparse ($p$), large or small ($N$), noisy or clean ($\eta$), and the number of clusters is few or many ($K$), DSGC maintains notably superior performance on all $20$ labeled signed graphs.
%As $\eta$ and $K$ increase, the performance of all methods decreases gradually with the performance gap remaining pronounced. As the $N$ and $p$ increase, the performance of all methods increases gradually and the extent of the performance gap becomes less pronounced. Regardless of whether the graph is dense or sparse ($p$), large or small ($N$), noisy or clean ($\eta$), and the number of clusters is few or many ($K$), DSGC exhibits notably superior performance on all $20$ labeled signed graphs. This observation highlights the effectiveness and robustness of our approach to four parameters $\eta$, $p$, $N$, and $K$. 
% Especially for SSBM ($1000$, $10$, $0.02$, $0.05$), our method DSGC achieved an ACC gain of $45.3\%$ than the runner-up ACC of $\mathbf{L}_{sym}$. 
(iii) \textit{Comparative analysis:} In terms of ARI and F1 score, while unsigned clustering methods (DAEGC, DFCN, and DCRN) generally outperform non-deep spectral methods due to their advanced representation learning capabilities, DSGC still maintains a significant advantage, confirming that the specialized design of DSGC is effective for the unique challenges of signed graph clustering.

\section{More evaluations on unlabeled datasets}~\label{app:unlabel_data}
To evaluate clusters on unlabeled datasets, we follow \cite{GhasemianHC20} to design a link prediction task by masking edges with a probability $p_{m}$. Using DSGC's output, we predict positive edges for nodes in the same cluster and negative edges otherwise. AUC is used to evaluate predictions on masked edges for the S\&P1500 and Rainfall datasets. Tables below present results for ranging from $0.1$ to $0.9$. DSGC achieves superior link prediction performance, indicating better and robust clustering results. SNEA~\cite{li2020learning} suffers from the collapse phenomenon ($50\%$ AUC) across S\&P1500 and Rainfall. That is because the convolution networks in SNEA adopt the “EEF” principle of Social Balance (suitable for 2-clustering) to aggregate neighbors, which is not appropriate for our K-way ($K>2$) clustering task.
\begin{table}[ht!]
    \centering
        \caption{AUC ($\%$) of link prediction on S\&P1500.}
    \scalebox{0.75}{
    \begin{tabular}{c|c|c|c|c|c|c|c|c|c}
        \toprule[1.3pt]
        \midrule[0.3pt]
          $p_{m}$ & 0.9 & 0.8 & 0.7 & 0.6 & 0.5 & 0.4 & 0.3 & 0.2 & 0.1\\
          \midrule[0.3pt]
          BNC & 75.32	& 74.85	& 74.96	& 75.19	& 75.11	& 75.23	& 74.90	& 78.65	& 74.32\\
          BRC & 51.92	& 51.96	& 51.97	& 51.98	& 51.91	& 51.90	& 51.91	& 51.86	& 51.85\\
          SPONGE & 57.83	& 57.71	& 59.29	& 58.38	& 59.66	& 59.88	& 59.68	& 58.48	& 65.41\\
          SPONGE$_{sym}$ & 58.52	& 60.18	& 60.93	& 59.27	& 60.98	& 59.63	& 64.15	& 65.67	& 67.02\\
          SiNE & 70.85	& 72.98	& 69.78	& 62.37	& 65.51	& 59.69	& 56.16	& 60.17	& 65.36\\
          SNEA & 50	& 50	& 50	& 50	& 50	& 50	& 50	& 50	& 50\\
          DSGC & 77.04	& 76.98	& 76.95	& 76.74	& 76.68	& 76.63	& 76.61	& 79.23	& 78.97\\
	\bottomrule[1.3pt]
    \end{tabular}}
\end{table}

\begin{table}[ht!]
    \centering
        \caption{AUC ($\%$) of link prediction on Rainfall.}
    \scalebox{0.75}{
    \begin{tabular}{c|c|c|c|c|c|c|c|c|c}
        \toprule[1.3pt]
        \midrule[0.3pt]
          $p_{m}$ & 0.9 & 0.8 & 0.7 & 0.6 & 0.5 & 0.4 & 0.3 & 0.2 & 0.1\\
          \midrule[0.3pt]
          BNC & 71.06	& 73.53	& 70.56	& 73.61	& 73.29	& 73.71	& 73.87	& 73.72	& 74.16\\
          BRC & 81.28	& 81.64	& 81.70	& 81.68	& 81.69	& 51.64	& 51.59	& 81.78	& 81.25\\
          SPONGE & 64.27	& 64.57	& 64.64	& 67.80	& 64.56	& 71.68	& 64.45	& 69.96	& 65.90\\
          SPONGE$_{sym}$ & 64.52	& 64.23	& 67.97	& 66.20	& 68.11	& 65.87	& 67.13	& 71.40	& 71.69\\
          SiNE & 61.10	& 72.55	& 67.47	& 63.81	& 73.63	& 73.67	& 72.32	& 70.40	& 57.86\\
          SNEA & 50	& 50	& 50	& 50	& 50	& 50	& 50	& 50	& 50\\
          DSGC & 82.43	& 82.44	& 82.49	& 82.41	& 82.51	& 82.49	& 82.40	& 82.53	& 82.46\\
	\bottomrule[1.3pt]
    \end{tabular}}
\end{table}

\settopmatter{printacmref=true}

\end{document}